\documentclass[11pt]{article}
\usepackage{geometry}                
\usepackage{setspace}
\usepackage{graphicx}
\usepackage{amssymb}
\usepackage{epstopdf}
\usepackage{placeins}
\usepackage[square,authoryear]{natbib}
\usepackage{lineno}
\usepackage[font=small,labelfont=bf]{caption}
\usepackage[titletoc,title]{appendix}
\usepackage{float}
\usepackage[caption = false]{subfig}
\newcommand{\ra}[1]{\renewcommand{\arraystretch}{#1}}

\mathcode`@="8000 
{\catcode`\@=\active\gdef@{\mkern1mu}}

\title{Tidal Love numbers of membrane worlds:\\ Europa, Titan, and Co.}

\author{Mikael Beuthe\\
\it Royal Observatory of Belgium,\\
\it Avenue Circulaire 3, 1180 Brussels, Belgium\\
\it E-mail: mikael.beuthe@observatoire.be}      
\date{}      

\begin{document}
\maketitle

\begin{abstract}
Under tidal forcing, icy satellites with subsurface oceans deform as if the surface were a membrane stretched around a fluid layer. `Membrane worlds' is thus a fitting name for these bodies and membrane theory provides the perfect toolbox to predict tidal effects. I describe here a new membrane approach to tidal perturbations based on the general theory of viscoelastic-gravitational deformations of spherically symmetric bodies. The massive membrane approach leads to explicit formulas for viscoelastic tidal Love numbers which are exact in the limit of zero crust thickness. Formulas for load Love numbers come as a bonus. The accuracy on $k_2$ and $h_2$ is better than one percent if the crust thickness is less than five percents of the surface radius, which is probably the case for Europa and Titan. The new approach allows for density differences between crust and ocean and correctly includes crust compressibility. This last feature makes it more accurate than the incompressible propagator matrix method. Membrane formulas factorize shallow and deep interior contributions, the latter affecting Love numbers mainly through density stratification. I show that a screening effect explains why ocean stratification typically increases Love numbers instead of reducing them. For Titan, a thin and dense liquid layer at the bottom of a light ocean can raise $k_2$ by more than ten percents. The membrane approach can also deal with dynamical tides in a non-rotating body. I show that a dynamical resonance significantly decreases the tilt factor and may thus lead to underestimating Europa's crust thickness. Finally, the dynamical resonance increases tidal deformations and tidal heating in the crust if the ocean thickness is of the order of a few hundred meters.
\end{abstract}

\vspace{\stretch{1}}

{\it
\noindent
Submitted to Icarus (www.elsevier.com/locate/icarus)}

\newpage

\tableofcontents
\newpage

\listoffigures
\listoftables
\newpage


\section{Introduction}
\label{Introduction}

Tidal Love numbers are three numbers quantifying the response of a spherically symmetric body to tides or to changes in rotation or orientation.
Their computation is required for all applications in which global deformations intervene: tidal or despinning tectonics, tidal heating, true polar wander, tidal currents in Titan's seas (applications discussed in \citet{beuthe2014} except for the last one, see \citet{tokano2014}).
Conversely, measuring Love numbers helps to constrain interior models.

Membrane worlds refer to planetary bodies with a thin shell floating on a liquid layer \citep{beuthe2014}.
`Thin shell' means here that deformations can be predicted with simple membrane equations instead of the more complicated thick shell theory.
In practice, membrane theory applies to shells having a thickness less than five to ten percents of the surface radius.
The term is thus perfectly suited to the large Galilean and Saturnian icy satellites for which electric, magnetic (including auroral), and gravity data point to the existence of a global ocean close to the surface (Table~\ref{TableOceans}).
Though observations are still lacking, Triton and Ceres are candidate membrane worlds \citep{nimmo2015,hand2015}; many smaller bodies could also enclose an ocean but are unlikely to satisfy the membrane assumption \citep{hussmann2006}.
In this paper, I choose Europa and Titan as case studies because of the available data, their potential for future missions, and their differences in internal structure and orbital period (Table~\ref{TableBulkOrbital}).

\begin{table}[ht]\centering
\ra{1.3}
\small
\caption[Crust thickness of large icy satellites]{\small
Large icy satellites: some constraints on their crust thickness $d$ (absolute and relative to the surface radius $R$) from gravity (G), magnetic (M), auroral (A), and electric (E) data.}
\begin{tabular}{@{}lllll@{}}
\hline
 & $d$ (km) &  $d/R$ (\%) & Data & Reference  \\
\hline
Europa &  $<170$ & $<11$ & G & \citet{anderson1998} \\
&  $<200$ & $<13$ & M & \citet{zimmer2000} \\
&  $<15$ & $<1$ & M & \citet{hand2007} \\
Ganymede &  $150-330$ & $6-13$& A & \citet{saur2015} \\
Callisto & $<300$ & $<12$ & M & \citet{zimmer2000} \\
Titan & $55-80$ & $2-3$ & E & \citet{beghin2012}
\vspace{0.5mm} \\
\hline
\end{tabular}
\label{TableOceans}
\end{table}%

\begin{table}[ht]\centering
\ra{1.3}
\small
\caption[Bulk and orbital parameters of Europa and Titan]{\small
Bulk and orbital parameters of Europa and Titan.
}
\begin{tabular}{@{}lllll@{}}
\hline
Parameter &  Symbol & Europa & Titan & Unit
\\
\hline
Spin rate${}^a$    & $\omega$ & $2.048$ & $0.456$ & $10^{-5}\rm\,s^{-1}$ \\
Surface radius${}^b$  & $R$ & $1560.8$ & $2574.76$ & km \\
GM${}^{a,c}$ & $GM$ & $3202.74$ & $8978.14$ & $\rm km^3 \, s^{-2}$ \\
Moment of inertia${}^d$ & MoI  & $0.346$ & $0.341$ & - \\
Bulk density${}^e$ & $\rho_b$ & $3013$ & $1881.5$ & $\rm kg \, m^{-3}$ \\
Surface gravity${}^e$ & $g$ & $1.315$ & $1.354$ & $\rm m \, s^{-2}$ \\
Dynamical parameter${}^f$ & $q_\omega$ & $4.98$ & $0.395$ & $10^{-4}$
\vspace{0.5mm} \\
\hline
\multicolumn{4}{l}{\scriptsize ${}^a$ JPL satellite ephemerides (http://ssd.jpl.nasa.gov/).}
\vspace{-1.5mm}\\
\multicolumn{4}{l}{\scriptsize ${}^b$ \citet{nimmo2007} for Europa, \citet{mitri2014} for Titan.}
\vspace{-1.5mm}\\
\multicolumn{4}{l}{\scriptsize ${}^c$ \citet{iess2010} for Titan.}
\vspace{-1.5mm}\\
\multicolumn{4}{l}{\scriptsize ${}^d$ \citet{anderson1998} for Europa, \citet{iess2012} for Titan.}
\vspace{-1.5mm}\\
\multicolumn{4}{l}{\scriptsize ${}^e$ computed from $GM$ and $R$ ($G=6.674\times10^{-11}\rm\,m^3kg^{-1}s^{-2}$).}
\vspace{-1.5mm}\\
\multicolumn{4}{l}{\scriptsize ${}^f$ computed from Eq.~(\ref{qomega}).}
\end{tabular}
\label{TableBulkOrbital}
\end{table}

In a previous paper, I obtained analytical formulas for tidal Love numbers using thin shell theory in the membrane limit \citep{beuthe2014}.
Although the method was by and large successful (especially regarding depth-dependent crustal rheology), it was lacking in some respects.
First, it required that the floating shell be of the same density as the underlying ocean.
In the membrane limit (shell of vanishing thickness), this is equivalent to assuming that the membrane is massless.
I will thus call this method the \textit{massless membrane approach}.
Second, accurate benchmarking of the tilt factor formula revealed a mismatch associated with shell compressibility.
Apparently, the classical equations of thin shell theory are not completely satisfactory regarding their dependence on compressibility.
For these two reasons, I develop in this paper an alternative membrane formalism, called the \textit{massive membrane approach}, which is based on the viscoelastic-gravitational equations used to predict tidal deformations in thick shell theory.
These equations have been extensively validated through their accurate prediction of the frequency spectrum of Earth normal modes \citep{dahlen1999}.

Though technically complex, the massive membrane approach is based on two simple ideas.
The first idea consists in using the viscoelastic-gravitational equations in order to propagate the three unknown Love numbers from the surface to the crust-ocean boundary, where they must satisfy two conditions called \textit{free-slip} and \textit{fluid constraint}.
This procedure results in two relations between tidal Love numbers, the $l_n-h_n$ and $k_n-h_n$ \textit{relations}, which depend on the \textit{effective viscoelastic parameters} of the crust.
The second idea consists in factorizing the shallow interior from the deep interior:
in the static limit of equilibrium tides, the Love numbers of the body with its viscoelastic crust are expressed in terms of the Love numbers of a simpler model (or \textit{fluid-crust model}) in which the crust is fluid-like.
Combining these two ideas leads to explicit formulas for Love numbers in terms of crustal parameters and of the deep interior structure. 
If tides are dynamical, fluid-crust models must be given up but it remains possible to derive membrane formulas for Love numbers in a model with an infinitely rigid mantle.
I will show that a dynamical resonance increases surface deformations and tidal heating in the crust as the ocean becomes shallower.

\begin{figure}
   \centering
   \includegraphics[width=7cm]{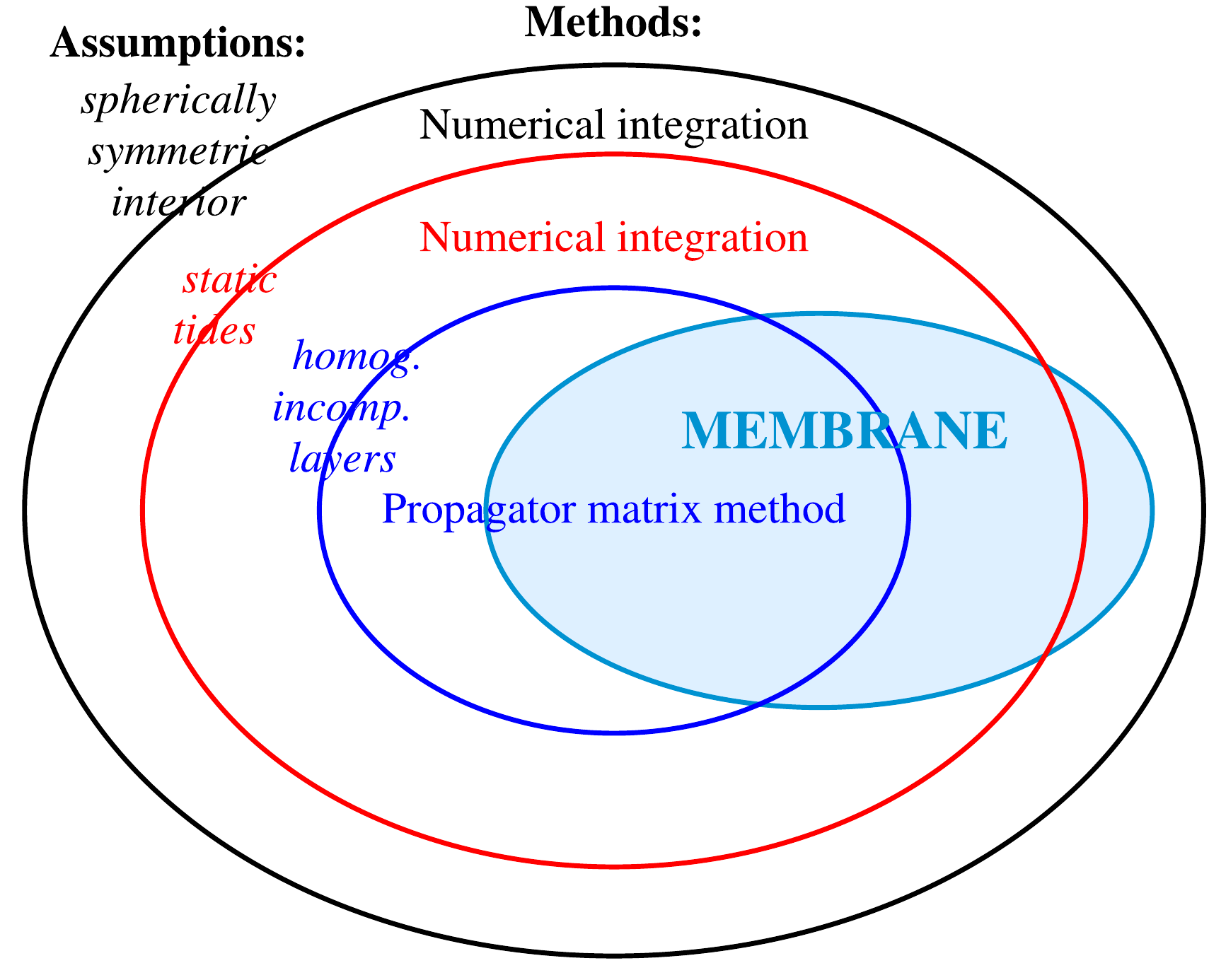}
   \caption[Domain of validity of the membrane approach]
   {Domain of validity of the membrane approach in comparison with other methods (`homog.\ incomp.' = homogeneous and incompressible).}
   \label{FigVenn}
\end{figure}

The massive membrane approach is more than `yet another method' for computing Love numbers.
It has the interesting feature that there is an overlap, but no coincidence, between the domains of validity of the membrane approach and of the standard methods (Fig.~\ref{FigVenn}).
To be more clear, three successive assumptions are common when computing Love numbers.
First, the interior structure of the undeformed body is assumed to be spherically symmetric.
Without this assumption, Love numbers do not really make sense although the Love number concept is sometimes extended to flattened bodies in rotation \citep{wahr1981}.
Numerical integration methods must be used if no other assumption is made \citep[e.g.][]{tobie2005}.
Second, the static limit is often applied because numerical codes tend to diverge at tidal periods if the body contains a fluid layer (tides are particularly slow on Titan).
For example, \citet{wahr2006}, \citet{rappaport2008}, and \citet{wahr2009} use a code assuming the static limit in all layers whereas \citet{mitri2014} only apply the static assumption to the ocean.
Numerical integration remains necessary in the static limit.
Third, the interior structure is often discretized as an onion-like superposition of incompressible and homogeneous layers.
These rather strong assumptions lead to the \textit{incompressible propagator matrix method} \citep[e.g.][]{sabadini2004} which provides analytic solutions for two- or three-layer models while models with more layers are easily solved numerically (the propagator matrix method also exists in a dynamical and compressible version which is seldom used for reasons explained in Appendix~\ref{DynamicalPropagationMatrix}).
The matrix method is stable when solid layers become fluid-like, contrary to most numerical codes.
These qualities make it popular in planetology \citep[e.g.][]{moore2000,hussmann2002,roberts2008,jaraorue2011}.
By contrast, the membrane approach is based on the thin shell approximation, but it does not require incompressible and homogeneous layers.
Compared to the propagator matrix method, the massive membrane approach is simultaneously more restrictive (requiring a thin shell) and more general (allowing for compressibility).
As dynamical effects can be included in some cases, one could say the same with respect to codes computing static Love numbers by numerical integration.

\section{Viscoelastic-gravitational theory}

This section reviews the basics of viscoelastic-gravitational theory that are used in the membrane approach.

\subsection{$y_i$ functions and Love numbers}

Viscoelastic-gravitational theory describes the deformations of a self-gravitating body with a spherically symmetric internal structure.
The deformations can result from tides, rotational flattening, surface loading or free oscillations due to an (earth)quake.
In the standard formalism of \citet{alterman1959}, the equations of motion and Poisson's equation form a set of six differential equations of first order, the solutions of which are six radial functions $y_i$ ($i=1,...,6$).
Following the conventions of \citet{takeuchi1972}, $y_1$ and $y_3$ are scalars associated with radial and tangential displacements, respectively,
\begin{equation}
\left( u_r , u_\theta , u_\phi \right) = \left( y^{}_1 \, , y^{}_3 \, \frac{\partial }{\partial \theta} \, ,  \frac{y^{}_3}{\sin\theta} \, \frac{\partial }{\partial \phi} \right) U \, ,
\end{equation}
while $y_2$ and $y_4$ are scalars associated with the stresses having a radial component:
\begin{equation}
\left( \sigma_{rr}, \sigma_{r\theta} , \sigma_{r\phi} \right) =  \left( y^{}_2 \, , y^{}_4 \frac{\partial}{\partial \theta} \,  ,  \frac{y^{}_4}{\sin\theta} \, \frac{\partial}{\partial \phi} \right) U \, .
\label{y2y4def}
\end{equation}
For tidal deformations, $U$ is the tidal potential component of harmonic degree $n$ at the surface and $(r,\theta,\phi)$ are the usual spherical coordinates (radius, colatitude, longitude).
Tangential stresses can be expressed in terms of $(y_1,y_3)$ \citep[][Eq.~(79)]{takeuchi1972}.
The total gravity potential perturbation inside the body is expressed as
\begin{equation}
\Phi = y_5 \, U \, ,
\label{y5def}
\end{equation}
while $y_6$ is related to the derivative of $y_5$ (see Eq.~(\ref{EG5}) below).
The $y_i$ are continuous within the body with one exception: $y_3$ is discontinuous at fluid/solid interfaces because solid layers freely slip on fluid layers.
The variables are nondimensionalized with the ocean density $\rho$, the surface gravity $g$, and the surface radius $R$:
\begin{equation}
\left( gy_1\, , y_2/\rho \, , gy_3 \, ,  y_4/\rho \, , y_5 \, , Ry_6 \right) \, .
\label{nondim}
\end{equation}

Though there are six independent solutions, only three are regular at the center, leaving only three degrees of freedom.
For tidal deformations of degree $n$, these regular solutions must be combined in order to satisfy three boundary conditions at the surface \citep{saito1974}:
\begin{equation}
\left( y_2(R) \, , y_4(R) \, , y_6(R) \right) = \left( 0 \, , 0 \, , \frac{2n+1}{R} \right) .
\label{boundcond}
\end{equation}
The boundary conditions on $y_2$ and $y_4$ result from imposing that the surface stress has no radial component,
whereas the condition on $y_6$ results from a boundary condition on the gradient of the gravity potential.
For surface loading of degree $n$, the boundary conditions are \citep{saito1974}:
\begin{equation}
\left( y_2^L(R) \, , y_4^L(R) \, , y_6^L(R) \right) = \left( - \frac{2n+1}{3}\, \rho_b \, , 0 \, , \frac{2n+1}{R} \right) .
\label{boundcondLoad}
\end{equation}
where $\rho_b$ is the bulk density and the superscript $L$ denotes surface loading.

The response of the satellite to a tidal perturbation of degree $n$ is parameterized by the \textit{radial}, \textit{tangential}, and \textit{gravitational tidal Love numbers}:
\begin{equation}
\left( h_n \, , l_n \, , k_n \right) = \left( g  y_1(R) \, , g y_3(R) \, , y_5(R)-1 \right) .
\label{hkl}
\end{equation}
By analogy, the response of the body to surface loading of degree $n$ is characterized by three \textit{load Love numbers}:
\begin{equation}
\left( h_n' \, , l_n' \, , k_n' \right) = \left( g  y_1^L(R) \, , g y_3^L(R) \, , y_5^L(R)-1 \right) .
\label{hklLoad}
\end{equation}

In pre-spatial geodesy, measurements at the surface yielded combinations of Love numbers \citep{melchior1978,hussmann2011}, three of which are the \textit{tilt factor} $\gamma_n$ (also called \textit{diminishing factor}), the \textit{gravimetric factor} $\delta_n$, and the \textit{strain factor} $s_n$:
\begin{eqnarray}
\gamma_n &=& 1 + k_n - h_n \, ,
\label{tiltfactor} \\
\delta_n &=& \left( n + 2 h_n - \left( n+1 \right) k_n \right)/n \, ,
\label{gravifactor} \\
s_n &=& 2 h_n - n \left( n+1\right) l_n \, .
\label{strainfactor}
\end{eqnarray}
Similar combinations $(\gamma_n',\delta_n',s_n')$ can be defined for load Love numbers.
Nowadays, Doppler and range tracking of an orbiting spacecraft and satellite laser altimetry lead to independent estimates of $k_2$ and $h_2$, respectively (measurements of $l_2$ with this technique are unsatisfactory).

In this paper, I only use $\delta_n$ and $s_n$ as compact notations.
By contrast, the tilt factor is amenable to direct physical interpretation even if surface measurements are unavailable.
First, $\gamma_2$ is required to predict the height of equilibrium tides in the seas of Titan \citep{tokano2014}:
it represents the reduction that must be applied to the surface tide because the bottom of the sea also deforms (in which case it is more appropriately called the diminishing factor).
Second, $\gamma_2$ is of great interest for estimating the crust thickness of icy satellites with a subsurface ocean.
On the one hand, $\gamma_2$ is known once the tidal Love numbers $k_2$ and $h_2$ have been determined from gravity and altimetry data. 
On the other, $\gamma_2$ is approximately proportional to the crust thickness $d$ if $d/R\ll1$ \citep{wahr2006}.
Measuring the tilt factor is thus one way of estimating the crust thickness.
Compared to the Love numbers taken separately, the tilt factor has the advantage that it is less affected by uncertainties about the interior density structure.

\subsection{Viscoelastic-gravitational equations}
\label{EOM}

Consider a body whose properties vary only with the radius: density $\rho_r(r)$, unperturbed gravitational acceleration $g_r(r)$, and Lam\'e constants $\lambda(r)$ and $\mu(r)$ (the latter is also called shear modulus).
In order to facilitate comparisons with thin shell theory, I use Poisson's ratio $\nu(r)$ instead of the first Lam\'e constant $\lambda(r)$ (see Table~\ref{TableCompressibility}).
For later use, I define the \textit{compressibility factor} $\chi$ as:
\begin{equation}
\chi = \frac{1-2\nu}{1-\nu} \, .
\label{chidef}
\end{equation}
For common materials, the compressibility factor ranges from $\chi=0$ (if $\nu=1/2$ i.e.\ incompressible limit) to $\chi=1$ (if $\nu=0$).
For elastic ice, $\nu\sim1/3$ so that $\chi\sim1/2$.
Table~\ref{TableCompressibility} summarizes the relations between the various compressibility parameters ($\lambda$, $K$, $\nu$, $\chi$) and common pairs of elastic constants.

\begin{table}[ht]\centering
\ra{1.3}
\small
\caption[Compressibility parameters]{\small
Various compressibility parameters}
\begin{tabular}{@{}cccc@{}}
\hline
 &  $(\mu,\lambda)$ & $(\mu,K)$ & $(\mu,\nu)$ \\
\hline
$\lambda$ & $\lambda$ & $K- \frac{2}{3} \mu$ & $\frac{2\mu\nu}{1-2\nu}$
\vspace{1mm}  \\
$K$ & $ \lambda +\frac{2}{3} \mu$ & $K$ & $\frac{2}{3}\frac{1+\nu}{1-2\nu}\mu$
\vspace{1mm}  \\
$\nu$ & $\frac{\lambda}{2(\lambda+\mu)}$ & $\frac{3K-2\mu}{6K+2\mu}$ & $\nu$
\vspace{1mm}  \\
$\chi$ & $\frac{2\mu}{\lambda+2\mu}$ & $\frac{6\mu}{3K+4\mu}$ & $\frac{1-2\nu}{1-\nu}$
\vspace{1mm}\\
\hline
\end{tabular}
\label{TableCompressibility}
\end{table}%

For deformations of harmonic degree $n$ and angular frequency $\omega$, the viscoelastic-gravitational equations (Eq.~(82) of \citet{takeuchi1972}) can be written as
\begin{eqnarray}
y_1' &=& \frac{\chi}{2\mu} \, y_2 - \frac{1}{r} \left(1-\chi \right) \left( 2 y_1 - n \left(n+1\right) y_3 \right) ,
\label{EG1} \\
y_2' &=& -\omega^2 \rho_r \, y_1 - \frac{2}{r} \, \chi \, y_2 + \frac{1}{r^2} \left( 2\mu \, \frac{1+\nu}{1-\nu} - \rho_r g_r r \right)  \left( 2 y_1 - n \left(n+1\right) y_3 \right)
\nonumber \\
&& +\,  \frac{n(n+1)}{r} \, y_4 - \rho_r \left( y_6 - \frac{n+1}{r} \, y_5 + \frac{2}{r} \, g_r \, y_1 \right) ,
\label{EG2} \\
y_3' &=& \frac{1}{\mu} \, y_4 + \frac{1}{r} \left( y_3 - y_1 \right) ,
\label{EG3} \\
y_4' &=&  -\omega^2 \rho_r \, y_3 - \frac{1}{r} \left( 1-\chi \right)y_2 - \frac{2}{r^2} \, \mu \left( \frac{1+\nu}{1-\nu} \, y_1 - \frac{x_n+1+\nu}{1-\nu} \, y_3 \right)
\nonumber \\
&& - \,  \frac{3}{r} \, y_4 -\frac{\rho_r}{r} \left( y_5 - g_r y_1 \right) ,
\label{EG4} \\
y_5' &=& y_6 + 4\pi G \rho_r \, y_1 - \frac{n+1}{r} \, y_5 \, ,
\label{EG5} \\
y_6' &=& \frac{n-1}{r} \, y_6 + \frac{4\pi G \rho_r}{r} \left(n+1\right) \left( y_1 - n y_3 \right) ,
\label{EG6}
\end{eqnarray}
where the prime denotes a derivative with respect to radius.
The radial dependence of the functions $y_i$ and of the parameters $(\rho_r,g_r,\mu,\nu,\chi)$ is implicit.
In Eq.~(\ref{EG4}), $x_n$ is defined by
\begin{equation}
x_n = \left( n-1 \right) \left( n+2 \right) ,
\label{xn}
\end{equation}
which is the degree-$n$ eigenvalue of the operator $-(\Delta+2)$, $\Delta$ being the spherical Laplacian.

\subsection{Static versus dynamic}
\label{StaticDynamic}

Dynamical (or inertial) terms are the terms of Eqs.~(\ref{EG2}) and (\ref{EG4}) that are proportional to $\omega^2$ and which result from the acceleration term in the equations of motion.
Compared to seismic perturbations, tidal deformations are slow so that dynamical terms are expected to be small with respect to other terms.
This assertion can be verified by nondimensionalizing Eq.~(\ref{EG2}) with Eq.~(\ref{nondim}) and comparing the different terms which are of three types: dynamical, viscoelastic, and gravitational.
Dynamical terms are parameterized by the dimensionless quantity $q_\omega$:
\begin{equation}
q_\omega = \frac{\omega^2 \, R}{g} \, .
\label{qomega}
\end{equation}
Eccentricity and obliquity tides on synchronously rotating satellites have a frequency $\omega$ equal to the spin rate of the body.
In that case, $q_\omega$ is equal to the ratio of the centrifugal acceleration at the surface to the gravitational acceleration, which is a parameter (denoted $q$ or $m$) commonly used in the theory of equilibrium figures \citep[e.g.][]{schubert2004}.
Viscoelastic terms are parameterized by the nondimensional shear modulus,
\begin{equation}
\hat\mu_{ice}= \frac{\mu_{ice}}{\rho_{ice} g R} \, ,
\end{equation}
where $\mu_{ice}\sim3.5\,\mbox{GPa}$ and $\rho_{ice}\sim1000\rm\,kg/m^3$ while $(g,R)$ are given in Table~\ref{TableBulkOrbital}.
Finally, the nondimensional quantity associated with gravity terms is 1.

In the crust, dynamical terms are small with respect to viscoelastic terms if $q_\omega \ll \hat\mu_{ice}$
whereas they are small with respect to gravitational terms if $q_\omega \ll 1$.
The former constraint means that the tidal velocity $\omega{}R$ is small with respect to the seismic S-wave velocity $\sqrt{\mu_{ice}/\rho_{ice}}$.
The latter means that the tidal frequency $\omega$ is small with respect to the free oscillation frequency of an incompressible homogeneous fluid sphere of radius $R$ and surface gravity $g$ (Eq.~(\ref{EqLamb}) with $z=0$ and $\xi=1$; or Eq.~(8.188) of \citet{dahlen1999}).
For large satellites, the two constraints are similar because $\hat\mu_{ice}\sim1.7$ and $1.0$ for Europa and Titan, respectively (for smaller bodies, the viscoelastic constraint is weaker).
Since $q_\omega\ll1$ for Europa and Titan (Table~\ref{TableBulkOrbital}), it is an excellent approximation to neglect these terms in the crust as long as there is a lithosphere, i.e.\ the crust is mechanically rigid close to the surface.
The same argument holds for the mantle and the core (if it is solid) by substituting $\mu_{ice}$ and $\rho_{ice}$ with the appropriate shear modulus and density of the layer.
Dynamical effects, however, can be significant in a liquid layer (see Section~\ref{LoveDynamic}).

In a given layer, the static limit consists in setting
\begin{equation}
\omega = 0
\end{equation}
in the viscoelastic-gravitational equations.
The static limit is typically applied to the whole body \citep[e.g.][]{wahr2006} or only to the ocean in order to stabilize the solution \citep{mitri2014}.
In Section~\ref{LoveDynamic}, I will do the opposite by taking the static limit in the solid layers but not in the ocean.
This choice has the advantage of preserving the dominant dynamical effect while considerably simplifying the solution.

\subsection{Fluid layer}
\label{FluidLayer}

The case of a fluid layer (here the subsurface ocean) deserves particular attention for two reasons.
First, free slip between fluid and solid layers leads to new boundary conditions at their interfaces.
In particular, free slip at the crust-ocean boundary (of radius $R_\varepsilon$) is a crucial condition in the membrane approach:
\begin{equation}
y_4(R_\varepsilon) = 0 \, .
\label{freeslip}
\end{equation}
Second, displacements within the fluid are undetermined in the static limit, in which case only gravity variables can be propagated through the fluid.
A fluid layer is characterized by vanishing shear stress and shear modulus ($y_4=y_4'=0$ and $\mu=0$) which implies that $\nu=1/2$ (from its definition in Table~\ref{TableCompressibility}).
Within the fluid, the fourth viscoelastic-gravitational equation (Eq.~(\ref{EG4})) becomes
\begin{equation}
y_2 =  \rho \left( g_r y_1 - y_5 \right) - \rho \, \omega^2 r \, y_3 \, ,
\label{fluideqDyn}
\end{equation}
where $\rho$ is the fluid density.

In the static limit ($\omega=0$), displacement and stress become indeterminate inside the fluid layer (though the radial variables $y_1$ and $y_2$ are well-defined at fluid/solid interfaces).
Nevertheless $y_1$ and $y_2$ are constrained by Eq.~(\ref{fluideqDyn}) with $\omega=0$:
\begin{equation}
y_2 = \rho \left( g_r y_1 - y_5 \right) .
\label{fluideq}
\end{equation}
In spite of this indeterminacy, knowing the potential and its derivative in the fluid is sufficient to solve the viscoelastic-gravitational problem.
As $y_6$ depends on the displacement $y_1$, it is replaced by the variable $y_7$ which depends only on the gravitational potential and is everywhere continuous:
\begin{equation}
y_7 = y_6 + \frac{4\pi G}{g_r} \, y_2 \, .
\label{y7def}
\end{equation}
The equation for $y_6'$ is replaced by an equation for $y_7'$ involving only the variables $y_5$ and $y_7$ \citep{saito1974}.
Within the fluid layer, the gravity potential (in the static limit) is thus a superposition of two general solutions decoupled from stress and strain. 

If the fluid layer reaches the surface (surface ocean or quasi-fluid crust), the boundary condition $y_2(R)=0$ (Eq.~(\ref{boundcond})) applied to the static fluid constraint (Eq.~(\ref{fluideq})) leads to a relation between Love numbers:
\begin{equation}
k_n^\circ + 1 = h_n^\circ \, ,
\label{knhnhydrostat}
\end{equation}
where the superscript ${}^\circ$ indicates that the surface layer of the body behaves as a fluid.
This equation means that tides are in hydrostatic equilibrium:  the surface of the satellite coincides with the geoid.

\section{The crust as a membrane}
\label{CrustMembrane}

In this section, I apply the membrane approximation to the viscoelastic-gravitational equations.
As a result, crustal rheology will be described by three (if tides) or four (if surface loading) effective viscoelastic parameters.
If there is no bulk dissipation, only two of them are independent.

\subsection{Principle}
\label{Principle}

\begin{figure}
   \centering
   \includegraphics[width=6cm]{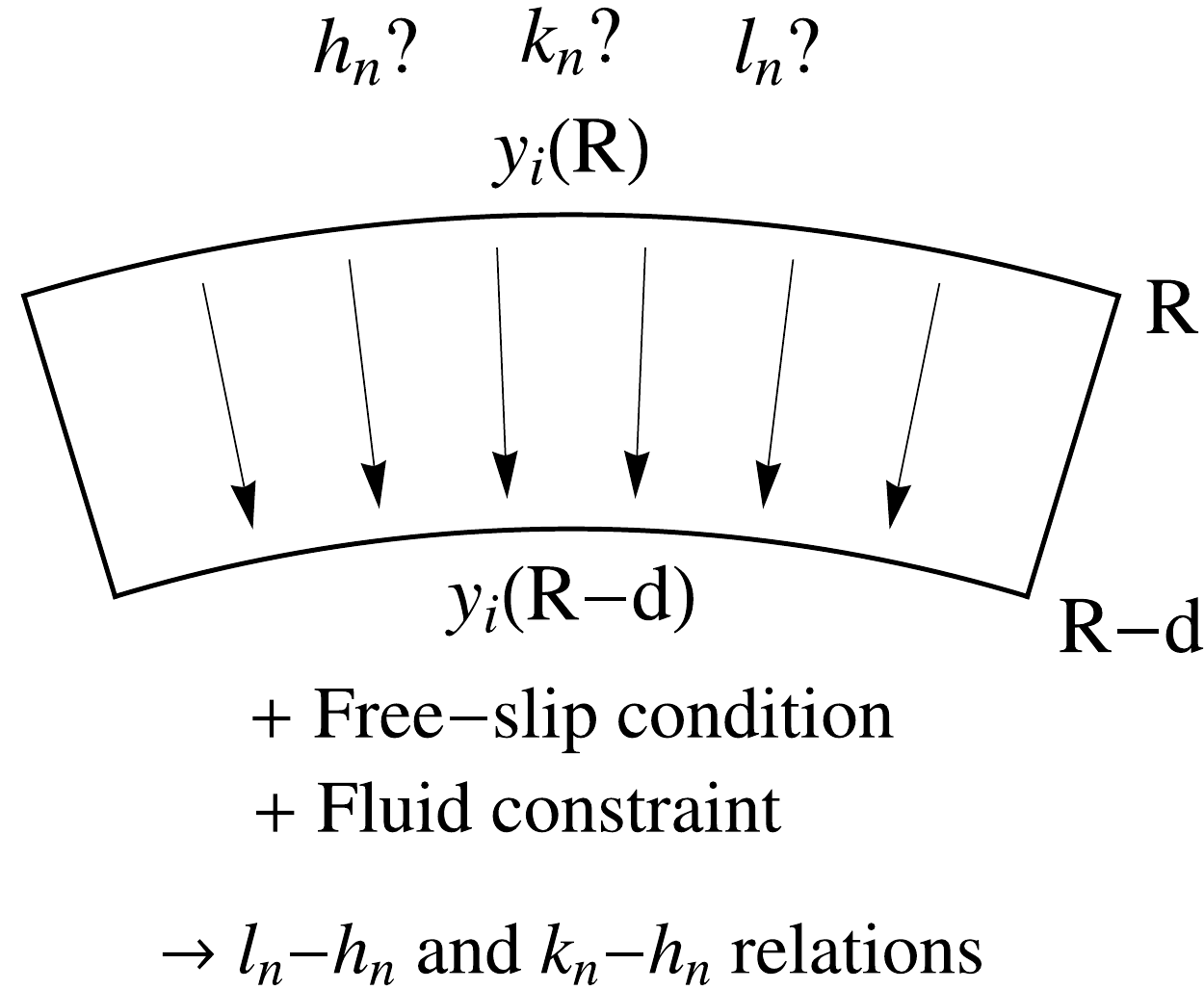}
   \caption[First basic idea of the membrane approach]
   {First basic idea of the membrane approach: propagate the $y_i$ variables from the surface (radius $R$), where they are either fixed by the boundary conditions or parameterized in terms of Love numbers, to the crust-ocean boundary (radius $R-d$) where the free-slip condition and the fluid constraint must be satisfied. The results are two relations between Love numbers.}
   \label{FigIdea1}
\end{figure}

As mentioned in the introduction,  the basic idea of the membrane approach consists in propagating the $y_i$ variables from the surface, where they are either fixed by the boundary conditions or parameterized in terms of Love numbers, to the crust-ocean boundary where two constraints hold: the free-slip condition and the fluid constraint (Eqs.~(\ref{freeslip})-(\ref{fluideqDyn})).
The idea is depicted in Fig.~\ref{FigIdea1}.
If the crust thickness $d$ is small with respect to the surface radius $R$, all equations can be expanded in the small parameter $\varepsilon$:
\begin{equation}
\varepsilon = \frac{d}{R} \, .
\label{defepsilon}
\end{equation}
The radius of the bottom of the crust (or crust-ocean boundary) is denoted
\begin{equation}
R_\varepsilon = R(1-\varepsilon) \, .
\label{defRepsilon}
\end{equation}
I will evaluate the variables $y_i$ at the bottom of the crust in the thin crust limit, that is at first order in $\varepsilon$ (denoted ${\cal O}(\varepsilon)$).

The value of $y_i$ at the bottom of the crust is related to its surface value by
\begin{equation}
y_i(R_\varepsilon) = y_i(R) - \int_{R_\varepsilon}^R y_i'(r) \, dr \, .
\label{yiRminus}
\end{equation}
The derivatives $y_i'$ are given by Eqs.~(\ref{EG1})-(\ref{EG6}).
The integral over the crust thickness being proportional to $\varepsilon$, I only need to keep terms in $y_i'$ that are ${\cal O}(1)$ (i.e.\ zeroth order in $\varepsilon$).
Thus I can make the following approximations in the right-hand sides of Eqs.~(\ref{EG1})-(\ref{EG6}):
\begin{enumerate}
\item
Evaluate all quantities at $r=R$, except the density and the viscoelastic parameters.
\item
Apply the surface boundary conditions for tides or surface loading (Eq.~(\ref{boundcond}) or (\ref{boundcondLoad})).
Express $y_1(R)$, $y_3(R)$, and $y_5(R)$ in terms of Love numbers (Eq.~(\ref{hkl}) or (\ref{hklLoad})).
Combinations of Love numbers can be compactly written with the factors $\gamma_n$, $\delta_n$, and $s_n$ (Eqs.~(\ref{tiltfactor})-(\ref{strainfactor})) or with the corresponding surface loading factors $(\gamma_n',\delta_n',s_n')$.
\item
Static limit in the crust: set $\omega=0$ in Eqs.~(\ref{EG2}) and (\ref{EG4}).
\end{enumerate}

\subsection{Density and gravity}

In the membrane approximation (Eq.~(\ref{yiRminus})), density terms are integrated over the crust thickness.
What remains is the mean density of the crust:
\begin{equation}
\bar\rho = \frac{1}{d} \int_d \rho_r \, dr \, .
\label{rhobar}
\end{equation}
In general, the ocean density increases with depth.
The density of the top layer of the ocean, in contact with the crust, is denoted $\rho$.
The density contrast at the crust-ocean boundary is denoted
\begin{equation}
\delta\rho = \bar\rho - \rho \, .
\label{deltaRho}
\end{equation}
The bulk density (or mean density of the whole body) is denoted $\rho_b$.
The ocean-to-bulk and crust-to-bulk density ratios are defined by
\begin{equation}
\left( \xi \, , \, \bar\xi \, \right) = \left( \frac{\rho}{\rho_b} \, , \, \frac{\bar\rho}{\rho_b} \right) .
\label{defxi}
\end{equation}
The surface gravity ($g$) and the gravity at the crust-ocean boundary ($g_\varepsilon$) are given by
\begin{eqnarray}
g &=& \frac{4\pi}{3} \, G \rho_b R \, ,
\label{gsurf} \\
g_\varepsilon &=& \left( 1 + \varepsilon \left( 2 - 3 \, \bar\xi \, \right) \right) g \, .
\label{gsub}
\end{eqnarray}
The second equation is valid at ${\cal O}(\varepsilon)$.
It results from $g_\varepsilon = (4\pi/3) G \rho_{\varepsilon} R_\varepsilon$, where $\rho_{\varepsilon}$ is the mean density of the body without its crust.
At ${\cal O}(\varepsilon)$, $\rho_{\varepsilon} = \rho_b + 3 ( \rho_b - \bar\rho ) \varepsilon$ so that Eq.~(\ref{gsub}) follows.

\subsection{Effective viscoelastic parameters}
\label{effectiveviscoparam}

As rheology depends on depth, viscoelastic parameters cannot be considered as constant when integrating $y_i'$ over the crust thickness (Eq.~(\ref{yiRminus})). 
In Eqs.~(\ref{EG1})-(\ref{EG6}), the parameters $\mu$, $\nu$ and $\chi$ appear in $y_i'$ in various combinations.
For tidal deformations, two of them ($\chi/\mu$ and $1/\mu$) are eliminated by the zero boundary conditions on $y_2$ and $y_4$.
It is thus only necessary to integrate on $\chi$ and on the parameters $(p,q)$ defined by
\begin{equation}
\left( p , q \right) =   \left( \frac{\mu}{1-\nu} ,  \frac{\mu\nu}{1-\nu} \right) \, .
\label{defpq}
\end{equation}
The inverse relations read
\begin{equation}
\left( \mu , \nu \right) = \left( p-q, \frac{q}{p} \right) \, .
\label{pqinverse}
\end{equation}
Integrating $\chi$ and $(p,q)$ on depth yields three effective parameters (denoted with a bar):
\begin{eqnarray}
\bar\chi &=& \frac{1}{d} \int_d \chi \, dr \, ,
\nonumber \\
\left( \bar p \, , \bar q \right) &=& \frac{1}{d} \int_d \left( \frac{\mu}{1-\nu} \, , \frac{\mu\nu}{1-\nu} \right) dr \, .
\label{EffectiveParam}
\end{eqnarray}
The \textit{effective shear modulus} $\bar\mu$ and the \textit{effective Poisson's ratio} $\bar\nu$ are then defined by relations similar to Eq.~(\ref{pqinverse}):
\begin{equation}
\left( \bar\mu , \bar\nu \right) = \left( \bar p - \bar q, \frac{\bar q}{\bar p} \right) \, ,
\label{munubar1}
\end{equation}
which can be inverted as
\begin{equation}
\left( \bar p , \bar q \right) =   \left( \frac{\bar\mu}{1-\bar\nu} ,  \frac{\bar\mu\bar\nu}{1-\bar\nu} \right) \, .
\label{defpqbar}
\end{equation}
In other words, I have defined the effective parameters so that integrating over the crust thickness amounts to put a `bar' on the parameters:
\begin{equation}
\left( \bar\chi \, , \frac{\bar\mu}{1-\bar\nu} \, ,  \frac{\bar\mu\bar\nu}{1-\bar\nu} \right) = \frac{1}{d} \int_d \left( \chi \, , \frac{\mu}{1-\nu} \, , \frac{\mu\nu}{1-\nu} \right) dr \, .
\end{equation}
In terms of $(\mu,\nu)$, the effective parameters $(\bar\mu,\bar\nu)$ are given by
\begin{eqnarray}
\bar\mu &=& \frac{1}{d} \int_d \mu \, dr \, ,
\label{mubar2} \\
\bar \nu &=& \left(  \int_d  \frac{\mu}{1-\nu} \, dr  \right)^{-1} \, \int_d \frac{\mu \nu}{1-\nu} \, dr \, .
\label{nubar2}
\end{eqnarray}
The interpretation of $\bar\chi$ and $\bar\mu$ is straightforward: $\bar\chi$ is the mean of $\chi$ and $\bar\mu$ is the mean of $\mu$.
The definition of $\bar\nu$ is a bit more complicated: $\bar\nu$ is the weighted mean of $\nu$ with weight $p$. 
Remarkably, $\bar\mu$ and $\bar\nu$ are identical to the effective shear modulus and effective Poisson's ratio of the massless membrane approach \citep{beuthe2014}.
The effective compressibility factor $\bar{\chi}$, however, does not appear in the massless membrane approach.

In general, the relation between $\chi$ and $(p,q)$ is not linear so that $\bar\chi$ cannot be written as a combination of $\bar p$ and $\bar q$.
The assumption of zero bulk dissipation, however, imposes the condition that $K=K_E$ (the subscript $E$ stands for `elastic').
Since $K=(2/3)(p+q)/\chi$ (see Table~\ref{TableCompressibility}), the condition $K=K_E$ is equivalent to
\begin{equation}
\frac{\chi}{\chi_E} = \frac{p+q}{p_E+q_E} \, .
\label{chipq}
\end{equation}
If the elastic parameters $(\mu_E,\nu_E)$ are constant with depth, integrating this equation on depth yields
\begin{eqnarray}
\bar\chi &=&  \frac{\chi_E}{p_E+q_E} \left( \bar p+\bar q \right)
\nonumber \\
&=& \frac{1-2\nu_E}{1+\nu_E} \, \frac{\bar\mu}{\mu_E} \, \frac{1+\bar\nu}{1-\bar\nu} \, .
\label{chibar}
\end{eqnarray}
This relation shows two things: 
\begin{itemize}
\item there are only two independent effective parameters: $\bar\mu$ and $\bar\nu$.
\item
$\bar\chi$ varies with viscosity approximately like $\bar\mu/2\mu_E$ (if $\nu_E\sim1/3$) because $\bar\nu$ is much less sensitive to viscosity than $\bar\mu$ (Fig.~\ref{FigEffectiveParam}).
\end{itemize}
If bulk dissipation is not zero or if $(\mu_E,\nu_E)$ vary with depth, $\bar\chi$ is an independent parameter which must be computed with Eq.~(\ref{EffectiveParam}).

In membrane equations, the effective shear modulus is always multiplied by the membrane thickness $d$.
It is thus often replaced by the extensional rigidity $D_{ex}=2\bar{p}d$ which relates integrated plane stresses to plane strains (Eq.~(5) of \cite{beuthe2014}).
The effective shear modulus and the extensional rigidity are nondimensionalized as follows:
\begin{eqnarray}
\hat\mu &=& \frac{\bar\mu}{\rho{}gR} \,,
\label{hatmu} \\
\hat D_{ex} &=&  \frac{2\hat\mu}{1-\bar\nu} \,  \varepsilon \, .
\label{hatDex}
\end{eqnarray}

Computing load Love numbers requires integrating on $\chi/\mu$ (Eq.~(\ref{EG1})) or, equivalently, on the nondimensional parameter $\chi_\mu=\chi/\hat\mu$:
\begin{eqnarray}
\bar\chi_\mu &=& \frac{1}{d} \int_d \chi_\mu \, dr
\nonumber \\
&=& \frac{1-2\nu_E}{1+\nu_E} \, \frac{\rho{}gR}{\mu_E} \left( 3 - 2 \bar\chi \right) .
\label{chimubar}
\end{eqnarray}
The second line holds if there is no bulk dissipation (same argument as above) and shows that $\bar\chi_\mu$ is not an independent parameter under that assumption.

\subsection{Example: stagnant lid regime}
\label{StagnantLidRegime}

\begin{figure}
   \hspace{-3mm}
   \includegraphics[width=15.3cm]{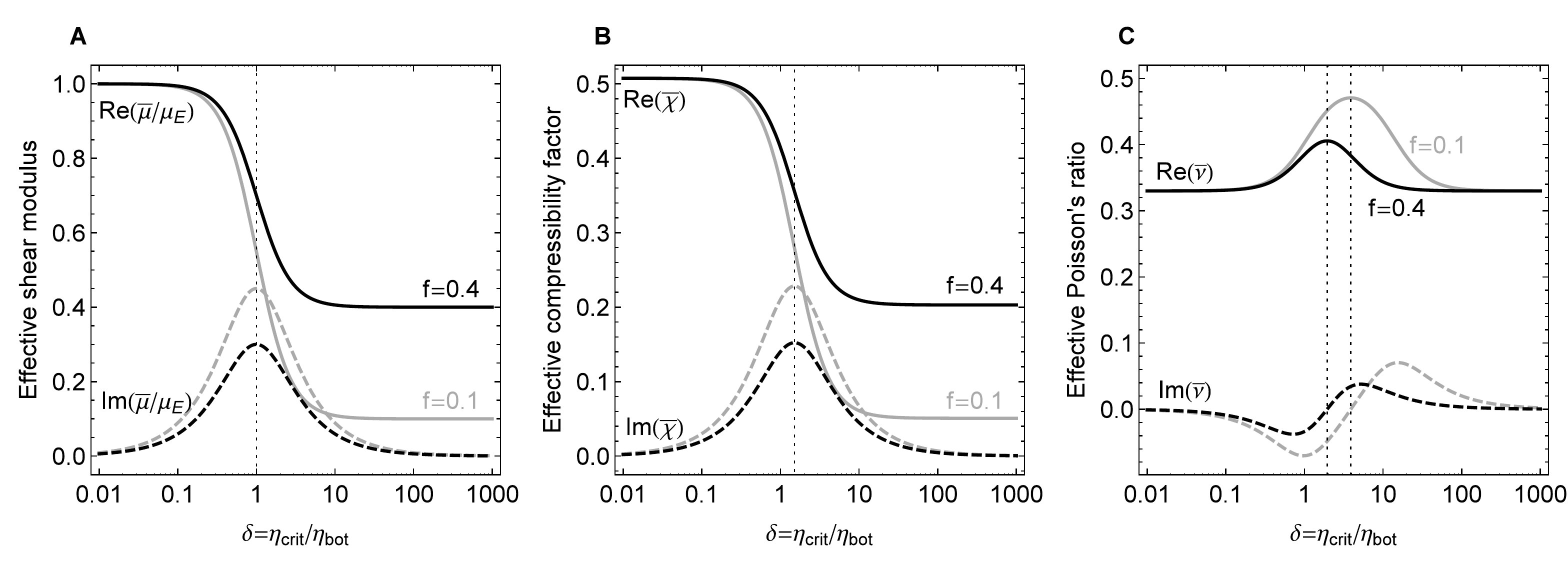}
   \caption[Effective viscoelastic parameters]
   {Effective viscoelastic parameters as functions of inverse viscosity for a conductive/convective icy crust: (A) shear modulus (normalized by its elastic value), (B) compressibility factor, (C) Poisson's ratio.
   The crust is divided into a top elastic layer and a bottom viscoelastic layer with $f$ being the relative thickness of the top layer.
   The elastic Poisson's ratio is $\nu_E=0.33$.
   Black (resp.\ gray) curves correspond to a top layer that makes 40\% (resp.\ 10\%) of the total crust thickness.
   Solid (resp.\ dashed) curves show the real part (resp.\ imaginary part).
   Vertical dotted lines indicate the positions of the critical transitions.
   See Section~\ref{effectiveviscoparam} for details.}
   \label{FigEffectiveParam}
\end{figure}

As an illustration, suppose that the crust is made of two uniform layers: the top layer is elastic (no viscous effects) while the bottom layer has a rheology of Maxwell type (Appendix~\ref{MaxwellRheology}).
If $d_{top}$ is the thickness of the top layer, its relative thickness is denoted $f=d_{top}/d$.
This two-layer structure can represent the rheology of either a conductive crust \citep{ojakangas1989a} or a conductive/convective crust in the stagnant lid regime \citep[e.g.][]{hussmann2002}.
In the former case, the crust is nearly elastic throughout and the two layers are not distinguished.
In the latter case, convection occurs under a conductive lid: the top layer is conductive and elastic whereas the bottom layer is convective and viscoelastic.
Fig.~\ref{FigEffectiveParam} shows the effective viscoelastic parameters in terms of the dimensionless number $\delta$ which is inversely proportional to the viscosity of the bottom layer (Eq.~(\ref{defdelta})).
On the $x$-axis, values from left to right correspond to a bottom layer which is elastic-like ($\delta\leq0.1$), critical ($0.1<\delta<10$), and fluid-like ($\delta\geq10$).
If the bottom layer is fluid-like, the absolute value of the effective shear modulus drops to a fraction $f$ of its elastic value: $|\bar\mu| \sim f \mu_E$ or, equivalently, $|\bar\mu| d \sim \mu_E \,d_{top}$.
In that case, the crust response is determined by the elastic top layer which acts as a lithosphere.
More generally, it makes sense to define the lithospheric thickness by
\begin{equation}
d_{litho} = \frac{|\bar\mu|}{\mu_E} \, d \, .
\label{dlitho}
\end{equation}

The classification elastic/critical/fluid-like is based on the dependence of $\bar\mu$ on viscosity and would be shifted to higher values of $\delta$ were it based on the behaviour of $\bar\nu$ (compare panels~A and C in Fig.~\ref{FigEffectiveParam}).
If the top layer is purely elastic, the critical transitions occur at
\begin{itemize}
\item $\delta=1$: middle point of $Re(\bar\mu)$ and maximum of \textit{Im}($\bar\mu$), 
\item $\delta=\delta'$: middle point of $Re(\bar\chi)$ and maximum of \textit{Im}($\bar\chi$),
\item $\delta=\delta''$: maximum of $Re(\bar\nu)$ and zero-crossing of \textit{Im}($\bar\nu$),
\end{itemize}
where $\delta'=3(1-\nu_E)/(1+\nu_E)$ and $\delta''=\sqrt{\delta'/f}$.

\subsection{Propagating $y_i$ functions through the crust}
\label{EqThinCrustLimit}

The $y_i$ functions can now be propagated from the top to the bottom of the crust.
Using Eqs.~(\ref{EffectiveParam})-(\ref{defpqbar}) and the approximations of Section~\ref{Principle}, I integrate Eqs.~(\ref{EG1})-(\ref{EG6}) and insert them into Eq.~(\ref{yiRminus}).
For tidal deformations, I get the following nondimensional equations:
\begin{eqnarray}
g y_1(R_\varepsilon) &=& h_n + \varepsilon \left( 1 - \bar\chi  \right) s_n \, ,
\label{y1crust} \\
\frac{1}{\rho} \, y_2(R_\varepsilon) &=& - \left( \left(1+\bar\nu\right) \hat D_{ex} -  \varepsilon \,  \frac{\bar\rho}{\rho}  \right) s_n + \varepsilon \, \frac{\bar\rho}{\rho} \, n \delta_n \, ,
\label{y2crust} \\
g y_3(R_\varepsilon) &=& \left( 1 - \varepsilon \right) l_n + \varepsilon \, h_n \, ,
\label{y3crust} \\
\frac{1}{\rho} \,  y_4(R_\varepsilon) &=& \hat D_{ex} \, \Big( \left( 1+\bar\nu \right) h_n - \left( x_n+1+\bar\nu \right) l_n \Big) +  \varepsilon \, \frac{\bar\rho}{\rho} \, \gamma_n \, ,
\label{y4crust} \\
y_5(R_\varepsilon) &=& \left( 1 + \left( n+1 \right) \varepsilon \right) \left( k_n+1 \right) -   3 \, \varepsilon \, \bar\xi \, h_n  - \left( 2 n +1 \right) \varepsilon \, ,
\label{y5crust} \\
R y_6(R_\varepsilon) &=& \left( 2 n+1\right) \left( 1 - \left( n -1 \right) \varepsilon \right) - 3 \, \varepsilon \, \bar\xi  \left( \left( n-1 \right) h_n + s_n \right) .
\label{y6crust}
\end{eqnarray}
I will also need the auxiliary gravity variable $y_7$ at the crust-ocean boundary.
Substituting Eqs.~(\ref{y2crust}) and (\ref{y6crust}) into Eq.~(\ref{y7def}), I get
\begin{equation}
R y_7(R_\varepsilon) = \left(2 n+1 \right) \left( 1 -  \left( n - 1 \right) \varepsilon \right)
- 3  \varepsilon \, \bar\xi  \left( \left( n-1 \right) h_n - n \delta_n \right)
- 3 \, \xi \, \hat D_{ex} \left(1+\bar\nu\right) s_n \, .
\label{y7crust}
\end{equation}
Finally, the quantity $g_r y_1 - y_5$ is of particular interest because it appears in the fluid constraint (Eq.~(\ref{fluideqDyn})).
Combining  Eqs.~(\ref{gsub}), (\ref{y1crust}), and (\ref{y5crust}), I can write
\begin{equation}
g_\varepsilon y_1(R_\varepsilon) - y_5(R_\varepsilon) = - \gamma_n + \varepsilon \, n \delta_n + \varepsilon \left( 1 - \bar\chi \right) s_n \, .
\label{tiltsub}
\end{equation}

\section{Relations between Love numbers}
\label{RelationsBetweenLoveNumbers}

In Section~\ref{CrustMembrane}, I propagated the $y_i$ functions from the surface to the crust-ocean boundary (Eqs.~(\ref{y1crust})-(\ref{tiltsub})).
I now examine the constraints imposed by the special kind of coupling between crust and ocean: the ocean exerts a radial push on the crust (continuity of the radial stress) but no lateral traction (shear stress vanishes, free slip occurs).
Table~\ref{TableNotation} summarizes the parameters relevant to the shallow interior which will be used repeatedly in the rest of the paper.

\begin{table}[ht]\centering
\ra{1.3}
\small
\caption[Parameters relevant to the shallow interior]{\small
Parameters relevant to the shallow interior.}
\begin{tabular}{@{}llll@{}}
\hline
& Parameter &  Symbol & Eq.  \\
\hline
Size & Surface radius & $R$ & - \\
& Crust thickness & $d$ & - \\
& Relative crust thickness & $\varepsilon$ & (\ref{defepsilon}) \\
& Radius of crust-ocean boundary & $R_\varepsilon$ & (\ref{defRepsilon}) \\
Density  & Mean crust density & $\bar\rho$ & (\ref{rhobar}) \\
& Ocean density (top layer) & $\rho$ & - \\
& Crust-ocean density contrast & $\delta\rho$ & (\ref{deltaRho}) \\
& Bulk density & $\rho_{b}$ &  - \\
& Ocean-to-bulk density ratio & $\xi$ &  (\ref{defxi}) \\
& Crust-to-bulk density ratio & $\bar\xi$ &  (\ref{defxi}) \\
Gravity & Surface gravity & $g$ & (\ref{gsurf}) \\
& Gravity at crust-ocean boundary & $g_\varepsilon$ & (\ref{gsub}) \\
Elasticity & Effective shear modulus & $\bar\mu$ & (\ref{mubar2}) \\
& Effective Poisson's ratio & $\bar\nu$ & (\ref{nubar2}) \\
& Effective compressibility factor & $\bar\chi$ & (\ref{chibar}) \\
& Effective shear modulus (nondim.) & $\hat\mu$ & (\ref{hatmu}) \\ 
& Extensional rigidity (nondim.) & $\hat D_{ex}$ & (\ref{hatDex}) \\
Varia & Eigenvalue of $-(\Delta+2)$ & $x_n$ & (\ref{xn}) \\
& Dynamical parameter & $q_\omega$ & (\ref{qomega})
\vspace{0.5mm}
\\
\hline
\end{tabular}
\label{TableNotation}
\end{table}%

\subsection{Magnitude of tilt factor}
\label{TiltFactor}

Let us start with the radial stress coupling at the crust-ocean boundary.
At the top of the ocean, the fluid constraint (Eq.~(\ref{fluideqDyn})) reads
\begin{equation}
\frac{1}{\rho} \, y_2(R_\varepsilon) =  g_\varepsilon y_1(R_\varepsilon) - y_5(R_\varepsilon)  - \frac{\omega^2 R_\varepsilon}{g} \, g y_3^F(R_\varepsilon) \, .
\label{fluideqcrust}
\end{equation}
As the crust freely slips on the ocean, the function $y_3$ is discontinuous at the crust-ocean boundary: the superscript $F$ indicates that $y_3$ must be evaluated on the fluid side of the boundary. 
I will assume for the moment that $y_3^F(R_\varepsilon)$ and $y_1(R_\varepsilon)$ are of the same order of magnitude (this assumption is discussed in Section~\ref{LoveDynamic}).
Since tides are slow (Section~\ref{StaticDynamic}), the last term in Eq.~(\ref{fluideqcrust}) is of ${\cal O}(\varepsilon)$ or smaller.

Contrary to the function $y_3$, the functions $y_1$, $y_2$ and $y_5$ are continuous at the crust-ocean boundary.
I can thus substitute Eqs.~(\ref{y2crust}) and (\ref{tiltsub}) into Eq.~(\ref{fluideqcrust}).
The result is a relation between Love numbers, or rather their linear combinations $\gamma_n$, $\delta_n$, and $s_n$:
\begin{equation}
\gamma_n = \left( \left(1+\bar\nu\right) \hat D_{ex} - \varepsilon \, \frac{\delta\rho}{\rho} - \varepsilon \, \bar\chi \right) s_n
- \varepsilon \, \frac{\delta\rho}{\rho} \, n \delta_n - q_\omega \, g y_3^F(R_\varepsilon) \, .
\label{gamma2raw}
\end{equation}
All terms in the right-hand side are of ${\cal O}(\varepsilon)$ except maybe the first one depending on the extensional rigidity.
Two cases must be considered:
\begin{itemize}
\item the crust is soft ($|\hat D_{ex}|\lesssim1$).
Love numbers are large and the tilt factor is small,
\begin{eqnarray}
\left( h_n , l_n , k_n , \delta_n , s_n \right) &\sim& {\cal O}(1) \, ,
\nonumber \\
\gamma_n &\sim& {\cal O}(\varepsilon) \, ,
\label{gammaneps}
\end{eqnarray}
which also means that
\begin{equation}
k_n+1 = h_n + {\cal O}(\varepsilon) \, .
\label{knhneps}
\end{equation}
\item the crust is hard ($|\hat D_{ex}|\gtrsim1$ and $\hat D_{ex}s_n\sim{\cal O}(1)$).
Love  numbers are small and the tilt factor is close to one,
\begin{eqnarray}
\left( h_n , l_n , k_n , \delta_n , s_n \right) &\ll& 1 \, ,
\nonumber \\
\gamma_n &\sim& 1 \, .
\label{gamman1}
\end{eqnarray}
\end{itemize}

From now on, I assume that the crust is soft so that Eq.~(\ref{gammaneps}) holds.
In this way, I will obtain formulas for $k_n$ and $h_n$ that are valid at ${\cal O}(\varepsilon)$.
If the crust is hard, the extensional rigidity $\hat D_{ex}$ must be considered as a parameter of ${\cal O}(1)$.
In that case, the expansions of the viscoelastic-gravitational equations (Eqs.~(\ref{y1crust})-(\ref{y6crust})) are not complete at ${\cal O}(\varepsilon)$: one should indeed include terms like $\hat D_{ex}\varepsilon$, which can only be obtained by formally expanding the viscoelastic-gravitational equations to ${\cal O}(\varepsilon^2)$.
All is not lost, however: up to ${\cal O}(1)$, the formulas for a soft crust also apply to a body with a hard crust.
This means that, if the crust is hard, only the dominant contribution of the crustal rigidity must taken into account, while density terms and other small corrections must be neglected for consistency.

\subsection{Relation between $l_n$ and $h_n$}
\label{FreeSlip}

Imposing the free-slip condition $y_4(R_\varepsilon)=0$ on Eq.~(\ref{y4crust}) yields $l_n$ in terms of $(h_n,\gamma_n)$ at the crust-ocean boundary:
\begin{equation}
l_n = \frac{1+\bar\nu}{x_n+1+\bar\nu} \, h_n + \frac{\bar\rho}{\rho} \, \frac{1}{2\hat\mu} \, \frac{1-\bar\nu}{x_n+1+\bar\nu} \, \gamma_n \, .
\label{y3y1y5}
\end{equation}
If the crust is soft, the first and second terms in the right-hand side are of ${\cal O}(1)$ and ${\cal O}(\varepsilon)$, respectively, because $\gamma_n\sim{\cal O}(\varepsilon)$ (Eq.~(\ref{gammaneps})).
Going back to Eq.~(\ref{y4crust}), one sees that the second term of Eq.~(\ref{y3y1y5}) comes from a term of ${\cal O}(\varepsilon^2)$ and should in principle be neglected because the equations were not expanded beyond ${\cal O}(\varepsilon)$.
This term, however, is not small in the limit of a quasi-fluid crust because of the prefactor $1/\hat\mu$.
I will exclude this possibility by imposing the following \textit{lithospheric condition} (a more precise form will be given later):
\begin{equation}
\hat\mu  \gg \gamma_n/h_n \, .
\label{LithoCond1}
\end{equation}
This constraint holds as long as there is a lithosphere, i.e.\ the upper part of the crust has nonzero rigidity.
Under that reasonable assumption, the free-slip condition yields a relation between the displacement Love numbers, or $l_n-h_n$ \textit{relation}:
\begin{equation}
l_n = \frac{1+\bar\nu}{x_n+1+\bar\nu} \, h_n \, ,
\label{lnhn}
\end{equation}
which is valid at ${\cal O}(1)$, i.e.\ in the limit of zero crust thickness.
The same relation holds for a hard crust because the second term of Eq.~(\ref{y3y1y5}) is smaller than the first one by a factor $\varepsilon$ ($h_n\sim{\cal O}(1/\hat\mu\varepsilon)$ and $\gamma_n\sim1$).

The $l_n-h_n$ relation coincides with the one derived in the massless membrane approach (Appendix~\ref{MasslessMembrane}).
Finite thickness corrections depend not only on the thickness and density of the crust but also on its rheology \citep[Fig.~4]{beuthe2014}.
If  the crust does not convect, it is nearly elastic and finite thickness corrections can be estimated with the homogeneous crust model of Appendix~\ref{HomogeneousCrustModel}.
If tides are static, the $l_2-h_2$ relation for this model, up to order ${\cal O}(\varepsilon)$, can be read from Table~\ref{TableGeoFac}:
\begin{equation}
l_2 = \frac{3}{11} \left( 1 - \frac{32}{33} \varepsilon \right) h_2 \, .
\label{l2h2corr}
\end{equation}
The factor 3/11 results from $(x_2+1+\bar\nu)/(5+\bar\nu)$ with $\bar\nu=1/2$.
Eq.~(\ref{l2h2corr}) is a good approximation if $\varepsilon\lesssim0.3$ \citep[Fig.~13]{beuthe2014}.

\subsection{Relation between $k_n$ and $h_n$}
\label{Relationk2h2}

In Sections~\ref{TiltFactor} and \ref{FreeSlip},  I obtained two independent relations between Love numbers:
the $\gamma_n-\delta_n-s_n$ and $l_n-h_n$ relations which are of ${\cal O}(\varepsilon)$ and ${\cal O}(1)$, respectively.
I will now combine them so as to relate $k_n$ and $h_n$ at ${\cal O}(\varepsilon)$.
If the crust is soft, the right-hand side of Eq.~(\ref{gamma2raw}) can be evaluated at ${\cal O}(\varepsilon)$ by knowing $s_n$ and $\delta_n$ at ${\cal O}(1)$.
Inserting Eqs.~(\ref{knhneps}) and (\ref{lnhn}) into Eqs.~(\ref{tiltfactor})-(\ref{gravifactor}), I get
\begin{eqnarray}
s_n &=& \frac{x_n \left(1-\bar\nu\right)}{x_n+1+\bar\nu} \, h_n +{\cal O}(\varepsilon) \, ,
\label{snhn} \\
n \delta_n &=& 2 n +1 - \left( n - 1 \right) h_n +{\cal O}(\varepsilon) \, .
\label{dnhn}
\end{eqnarray}
Note that the lithospheric condition, Eq.~(\ref{LithoCond1}), is required for the first approximation to be valid whereas the second approximation depends on the assumption of a soft crust.
If the crust is hard, Eq.~(\ref{dnhn}) has an additional term $-(n+1)\gamma_n$ in the right-hand side.
This modification does not matter because the right-hand side of Eq.~(\ref{gamma2raw}) is then of ${\cal O}(1)$ and all terms of ${\cal O}(\varepsilon)$, including the one depending on $n\delta_n$, are neglected in that case.

The substitution of Eqs.~(\ref{snhn})-(\ref{dnhn}) into Eq.~(\ref{gamma2raw}) yields a simple relation between the tilt factor and the radial Love number:
\begin{equation}
\gamma_n =  \Lambda_T \, h_n - \left( 2 n+1 \right) \frac{\delta\rho}{\rho} \, \varepsilon \, .
\label{tiltn}
\end{equation}
In the right-hand side, the last term is the \textit{major density correction} while the factor $\Lambda_T$ in the first term is the sum  of four contributions:
\begin{equation}
\Lambda_T = \Lambda + \Lambda_\chi + \Lambda_\rho + \Lambda_\omega \, .
\label{Lambdatot}
\end{equation}
The first (and generally dominant) contribution is the \textit{membrane spring constant} $\Lambda$, which vanishes if the crust is fluid-like:
\begin{equation}
\Lambda = f_\mu \, \hat\mu \, \varepsilon \, .
\label{springconst}
\end{equation}
The dimensionless coefficient $f_\mu$ is  defined in Table~\ref{TableCoeffLambdaT}, together with the coefficients $f_\chi$ and $f_\rho$ appearing below.
The name `membrane spring constant' comes from the observation that the membrane radial response follows Hooke's law (Appendix~\ref{MasslessMembrane}).

The second contribution to $\Lambda_T$ is the \textit{compressibility correction} $\Lambda_\chi$, which vanishes if the crust is incompressible:
\begin{equation}
\Lambda_\chi = f_\chi \, \bar\chi \, \varepsilon \, .
\label{LambdaChi}
\end{equation}
The third one is the \textit{minor density correction} $\Lambda_{\rho}$,
which vanishes if there is no density contrast at the crust-ocean boundary:
\begin{equation}
\Lambda_\rho = f_\rho \, \frac{\delta\rho}{\rho} \, \varepsilon \, .
\label{LambdaRho}
\end{equation}
For tides of degree two, the qualification `minor' is justified because this term is about ten times smaller than the major density correction ($f_\rho\sim0.5$ if $n=2$).
The fourth one is the \textit{dynamical correction} $\Lambda_\omega$, which vanishes in the static limit:
\begin{equation}
\Lambda_\omega = - q_\omega \, \frac{y_3^F(R_\varepsilon)}{y_1(R_\varepsilon)} \, .
\label{LambdaOmega}
\end{equation}
The ratio $y_3^F(R_\varepsilon)/y_1(R_\varepsilon)$ is unknown at this stage but is supposed to be of order unity so that $\Lambda_\omega\sim{\cal O}(\varepsilon)$ (Section~\ref{TiltFactor}).
Dynamical corrections are discussed in more detail in Section~\ref{LoveDynamic}.

The tilt factor formula (Eq.~(\ref{tiltn})) can be written as a relation between Love numbers, or $k_n-h_n$ \textit{relation}:
\begin{equation}
k_n + 1 = \left( 1 + \Lambda_T \right) h_n - \left( 2 n+1 \right) \frac{\delta\rho}{\rho} \, \varepsilon \, .
\label{knhn}
\end{equation}
When the crust becomes fluid-like,  Eq.~(\ref{knhn}) does not tend in the static limit to the hydrostatic $k_n^\circ-h_n^\circ$ relation (Eq.~(\ref{knhnhydrostat})) because of the terms proportional to $\delta\rho$.
This is not surprising because the derivation of the $k_n\,$-$\,h_n$ relation requires the lithospheric condition, Eq.~(\ref{LithoCond1}).
This condition can be reformulated with the help of Eq.~(\ref{tiltn}):
\begin{equation}
\hat\mu \gg \frac{\delta\rho}{\rho} \, \varepsilon \, .
\label{LithoCond2}
\end{equation}
How does the above $k_n-h_n$ relation compare to the one obtained with the massless membrane approach \citep{beuthe2014}?
In the latter approach, the crust and ocean have the same density ($\delta_\rho=0$) and tides are static ($\Lambda_\omega=0$).
With these assumptions, Eq.~(\ref{knhn}) becomes $k_n + 1 = \left( 1 + \Lambda + \Lambda_\chi \right) h_n$.
This relation differs from the massless membrane relation by the compressibility factor $\Lambda_\chi$, which accounts for the mismatch found by \citet{beuthe2014} when benchmarking the tilt factor formula.
The differences between the massive and massless approaches are analyzed in more detail in Appendix~\ref{MasslessMembrane}.

\begin{table}[ht]\centering
\ra{1.3}
\small
\caption[Dimensionless coefficients appearing in the $k_n-h_n$ and $k_n'-h_n'$ relations]{\small
Dimensionless coefficients appearing in the $k_n-h_n$ and $k_n'-h_n'$ relations (Eqs.~(\ref{springconst})-(\ref{LambdaRho}) and Eq.~(\ref{defpsi})).
}
\begin{tabular}{@{}ccc@{}}
\hline
&  arbitrary $n$ & $n=2$ \\
\hline
$f_\mu$ & $\,\,\,\,\frac{2 x_n \left(1+\bar\nu\right)}{x_n+1+\bar\nu}$ & $\,\,\,\,\frac{8 \left(1+\bar\nu\right)}{5+\bar\nu}$
\vspace{1mm}  \\
$f_\chi$ & $-\frac{x_n \left(1-\bar\nu\right)}{x_n+1+\bar\nu}$ &  $-\frac{4 \left(1-\bar\nu\right)}{5+\bar\nu}$
\vspace{1mm}  \\
$f_\rho$ & $ \frac{(n -1)(n^2-3 + (n+3) \bar\nu)}{x_n+1+\bar\nu}$ & $\,\,\frac{1 + 5 \bar\nu}{5+\bar\nu}$
\vspace{1mm} \\
$f_\rho'$ & $ \frac{(n^2-1)(n+1- \bar\nu)}{x_n+1+\bar\nu}$ & $\,\,\frac{3(3- \bar\nu)}{5+\bar\nu}$
\vspace{1mm} \\
$f_{\chi\rho}$ & $ \frac{n(n+1)(1- \bar\nu)}{2(x_n+1+\bar\nu)}$ & $\,\,\frac{3(1- \bar\nu)}{5+\bar\nu}$
\vspace{1mm} \\
\hline
\end{tabular}
\label{TableCoeffLambdaT}
\end{table}%

\subsection{Load Love numbers}
\label{RelationsLoveLoad}

Similarly to tidal Love numbers, relations between load Love numbers can be obtained by propagating the $y_i$ functions through the crust and applying the fluid constraint plus the free-slip condition at the crust-ocean boundary.
The difference is that the nonzero boundary condition on $y_2$ (Eq.~(\ref{boundcondLoad})) introduces new terms in the equations of propagation.
The $l_n'-h_n'$ relation reads
\begin{equation}
l_n' = \frac{1+\bar\nu}{x_n+1+\bar\nu} \, h_n' + \frac{2n+1}{3\xi} \, \frac{1}{2\hat\mu} \, \frac{1-\bar\nu}{x_n+1+\bar\nu} \left( \bar\chi + \frac{\delta\rho}{\rho} \right) \, .
\label{lnhnLoad}
\end{equation}
The second term in the right-hand side results from the nonzero radial stress at the surface.
As in Eq.~(\ref{y3y1y5}), the term $1/\hat\mu$ diverges if there is no lithosphere.
The $k_n'-h_n'$ relation reads 
\begin{equation}
k_n' + 1 = \left( 1 + \Lambda_T \right) h_n' - \left( 2 n+1 \right) \frac{\delta\rho}{\rho} \, \varepsilon + \left( 1 + \psi \right) \frac{2n+1}{3\xi} \, .
\label{knhnLoad}
\end{equation}
This relation has the same form as the $k_n-h_n$ relation except for the term proportional to $(2n+1)/(3\xi)$ which results from the nonvanishing radial stress at the surface.
The dimensionless number $\psi$ gathers new compressibility and density corrections of ${\cal O}(\varepsilon)$:
\begin{equation}
\psi = \left( -f_\chi \, \bar\chi + f_\rho' \, \frac{\delta\rho}{\rho} + \frac{1}{2} \, \bar\chi_\mu + f_{\chi\rho} \, \frac{1}{\hat\mu}  \left( \bar\chi + \frac{\delta\rho}{\rho} \right)^2 \right) \varepsilon \, ,
\label{defpsi}
\end{equation}
where the coefficients $f_\chi$, $f_\rho'$, and $f_{\chi\rho}$ are defined by Table~\ref{TableCoeffLambdaT} while $\bar\chi_\mu$ is given by Eq.~(\ref{chimubar}).

Now the gravitational load Love number is related to tidal Love numbers by the Saito-Molodensky relation \citep{molodensky1977,saito1978,lambeck1980}:
\begin{eqnarray}
k_n' &=& k_n - h_n
\nonumber \\
&=& \gamma_n - 1 \, .
\label{SaitoMolo}
\end{eqnarray}
Combining this equation with the tilt factor formula (Eq.~(\ref{tiltn})) and the $k_n'-h_n'$ relation (Eq.~(\ref{knhnLoad})), I can express the radial load Love number in terms of the radial tidal Love number:
\begin{equation}
h_n' = \frac{\Lambda_T \, h_n - \left( 1 + \psi \right) \frac{2n+1}{3\xi}}{1+\Lambda_T} \, ,
\label{hnLoad}
\end{equation}
where $\psi$ is defined by Eq.~(\ref{defpsi}).
The computation of load Love numbers is thus reduced to the evaluation of tidal Love numbers.

\subsection{Rigid mantle model}
\label{RigidMantleLimit}

In this section, I check the validity of the $k_n-h_n$ relation against a semi-analytical model derived with thick shell theory.
Consider static tides of degree two acting on a body described by the \textit{rigid mantle model}: the mantle is infinitely rigid and the ocean is homogeneous and incompressible.
\citet{wahr2006} solved the degree-two elastic-gravitational equations of this model assuming that the crust is incompressible and of uniform elasticity.
Since the resulting analytical formulas are very complicated, they expand them at ${\cal O}(\varepsilon)$ so as to obtain simple formulas for Love numbers.
Finally, they fit corrections due to crust compressibility with a numerical code.
I will now check that membrane formulas reproduce these results if the crust is homogeneous, i.e.\ in the limit $(\bar\mu,\bar\nu,\bar\chi)=(\mu,\nu,\chi)$.

I start by evaluating the tilt factor for an incompressible crust ($\nu=1/2$) before estimating compressibility corrections.
At ${\cal O}(1)$, $h_2$ is given by the formula for a two-layer body with rigid mantle and surface ocean:
$h_{2r}^\circ=5/(5-3\xi)$ (Eq.~(\ref{LoveRigid}) in which $\xi^\circ\rightarrow\xi$).
Substituting this formula in the right-hand side of Eq.~(\ref{tiltn}) and setting $\nu=1/2$ yields
\begin{equation}
\gamma_2^{inc}=  \left( \frac{24}{11} \, \frac{\mu}{ \rho{g}R} - 3 \left( \frac{16}{11} - \xi \right) \frac{\delta\rho}{\rho} \right) h_{2r}^\circ \, \varepsilon \, ,
\end{equation}
where the superscript `\textit{inc}' denotes that the crust is incompressible.
This formula coincides with Eq.~(7) of \citet{wahr2006}.

Next, I express corrections due to crust compressibility in terms of the parameter $\mu/K$ which is related to Poisson's ratio $\nu$ (see Table~\ref{TableCompressibility}).
In particular, $\nu=1/3$ is equivalent to $\mu/K=3/8$.
Using the expansion $\nu\sim1/2-\mu/2K$, I get
\begin{equation}
\gamma_2 = \gamma_2^{inc} + C \, \frac{\mu}{K} \, \varepsilon \, ,
\label{gamma2comp}
\end{equation}
where
\begin{equation}
C = - \frac{8}{121} \left( 8 \,\frac{\mu}{ \rho{g}R} + 11 + 6\, \frac{\delta\rho}{ \rho} \right) h_{2r}^\circ \, .
\label{factorC}
\end{equation}
The terms within brackets result from $\Lambda$, $\Lambda_\nu$, and $\Lambda_\rho$, respectively.
For Europa, \citet{wahr2006} assume that $\xi\sim1/3$, $\mu/(\rho{g}R)\sim1$ and $\delta\rho/\rho=-0.08$.
These parameters yield $\gamma_2^{inc}\sim3.1\varepsilon$ and $C\sim-1.5$, i.e.\ a correction of 18\%.
This result is close to the correction of $C=-1.4$ that \citet{wahr2006} found by fitting Eq.~(\ref{gamma2comp}) to their numerical model (see their Eq.~(12)).
For Titan and Ganymede, the correction can be smaller or larger, depending on the choice of $\mu$ and $\delta\rho$, while it goes down to 9\% for small bodies with $\mu/(\rho{}gR)\gg1$.
Corrections due to crust compressibility are larger than the error due to the membrane approximation, unless the crust is very thick (Section~\ref{AccuracyLove}).
They are also larger than corrections due to core-mantle elasticity (Section~\ref{DeepInterior}).
If the crust is thin, the membrane approach is thus more accurate than the incompressible propagator matrix method.

\section{Static Love numbers}
\label{LoveStatic}

\subsection{Outline of the method}

\begin{figure}
   \centering
   \includegraphics[width=12cm]{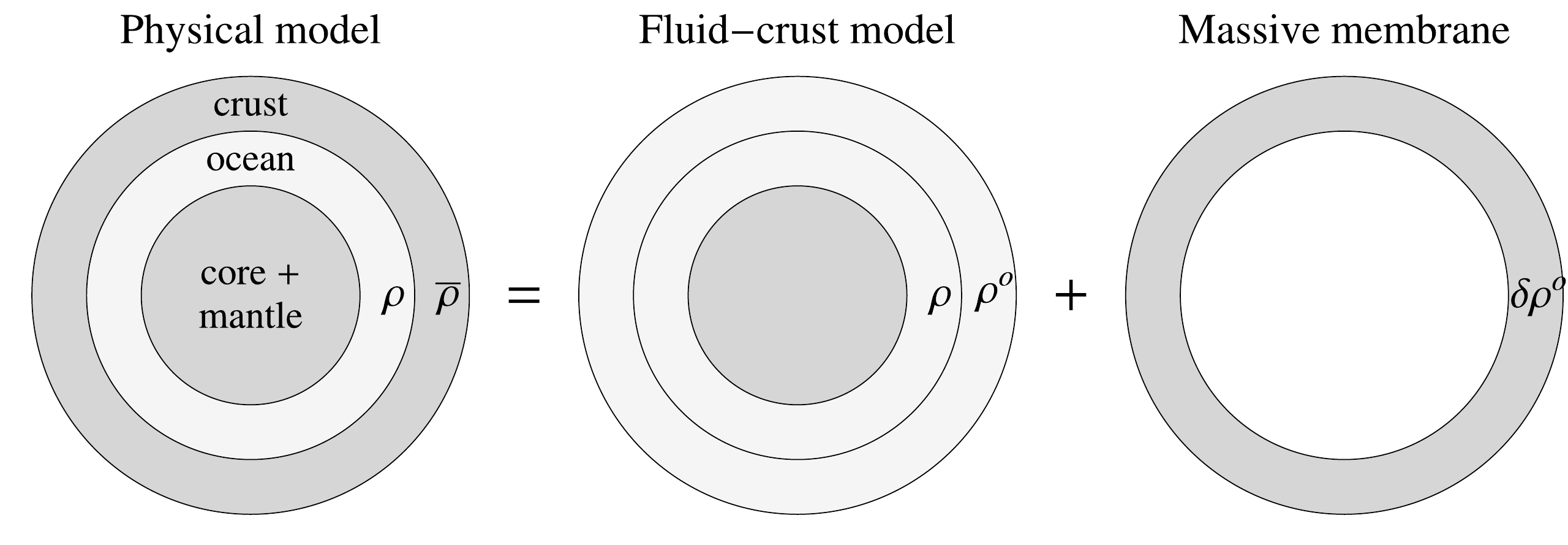}
   \caption[Second basic idea of the membrane approach]
   {Second basic idea of the membrane approach (for static tides): decomposition of the physical model into a fluid-crust model to which is added the contribution of the crust (`addition' should be taken in a figurative sense).
   Darker (resp.\ lighter) shades represent solid (resp.\ liquid) layers.
   See Section~\ref{FluidCrustModel} for details.}
   \label{FigFluidCrust}
\end{figure}

The $l_n-h_n$ and $k_n-h_n$ relations of Section~\ref{RelationsBetweenLoveNumbers} depend explicitly on the properties of the shallow interior (crust and top layer of the ocean).
By contrast, the deep interior structure appears only implicitly  through the Love numbers themselves.
This separation between the contributions of the shallow and deep interior can be taken further with the second basic idea of the membrane approach (Fig.~\ref{FigFluidCrust}):
determine the Love numbers of the body with its viscoelastic crust (or \textit{physical model}) in terms of the Love numbers of a simpler model (or \textit{fluid-crust model}) in which the crust behaves as a fluid \citep{beuthe2014}.
In a sense, this method factorizes the effect of the membrane from the deep interior.
I will here extend the method of \citet{beuthe2014} to a thin crust of finite thickness with a density jump at the crust-ocean boundary.

The static limit is a crucial requirement of the factorization between membrane and deep interior.
In this limit, gravity decouples from stress and strain within the fluid layer, so that one can solve for the gravity variables $(y_5,y_7)$ independently of the displacement and stress variables $(y_1,...,y_4)$ \citep{saito1974}.
The method consists in solving a linear system of two equations consisting of
\begin{itemize}
\item a constraint on the gravity variables at the crust-ocean boundary,
\item a scaling relation between the gravity solutions of the physical and fluid-crust models.
\end{itemize}
Once $k_n$ is determined, radial and tangential Love numbers result from the $k_n-h_n$ and $l_n-h_n$ relations. 

\subsection{Fluid-crust model}
\label{FluidCrustModel}

The fluid-crust model is defined as having the same internal structure as the body to be modeled (called the physical model), except that the crust is fluid and of density $\rho^\circ$.
There are two obvious choices for the fluid-crust density $\rho^\circ$:
it is equal either to the original crust density ($\rho^\circ=\bar\rho$), or to the density of the top layer of the ocean ($\rho^\circ=\rho$).
The latter choice makes it simpler to compute the Love numbers of the fluid-crust model, but it changes the bulk density of the body from $\rho_b$ to $\rho^\circ_b$.
Conservation of mass yields at ${\cal O}(\varepsilon)$
\begin{equation}
\rho_b = \rho^\circ_b + 3\, \delta\rho^\circ \,  \varepsilon \, ,
\label{rho0}
\end{equation}
where
\begin{equation}
\delta\rho^\circ = \bar\rho- \rho^\circ \, .
\label{delrhocirc}
\end{equation}
In analogy with Eq.~(\ref{defxi}), the ocean-to-bulk and crust-to-bulk density ratios for the fluid-crust model are denoted
\begin{equation}
\left( \xi^\circ \, , \bar\xi^\circ \right)  = \left( \frac{\rho}{\rho^\circ_b} \, , \frac{\rho^\circ}{\rho^\circ_b} \right) \, .
\label{defbarxicirc}
\end{equation}
Following the notation of Section~\ref{FluidLayer}, I define
\begin{itemize}
\item $y_i$ as the solutions of the physical model: viscoelastic crust ($\Lambda\neq0$) of density $\bar\rho$.
\item $y_i^\circ$ as the solutions of the fluid-crust model: fluid crust ($\Lambda=0$) of density $\rho^\circ$.
\end{itemize}
I adopt the same conventions for the corresponding Love numbers.
For example,
\begin{eqnarray}
k_n+1 = y_5(R) \, ,
\nonumber \\
k_n^\circ+1 = y_5^\circ(R) \, .
\end{eqnarray}
Simple fluid-crust models are given in Appendix~\ref{FluidCrustModels}.

\subsection{Gravity at the crust-ocean boundary}
\label{COBcondition}

First, I relate the gravity perturbation at the crust-ocean boundary to its value at the surface (equal to $k_n$):
Substituting Eq.~(\ref{knhneps}) into Eq.~(\ref{y5crust}), I get
\begin{equation}
y_5(R_\varepsilon) =  \left( 1 +  \left( n+1 - 3 \, \bar\xi \, \right)  \varepsilon \right) \left( k_n+1 \right) - \left( 2 n +1 \right) \varepsilon \, .
\label{y5propag}
\end{equation}
In the fluid-crust limit, this equation reads
\begin{equation}
y_5^\circ(R_\varepsilon) =  \left( 1 + \left( n+1 - 3 \, \bar\xi^\circ \right)  \varepsilon \right) \left( k_n^\circ+1 \right) -  \left( 2 n +1 \right) \varepsilon \, .
\label{y5propagcirc}
\end{equation}

Next, I relate the auxiliary gravity variable $y_7$ at the crust-ocean boundary to $y_5$ at the surface.
Starting with Eq.~(\ref{y7crust}), I express the right-hand side of this equation in terms of $h_n$ with the help of Eqs.~(\ref{snhn})-(\ref{dnhn}).
Then I eliminate $h_n$ in favor of $k_n$ with the $k_n-h_n$ relation (Eq.~(\ref{knhn})).
Terms of ${\cal O}(\varepsilon^2)$ are neglected except in the term in $\Lambda/(1+\Lambda)$ so that the limit of large $\Lambda$ is well-behaved.
At the crust-ocean boundary, the gravity variable $y_7$ is thus given by
\begin{equation}
R y_7(R_\varepsilon) + 3 \, \xi \left( \frac{ \Lambda}{1+\Lambda} + 2 \left(n-1\right) \frac{\bar\rho}{\rho} \, \varepsilon \right) \left( k_n+1 \right) = \left( 2 n +1 \right) \left( 1 - \varepsilon \left( n-1 - 3 \, \bar\xi \, \right)  \right) .
\label{y7y5}
\end{equation}
The fluid-crust limit of this equation does not pose any problem because its derivation does not require the lithospheric condition (Eq.~(\ref{LithoCond2})).
Indeed, the last term of Eq.~(\ref{y7crust}) does not diverge in the fluid-crust limit.
The fluid-crust limit of Eq.~(\ref{y7y5}) reads
\begin{equation}
R y_7^\circ(R_\varepsilon) + 6 \left(n-1\right) \bar\xi^\circ \,  \varepsilon \left( k_n^\circ+1 \right) = \left( 2 n +1 \right) \left( 1 - \varepsilon \left( n-1 - 3 \, \bar\xi^\circ \right) \right) .
\label{y7y50}
\end{equation}

\subsection{Gravity scaling with crustal parameters}

I recall here the scaling argument explained in \citet{beuthe2014}, extending it to a crust of finite thickness and allowing for a possible density contrast at the crust-ocean boundary.
At the mantle-ocean boundary ($r{=}R_m$), the variables $(y_5,y_7)$ can be related by continuity to the six $y_i$ solutions within the mantle.
In the mantle, the $y_i$-vector is a linear combination of three independent solutions because there are only three regular solutions at the center of the body (if there is a liquid core, the solutions within the mantle can be expressed in terms of three unknown constants at the core-mantle boundary \citep[e.g.][]{sabadini2004}).

The three constants of this linear combination reduce to one after applying the free-slip condition ($y_4(R_m)=0$) and the fluid condition taken in the static limit (Eq.~(\ref{fluideq})) at the mantle-ocean boundary.
Both conditions are homogeneous in the sense that they do not introduce a constant term that would be independent of the $y_i$ (by contrast Eq.~(\ref{y7y5}) is not homogeneous).
Therefore, the six $y_i(R_m)$ at the mantle-ocean boundary and $(y_5,y_7)$ at any radius within the fluid linearly depend on one free constant ${\cal C}$, with proportionality factors $f_i(r)$ depending on the radius and on the structure of the body below the crust (densities, radii of interfaces, rheology) but not on the crustal parameters $(\Lambda,\delta\rho^\circ)$ themselves:
\begin{equation}
y_i(r) = {\cal C} f_i(r) \, .
\end{equation}
The variables $y_i(R_m)$ and $(y_5,y_7)$ become dependent on $(\Lambda,\delta\rho^\circ)$ when the constant ${\cal C}$ is determined from the crust-ocean boundary condition on gravity (Eq.~(\ref{y7y5})).
Now suppose that the solutions $y_{i,a}={\cal C}_a f_i(r) $ and $y_{i,b}={\cal C}_b f_i(r) $ are associated with the crustal parameters $(\Lambda_a,\delta\rho_a^\circ)$ and $(\Lambda_b,\delta\rho_b^\circ)$, respectively.
Then the ratios $y_{i,a}/y_{i,b}$ are equal (for any $i$) to the ratio $\zeta={\cal C}_a/{\cal C}_b$.

From the general argument explained above, I can write
\begin{equation}
\zeta = \frac{y_5 \, (r)}{y_5^\circ(r)} = \frac{y_7 \, (r)}{y_7^\circ(r)}  \;\;  \mbox{for} \;\;  R_m<r<R_\varepsilon \, .
\label{scaling1}
\end{equation}
The ratio $\zeta$ depends on $(\Lambda,\delta\rho^\circ)$ and is related to the $y_i$ solutions within the mantle by
\begin{equation}
\zeta = \frac{y_1(R_m)}{y_1^\circ(R_m)} \, .
\label{zeta1}
\end{equation}
The ratio $\zeta$ can be related to $(k_n,k_n^\circ)$ by propagating $(y_5,y_5^\circ)$ to the surface.
Using Eqs.~(\ref{y5propag})-(\ref{y5propagcirc}), I write $\zeta$ at ${\cal O}(\varepsilon)$ as
\begin{equation}
\zeta =  \left( 1 - 3 \varepsilon \, \frac{\delta\rho^\circ}{\rho_b} \right) \frac{k_n+1}{k_n^\circ+1} \, .
\label{zeta2}
\end{equation}
Eqs.~(\ref{zeta1}) and (\ref{zeta2}) are useful when decomposing tidal heating into crustal and mantle contributions (Section~\ref{MicroMacro}).

\subsection{Explicit formulas for $k_n$ and $h_n$}
\label{ExplicitFormula}

The scaling relation given by Eq.~(\ref{scaling1}) evaluated at the crust-ocean boundary reads
\begin{equation}
\frac{y_5 \, (R_\varepsilon)}{y_5^\circ(R_\varepsilon)} = \frac{y_7 \, (R_\varepsilon)}{y_7^\circ(R_\varepsilon)} \, .
\label{scaling3}
\end{equation}
Substituting Eqs.~(\ref{y5propag})-(\ref{y7y50}) into Eq.~(\ref{scaling3}), I can solve to ${\cal O}(\varepsilon)$ for $k_n$ in terms of $k_n^\circ$.
Next, I combine this result with the $k_n-h_n$ relation in order to express $h_n$ in terms of $h_n^\circ$.
The resulting formulas read
\begin{eqnarray}
k_n +1 &=& \frac{ k_n^\circ + 1 }{1 + \frac{3\,\xi^\circ}{2n+1} \left( \left( k_n^\circ + 1 \right) \frac{\Lambda}{1+\Lambda} + K_\rho \right) } \, ,
\nonumber \\
h_n  &=& \frac{ h_n^\circ }{1 + \left( 1 + \frac{3\, \xi^\circ}{2n+1} \, h_n^\circ \right) \Lambda + \Lambda_\chi + H_\rho } \, ,
\label{knhnstatic}
\end{eqnarray}
where all terms of ${\cal O}(\varepsilon)$ are in the denominator ($\xi$ and $\xi^\circ$ are interchangeable in these terms).
The density corrections $K_\rho$ and $H_\rho$ are defined below.
As the compressibility factor $\Lambda_\chi$ does not appear in the formula for $k_n$, crust compressibility has a larger effect on $h_n$ than on $k_n$ ($k_n$ weakly depends on crust compressibility through $\Lambda$).

Density corrections are given by
\begin{eqnarray}
K_\rho &=&  2 \, \Big[ (n-1) \left( k_n^\circ + 1 \right) - (2 n + 1) \Big] \, \frac{\delta\rho^\circ}{\rho} \, \varepsilon \, ,
\nonumber \\
H_\rho &=& \Lambda_\rho - \frac{2n+1}{h_n^\circ} \, \frac{\delta\rho}{\rho} \, \varepsilon + \frac{3\,\xi^\circ}{2n+1} \, K_\rho \, .
\label{KHrho}
\end{eqnarray}
In the last equation, the first two terms in the right-hand side depend on $\delta\rho$ whereas the last term depends on $\delta\rho^\circ$.

What is the impact of choosing the ocean density or the physical crust density as the fluid-crust density?
With the former choice ($\rho^\circ=\rho$), the fluid-crust Love numbers can, for example, be computed with the two-layer incompressible body of bulk density $\rho^\circ_b$, made of a viscoelastic core and a homogeneous ocean of density $\rho$  reaching the surface (Eq.~(\ref{hn0visco})).
Beware that it is not correct (if $\rho^\circ=\rho$) to compute the fluid-crust Love numbers with a bulk density equal to $\rho_b$.
Expanding the solution about a zeroth-order configuration of bulk density $\rho_b$ affects density corrections of ${\cal O}(\varepsilon)$.
This procedure is illustrated in Appendix~\ref{LoveFormulasRigid}, where I prove that the membrane formulas for Love numbers agree with the analytical model of \citet{wahr2006} in the rigid mantle limit.

If the fluid-crust density is equal to the physical crust density ($\rho^\circ=\bar\rho$), then $\delta\rho^\circ=0$ and $K_\rho=0$.
The resulting formula for $k_n$ is formally identical to the one derived in the massless approach (Eq.~(57) of \citet{beuthe2014}).
Density corrections, however, are now hidden in $k_n^\circ$: when computing $k_n^\circ$, one must take into account an ocean of density $\rho$ and a fluid-like crust of density $\bar\rho$.
For example, Eq.~(\ref{h20threelayers}) gives the fluid-crust Love numbers for the three-layer incompressible body made of an infinitely rigid mantle, a homogeneous ocean, and a homogeneous fluid crust differing in density.

\subsection{Micro-macro equivalence in tidal dissipation}
\label{MicroMacro}

The imaginary part of Love numbers is much smaller than their real part and is thus difficult to measure with geodetic methods (altimetry or gravity).
The imaginary part of $k_2$, however, manifests itself in a very different way: the tidal energy dissipated by viscoelastic friction in the whole body is proportional to \textit{Im}($k_2$).
Consider a synchronously rotating body with spin rate $\omega$, orbital eccentricity $e$ and obliquity $I$.
The global heat flow due to tidal dissipation is given by 
\begin{equation}
\dot{E} = - \frac{5}{2} \, \frac{(\omega{}R)^5}{G} \,  \mbox{\textit{Im}}(k_2) \, \Psi^{}_0 \, ,
\label{Edot2}
\end{equation}
where $\Psi^{}_0 = (21/5) e^2 + (3/5) \sin^2 I$ (Eqs.~(41)-(42) of \citet{beuthe2013}).
This \textit{macro approach} to tidal dissipation is equivalent to the \textit{micro approach}, in which the dissipated energy is computed by integrating over the whole body the product of the microscopic stress and strain rate.
In particular, the global heat flow due to dissipation in a thin floating crust is given by
\begin{equation}
\dot{E}_{crust} = \frac{3}{2} \,  \frac{(\omega{}R)^5}{G} \, \xi \,  |h_2|^2 \, \mbox{\textit{Im}} (\Lambda) \, \Psi^{}_0 \, .
\label{Edot1}
\end{equation}
This formula was derived in the massless membrane approach (Eq.~(98) of \citet{beuthe2014}) but one can prove that it still holds in the massive membrane approach.

If the core and mantle are purely elastic, the global heat flow should be equal to the heat flow coming from the crust: $\dot{E}=\dot{E}_{crust}$.
Applying the method of \citet{beuthe2014} to Eq.~(\ref{knhnstatic}), I decompose \textit{Im}($k_2$) into contributions from the crust and core-mantle system (denoted \textit{c-m}):
\begin{equation}
\mbox{\textit{Im}}( k_2) = \left[ \mbox{\textit{Im}}( k_2 )\right]_{crust} + \left[ \mbox{\textit{Im}}( k_2) \right]_{c-m} \, ,
\label{decompMain}
\end{equation}
where
\begin{eqnarray}
\left[  \mbox{\textit{Im}}( k_2 )\right]_{crust} &=& -\frac{3}{5} \, \xi \, \left| h_2 \right|^2 \mbox{\textit{Im}}( \Lambda ) \, ,
\nonumber \\
\left[ \mbox{\textit{Im}}( k_2) \right]_{c-m} &=& |\zeta|^2 \mbox{\textit{Im}}( k_2^\circ ) \, ,
\label{Imk2}
\end{eqnarray}
in which $\zeta$ is defined by Eq.~(\ref{zeta1}).
That $[ \mbox{\textit{Im}}( k_2 )]_{crust}$ is indeed the crustal contribution can be checked by substitution into Eq.~(\ref{Edot2}): the result is Eq.~(\ref{Edot1}) as it should be.
With respect to the fluid-crust model, the core-mantle contribution of the physical model is reduced by the factor $|\zeta|^2$, where $\zeta$ is the reduction in radial displacement of the mantle-ocean boundary due to the viscoelasticity and to the density contrast $\delta\rho^\circ$ of the crust.

\section{Numerical benchmarking}
\label{AccuracyLove}

In membrane formulas, there is a clear distinction between the contributions of the shallow interior (crust and crust-ocean boundary) and of the deep interior (the rest).
The latter influences Love numbers through the fluid-crust Love numbers $h_n^\circ$ and $k_n^\circ$.
The accuracy of the membrane formulas thus depends on two different things:
\begin{enumerate}
\item the crust thickness. This type of error is intrinsic (membrane approach = perturbative expansion) and unavoidable (higher-order corrections due to finite crust thickness).
\item the choice of a particular fluid-crust model. This type of error can be reduced to zero; it depends on the degree of complication one is willing to accept in order to compute fluid-crust Love numbers.
\end{enumerate}
In this section, I quantify the intrinsic error of the membrane formulas in specific models of the shallow interior of Europa and Titan, postponing the error analysis of fluid-crust models to Section~\ref{DeepInterior}.
Benchmarking is done with the program \textit{love.f} included in the software SatStress (available at http://code.google.com/p/satstress) \citep{wahr2009}.
I also use the propagator matrix method in the Fourier domain.

As a first step, I benchmark the membrane formulas for tidal Love numbers (radial and gravitational).
Titan is a good choice for this purpose because its Love number $k_2$ was recently estimated with Cassini data \citep{iess2012}; this is actually the only measurement of a tidal Love number for an icy satellite.
As a second step, I will benchmark the membrane formula for the tilt factor.
Europa is chosen here for two reasons: (1) the tilt factor can be used to estimate the crust thickness \citep{wahr2006} which is a key parameter regarding Europa's habitability, and (2) convection may occur in Europa's crust, making it a good laboratory for the analysis of the influence of crustal rheology on Love numbers.

\subsection{Models of Europa and Titan: shallow interior}
\label{InteriorModelsEuropaTitan}

\subsubsection{Density of crust and ocean}

Testing the intrinsic error of membrane formulas does not call for detailed models of the deep interior:
core and mantle are treated together as an incompressible homogeneous layer with fixed radius (\textit{core-mantle system}) while the ocean is approximated as a layer of uniform density.
The parameters describing the shallow interior are the crust-ocean density contrast and the thickness, density, and rheology of the crust.
For each set of shallow interior parameters, I adjust the density of the core-mantle system so that the bulk density is equal to the physical value (Table~\ref{TableBulkOrbital}).
I do not try to fit the moment of inertia because it does not lead to more realistic models as long as the core is not distinguished from the mantle.

What do we know about the density of the ocean?
An ocean made of pure water has a density of $1000\rm\,kg/m^3$ at $273\rm\,K$ and atmospheric pressure.
Pressure effects increase the density by about $0.45\rm\,kg/km$ for Titan (e.g.\ Fig.~1 of \citet{mitri2014}), with a comparable effect on Europa due to the similar surface gravity.
For Titan, this means that the mean density of a $400\rm\,km$-thick pure water ocean below a $50\rm\,km$-thick crust is close to $1100\rm\,kg/m^3$.
For Europa, the ocean is not as deep and the mean density of a pure water ocean is less than $1050\rm\,kg/m^3$.
Besides the pressure effect, the ocean density is affected by the presence of solutes \citep{sohl2010}: ammonia can lower the uncompressed density down to $950\rm\,kg/m^3$, whereas dissolved salts increase the uncompressed density up to $1200\rm\,kg/m^3$.
\citet{fortes2012} uses these bounds to construct two models of Titan in which the compressed mean density of the ocean is either $1020\rm\,kg/m^3$ (`light-ocean') or $1280\rm\,kg/m^3$ (`dense-ocean'), assuming a $100\rm\,km$-thick crust and a $250\rm\,km$-thick ocean.
\citet{mitri2014} obtain a similar upper bound by imposing that the density at the bottom of Titan's ocean is less than the density of the high-pressure ice layer below.
While the compression effect is smaller for Europa, \citet{kargel2000} consider even higher density solutions for Europa's ocean.
It is thus a reasonable choice to adopt $1020$ and $1280\rm\,kg/m^3$ as lower and upper bounds for the ocean density in Europa and Titan.

What do we know about the density of the crust?
Pure water ice at atmospheric pressure has a density varying between $917\rm\,kg/m^3$ at $273\rm\,K$ and $933\rm\,kg/m^3$ at $100\rm\,K$ \citep{feistel2006}, providing a lower bound on the crust density (set here at  $930\rm\,kg/m^3$).
Porosity is not taken into account because it probably only affects a thin layer near the surface.
The crust, however, is probably highly impure and chemically layered:  ice including silicate dust or salt hydrates could have a density of $1050\rm\,kg/m^3$ \citep{schubert2009}, while
\citet{kargel2000} and \citet{spaun2001} consider eutectic mixtures with magnesium sulfates having a density of $1144\rm\,kg/m^3$.
Since the membrane formulas depend on the crust density through the density contrast $\delta\rho/\rho$, I adopt a slightly higher upper bound of $1167\rm\,kg/m^3$ for the crust density.
With this choice,  the density contrast $\delta\rho/\rho$ is the same whether the crust and ocean densities take the lower values ($930$ and $1020\rm\,kg/m^3$) or the higher values ($1167$ and $1280\rm\,kg/m^3$).
Table~\ref{TableCrustOceanDensity} summarizes the three test cases (`Light', `Mixed', and `Dense').

\begin{table}[ht]\centering
\ra{1.3}
\small
\caption[Density models for the crust and ocean of Europa and Titan]{\small
Density models for the crust and ocean of Europa and Titan.
The density of the core-mantle system is adjusted so that the bulk density remains the same.
The top of the mantle is at a depth of $170\rm\,km$ for Europa and $350\rm\,km$ for Titan.}
\begin{tabular}{@{}llll@{}}
\hline
Density model & Crust  & Ocean & Contrast \\
 & $\bar\rho$ ($\rm kg/m^3$) & $\rho$ ($\rm kg/m^3$) & $\delta\rho/\rho$  \\
\hline
Light & 930 & 1020 & $-0.09$  \\
Mixed & 930 & 1280 & $-0.27$  \\
Dense & 1167 & 1280 & $-0.09$   \\
\hline
\end{tabular}
\label{TableCrustOceanDensity}
\end{table}%

\subsubsection{Rheology}
\label{RheologyTitanEuropa}

Table~\ref{TableBodyRheology} gives the values adopted for the viscoelastic parameters in the core-mantle system (purely elastic and incompressible) and in the crust.
The value chosen for the shear modulus of the core-mantle system is justified in Section~\ref{DeepInterior}.
The critical viscosity $\eta_{crit}$ is defined by Eq.~(\ref{etacrit}).
The rheology of the crust is discussed in more detail below.
Although ocean compressibility is listed as an input of SatStress, \citet{saito1974} showed that the stress-strain relation of a liquid layer is irrelevant when computing static deformations.
Thus SatStress output is in principle independent of ocean compressibility which can be set either to its physical value (about 2~GPa) or to a much larger value simulating incompressibility.

\begin{table}[ht]\centering
\ra{1.3}
\small
\caption[Viscoelastic parameters of Europa and Titan]{\small
Viscoelastic parameters (viscosity is only relevant to Europa; Titan is treated as an elastic body).}
\begin{tabular}{@{}llll@{}}
\hline
Parameter &  Symbol & Value & Unit \\
\hline
Shear modulus of core-mantle system & $\mu_m$ & 40 & GPa \\
Bulk modulus of core-mantle system (for SatStress) & $K_m$ & $10^{20}$ & Pa \\
Shear modulus of elastic ice & $\mu_E$ & 3.5 & GPa \\
Poisson's ratio of elastic ice & $\nu_E$ & 0.33 & - \\
Viscosity of top ice layer${}^a$ & $\eta_{top}$ & $10^7$ & $\eta_{crit}$ \\
Viscosity of bottom ice layer${}^a$ & $\eta_{bot}$ & 10/1/0.1 & $\eta_{crit}$ \\
\hline
\multicolumn{4}{l}{\scriptsize ${}^a$ Europa: $\eta_{crit}=1.71\times10^{14}\rm\,Pa.s$; Titan: $\eta_{crit}=7.67\times10^{14}\rm\,Pa.s$ (Eq.~(\ref{etacrit})).} \\
\end{tabular}
\label{TableBodyRheology}
\end{table}%

Viscous effects are probably small within Titan's crust.
Uncompensated gravity anomalies indeed suggest that the crust is not convecting \citep{hemingway2013,mitri2014,lefevre2014}.
I will thus assume that Titan's crust is purely elastic.
There are no comparable gravity data constraining the state of Europa's crust which could be either in a conductive regime or in a stagnant lid regime (Section~\ref{StagnantLidRegime}).
In \citet{moore2006}'s study, Europa's crust always convects and the total crust thickness $d$ varies between 20~km and 120~km, depending on the  viscosity $\eta_{bot}$ of the convecting layer.
The three following cases can be distinguished (recall that the critical state is defined by the critical viscosity $\eta_{crit}=\mu_E/\omega$; see Table~\ref{TableBodyRheology} and Appendix~\ref{MaxwellRheology}).
If the convecting layer is close to its critical state, the crust is thin ($d\sim20\,$km) and the top ice layer is relatively thick (say 2/5 of $d$).
If the convecting layer is elastic-like ($\eta_{bot}\gg\eta_{crit}$), the crust can be thick ($d$ up to 100~km) and the thickness of the top ice layer scales with the total thickness (it is thus relatively thick as in the critical case).
If the convecting layer is fluid-like ($\eta_{bot}\ll\eta_{crit}$), the crust can be thick ($d$ up to 120~km) but the top ice layer does not scale with $d$: its thickness remains close to its value at critical viscosity.

Table~\ref{TableCrustRheology} summarizes the three test cases for crustal rheology.
The values of the effective viscoelastic parameters $(\bar\mu,\bar\nu,\bar\chi)$ are computed with Eqs.~(\ref{mubar2}), (\ref{nubar2}) and (\ref{chibar}) using Maxwell rheology (Eq.~(\ref{munuMaxwell})).
In the elastic-like case, the effective parameters are nearly equal to their elastic value.
In the critical model, the shear moduli of the conductive and convective layers are equal to $\mu_E$ and $\mu_E(1+i)/2$, respectively, so that their weighted average is $0.7+0.3i$.
In the fluid-like case, the real part of the effective shear modulus is approximately equal to the elastic shear modulus multiplied by the relative thickness of the lithosphere (Eq.~(\ref{dlitho})), while the effective Poisson's ratio is getting closer to its value in the fluid limit ($\nu=1/2$).

\begin{table}[ht]\centering
\ra{1.3}
\small
\caption[Rheology models for Europa's crust]{Rheology models for Europa's crust.
The crust (of thickness $d$) is made of two homogeneous layers: the top layer (of thickness $d_{top}$) is purely elastic while the bottom layer is viscoelastic.}
\begin{tabular}{@{}llllll@{}}
\hline
Rheology model & $\eta_{bot}/\eta_{crit}$ & $d_{top}/d$ & $\bar\mu/\mu_E$ &  $\bar\nu$ & $\bar\chi$ \\
\hline
Elastic-like & 10 & 0.4 &  $0.994+0.059\,i$  & $0.331-0.009\,i$ & $0.506+0.020\,i$ \\
Critical & 1 & 0.4 & $0.700+0.300\,i$  & $0.385-0.034\,i$ & $0.415+0.140\,i$ \\
Fluid-like & 0.1 & 0.1 &  $0.109+0.089\,i$ & $0.435+0.061\,i$ & $0.061+0.067\,i$ \\
\hline
\end{tabular}
\label{TableCrustRheology}
\end{table}%

\subsection{Love numbers of Titan}
\label{NumericalBenchmarkLove}

In this section, I compute the gravitational and radial Love numbers for models of Titan differing in ocean and crust densities (Table~\ref{TableCrustOceanDensity}).
The ocean density is a crucial parameter which could account for the large measured value of $k_2$ \citep{iess2012}.
I assume that the crust is purely elastic (Section~\ref{RheologyTitanEuropa}).

\begin{figure}
\hspace{0mm}
    \includegraphics[width=14cm]{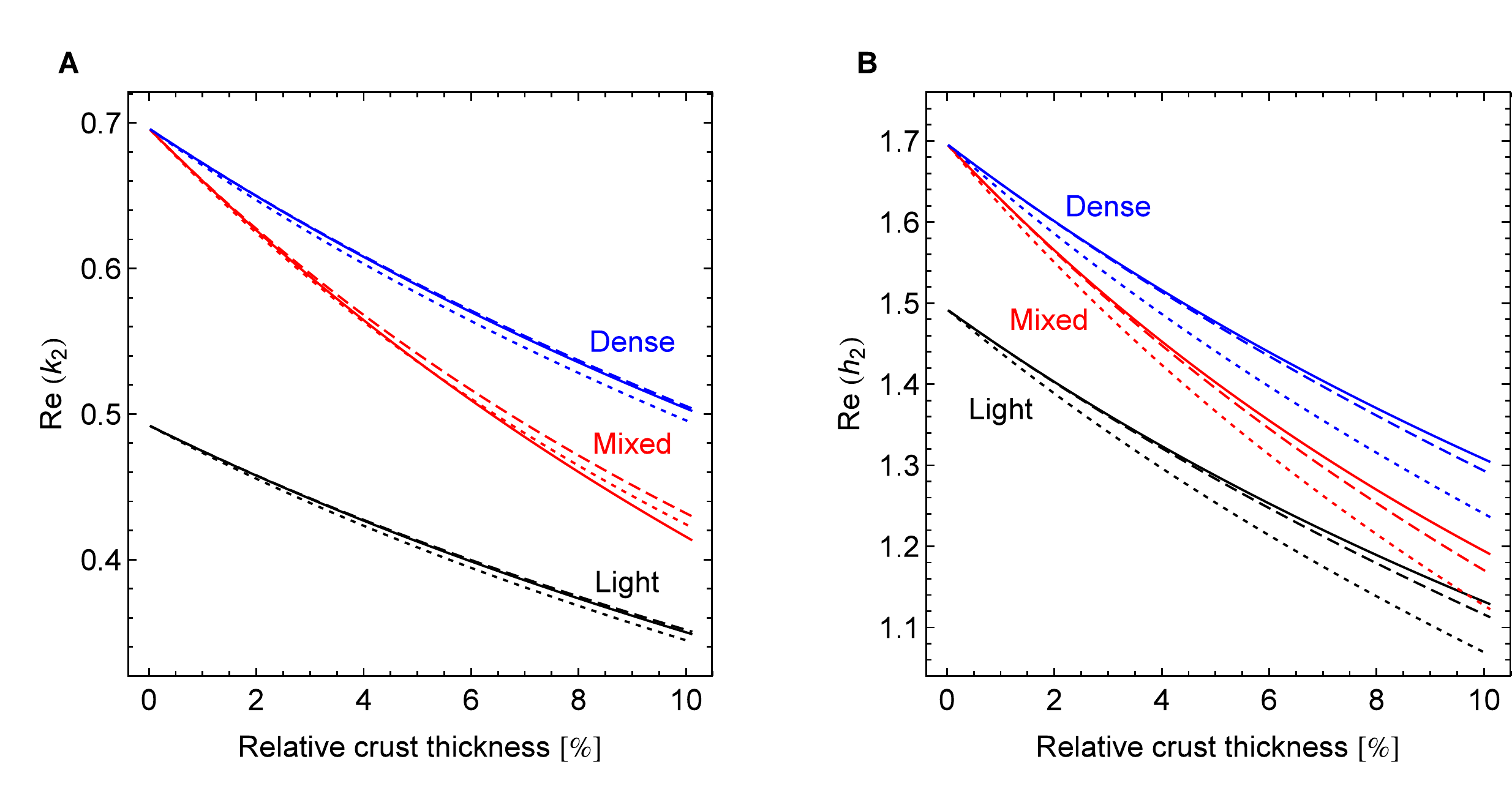}
   \caption[Tidal Love numbers of Titan]
   {Tidal Love numbers of Titan as functions of relative crust thickness:
   (A) $k_2$, and (B) $h_2$.
   Solid curves are membrane estimates.
   Dashed curves are SatStress predictions.
   Dotted curves are predictions made with the propagator matrix method.
   Light/Mixed/Dense refer to the density models of Table~\ref{TableCrustOceanDensity}.
   See Section~\ref{NumericalBenchmarkLove} for details.}
   \label{FigLoveTitan}
\end{figure}

Fig.~\ref{FigLoveTitan} shows the Love numbers $k_2$ and $h_2$ of Titan in terms of the relative crust thickness.
Solid curves show the membrane predictions (Eq.~(\ref{knhnstatic}) with an incompressible purely elastic core; the fluid-crust model is given by Eq.~(\ref{hn0visco})).
Dashed curves show SatStress predictions if the core-mantle system is incompressible.
Dotted curves show the results of the propagator matrix method (equivalent to SatStress predictions for an incompressible body).
Membrane and SatStress predictions agree perfectly in the limit of zero crust thickness.
For non-zero crust thickness, the agreement is excellent up to a relative crust thickness of 5\% and reasonable up to 10\%.
Compared to the membrane approach, the accuracy of the propagator matrix method is similar for $k_2$ but much worse for $h_2$, because crust compressibility has a much larger on $h_2$ (up to several percents) than on $k_2$.

\begin{figure}
\hspace{0mm}
    \includegraphics[width=14cm]{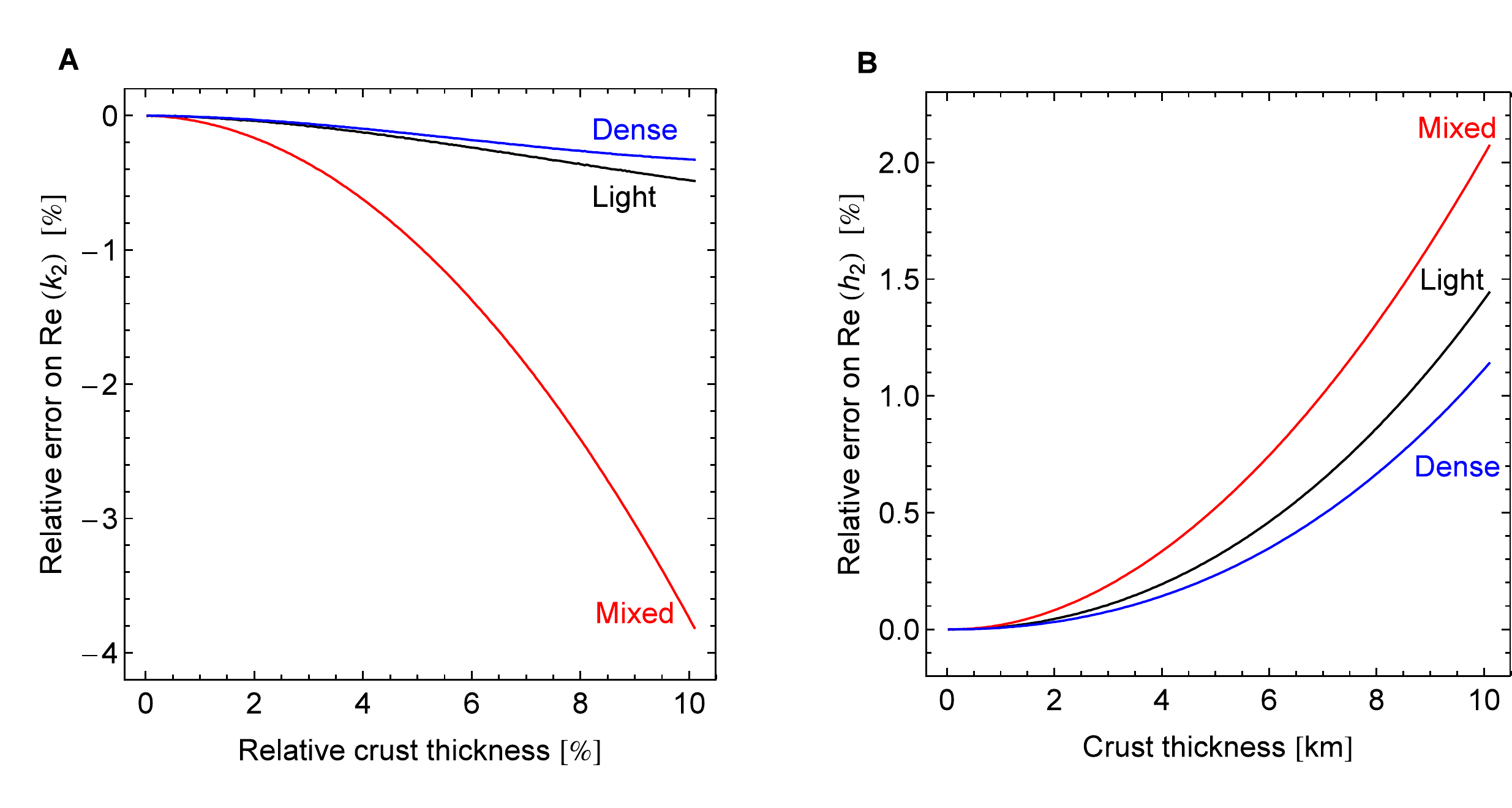}
   \caption[Relative error on the tidal Love numbers of Titan]
   {Tidal Love numbers of Titan: relative error of the membrane prediction for (A) $k_2$, and (B) $h_2$.
   The error corresponds to the difference between the solid and dashed curves of Fig.~\ref{FigLoveTitan}.}
   \label{FigLoveTitanError}
\end{figure}

Fig.~\ref{FigLoveTitanError} shows the relative error between the membrane and SatStress predictions.
Solid curves show the error for the Love numbers of Fig.~\ref{FigLoveTitan} (for which the core-mantle system is incompressible).
The error is less than 1\% (for $k_2$) and 0.5\% (for $h_2$) if the relative crust thickness is less than  5\%.


\subsection{Tilt factor of Europa}
\label{NumericalBenchmarkTilt}

In Section~\ref{NumericalBenchmarkLove}, I showed that membrane formulas approximate very well $k_2$ and $h_2$, and that the accuracy improves as the crust thickness $d$ decreases.
These observations, however, do not prove that the membrane formulas are the correct perturbative expansions of thick shell theory: it only shows that all crustal contributions to Love numbers are proportional to $d$.
The best way to test the dependence of Love numbers on crustal parameters is to study the tilt factor $\gamma_2$, because this quantity vanishes in the limit $d\rightarrow0$.
In particular, $\gamma_2'(0)$ (the slope of the tilt factor at $d=0$) is a good indicator of the correctness and completeness of membrane formulas.
For example, \citet{beuthe2014} concluded from benchmarking $\gamma_2'(0)$ that compressibility terms are missing in the $k_2-h_2$ relation derived from classical thin shell theory.

\begin{figure}
\hspace{-5mm}
   \includegraphics[width=15.3cm]{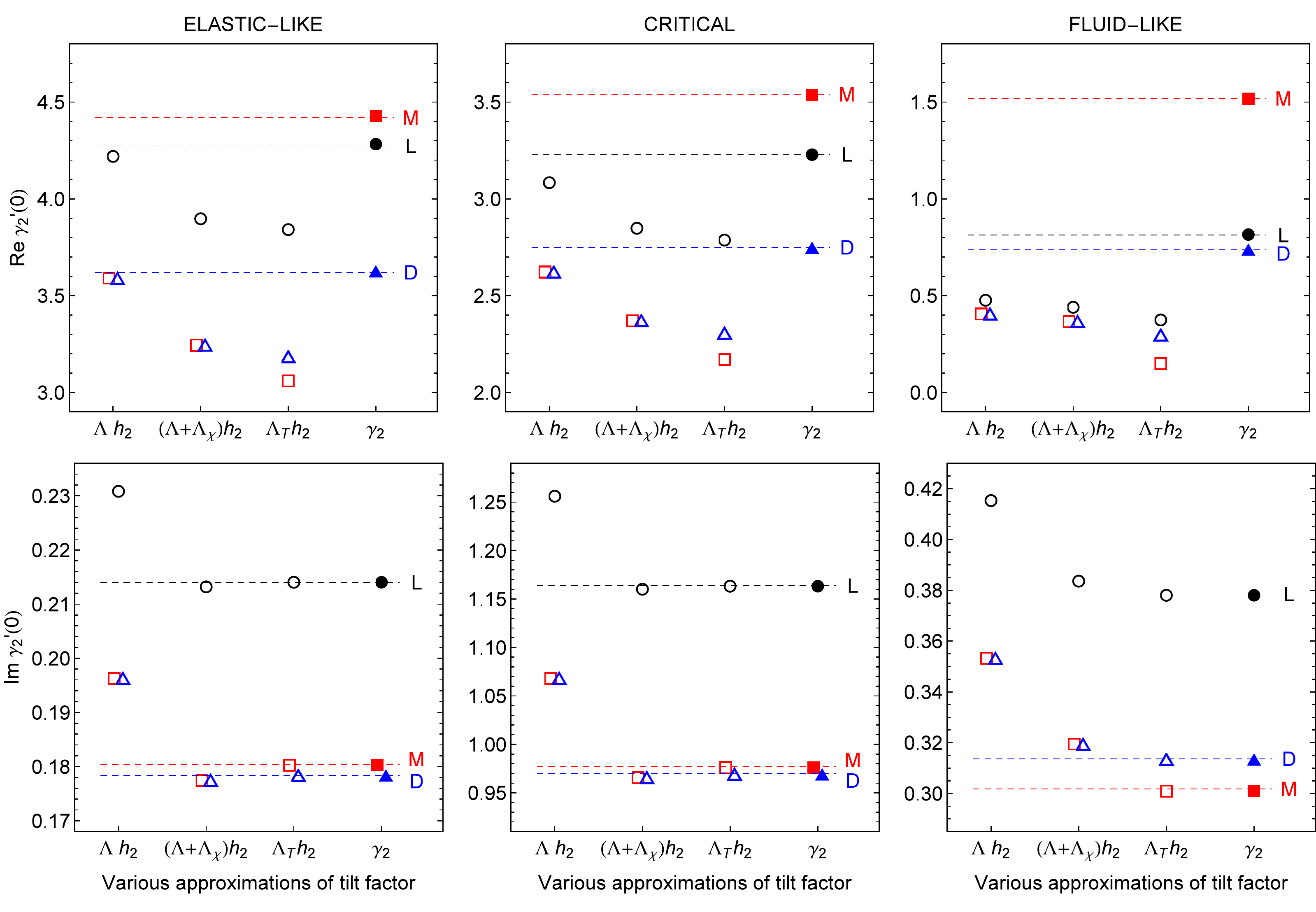}
   \caption[Slope of Europa's degree-two tilt factor]
   {Slope (at zero crust thickness) of Europa's degree-two tilt factor: Re$(\gamma_2'(0))$ (upper panels), and Im$(\gamma_2'(0))$ (lower panels).
   The different columns correspond to the crustal rheology models of Table~\ref{TableCrustRheology}.
   Circles/squares/triangles are membrane estimates for the Light/Mixed/Dense models of Table~\ref{TableCrustOceanDensity}.
   Empty symbols show partial membrane estimates (with or without compressibility and density corrections).
   Filled symbols show the full membrane estimates (Eq.~(\ref{tiltn})).
   Dashed lines show SatStress predictions ($d=1\rm\,km$), with L/M/D standing for Light/Mixed/Dense models.
    For membrane estimates, $h_2$ is evaluated with Eq.~(\ref{hn0visco}): $h_2=1.27$ (Model L) or $h_2=1.35$ (Models M and D).
    See Section~\ref{NumericalBenchmarkTilt} for details.}
   \label{FigTiltMembrane}
\end{figure}

In the membrane approach, $\gamma_2'(0)$ is given by the right-hand side of Eq.~(\ref{tiltn}) divided by the relative crust thickness $\epsilon$.
Fig.~\ref{FigTiltMembrane} shows the real and imaginary parts of $\gamma_2'(0)$ for the critical rheology model of Europa (Table~\ref{TableCrustRheology}).
The ticks on the $x$-axis correspond to four different approximations of the tilt factor (denoted by circles, squares and triangles):
\begin{enumerate}
\item $\Lambda{}h_2$: dominant elastic term (thin shell theory prediction of \citet{beuthe2014}),
\item $(\Lambda+\Lambda_\chi)h_2$: full elastic term, including the compressibility correction,
\item $\Lambda_T{}h_2$: terms proportional to $h_2$, including the minor density correction $\Lambda_\rho$,
\item $\gamma_2$: full membrane prediction, including the major density correction $-5(\delta\rho/\rho$).
\end{enumerate}
Fig.~\ref{FigTiltMembrane} demonstrates that all contributions are required in order to reach a near perfect agreement with the predictions of the numerical benchmark (horizontal dashed lines).
Regarding the real part of $\gamma_2'(0)$, one can say that
\begin{itemize}
\item the slope of the tilt factor decreases as the crust becomes softer (elastic-like $\rightarrow$ critical $\rightarrow$ fluid-like);
\item the elastic term is generally dominant though the density term can become large if the lower crust is fluid-like, at least for large satellites ($\mu_E/(\rho{}gR)\sim1$);
\item compressibility and minor density corrections are negative while the major density correction is positive: the various corrections thus partially cancel each other;
\item if the rheology is elastic-like, the tilt factor is not very sensitive to the ocean density (Light and Mixed models yield similar results); this is also true of Titan, but not of all icy satellites because it depends on the values of the crust parameters and of the surface gravity;
\item if the rheology is fluid-like, the tilt factor is sensitive to the crust-ocean density contrast, but does not depend much on the crust and ocean densities taken separately.
\end{itemize}
Regarding the imaginary part of $\gamma_2'(0)$, it is significant (with respect to the real part) if the rheology is critical or fluid-like.
The compressibility correction has a 10\% effect while density corrections contribute little to \textit{Im}($\gamma_2'(0)$).

Consider now the tilt factor at non-zero values of the crust thickness.
Theoretically, the right-hand side of Eq.~(\ref{tiltn}) could be evaluated with $h_2$ equal to its fluid-crust value (of ${\cal O}(1)$) because it is already multiplied by a term of ${\cal O}(\varepsilon)$.
However, the ${\cal O}(\varepsilon)$ terms in $h_2$ have a large multiplying factor so that $h_2$ is rather sensitive to crustal thickness (see Fig.~\ref{FigLoveTitan} for Titan).
By contrast, next-to-leading corrections to $\Lambda$ are relatively much smaller.
I can quantify this assertion with the homogeneous crust model of Appendix~\ref{HomogeneousCrustModel}.
For static tides, $\gamma_2=\Lambda{}h_2$ and $h_2=h_{2r}^\circ/(1+h_{2r}^\circ{}\Lambda)$ where $\Lambda=X_a\hat\mu$ and $X_a\sim(24\varepsilon/11)(1+4\varepsilon/11)$ (see Table~\ref{TableGeoFac}).
At next-to-leading order, $\Lambda$ is modified by the factor $1+4/11\varepsilon\sim1+0.4\varepsilon$ whereas $h_2$ is modified by the factor $1+h_{2r}^\circ{}\Lambda\sim1+4.7\varepsilon$ (assuming $h_{2r}^\circ\sim1.25$ and $\hat\mu\sim1.71$ for Europa).
If Europa's crust is $40\rm\,km$ thick, next-to-leading order corrections modify $\Lambda$ and $h_2$ by 1\% and 12\%, respectively.
Therefore the membrane estimate of the tilt factor is much more accurate if $h_2$ is evaluated at ${\cal O}(\varepsilon)$ in the right-hand side of Eq.~(\ref{tiltn}).

\begin{figure}
   \hspace{-5mm}
   \includegraphics[width=15.3cm]{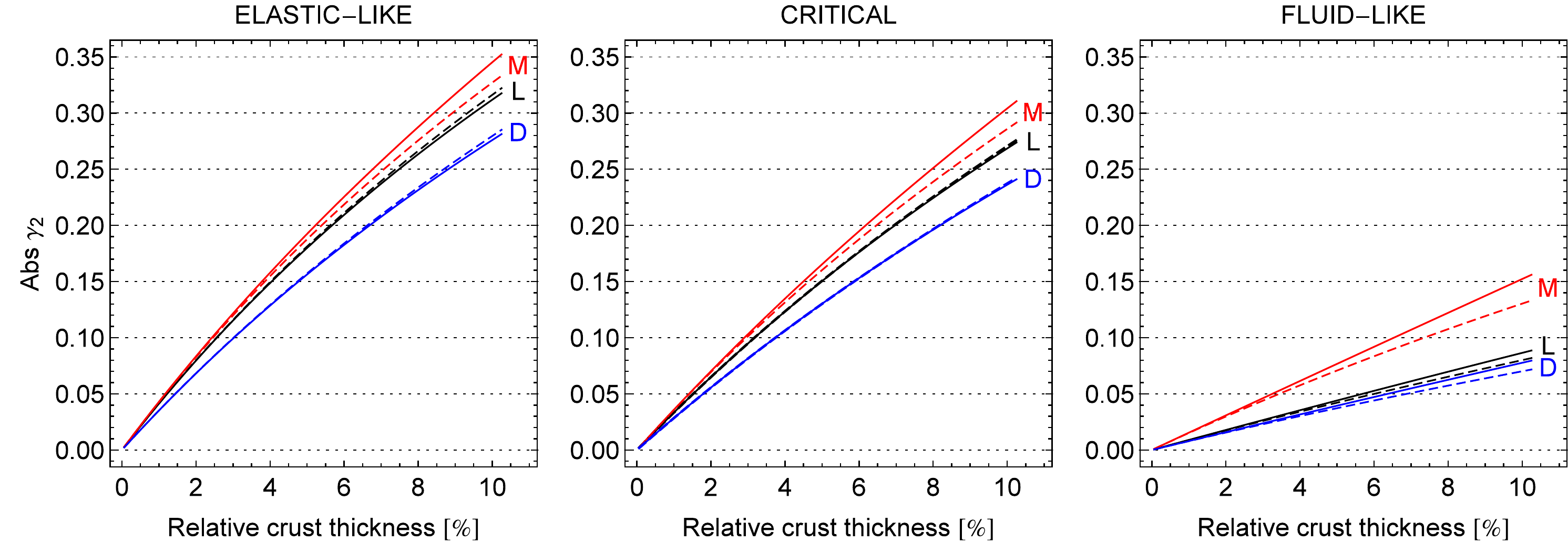}
   \caption[Europa's degree-two tilt factor]
   {Europa's degree-two tilt factor (absolute value) as a function of crust thickness.
   Solid curves are the membrane estimates (Eq.~(\ref{tiltn}) in which $h_2$ is evaluated with Eq.~(\ref{knhnstatic})).
   Dashed curves are SatStress predictions.
   The different panels correspond to the crustal rheology models of Table~\ref{TableCrustRheology}.
    L/M/D stand for Light/Mixed/Dense denoting the density models of Table~\ref{TableCrustOceanDensity}.
    See Section~\ref{NumericalBenchmarkTilt} for details.}
   \label{FigTiltThickness}
\end{figure}

Fig.~\ref{FigTiltThickness} shows the tilt factor of Europa as a function of the relative crust thickness for the density and rheology models of Tables~\ref{TableCrustOceanDensity} and \ref{TableCrustRheology}.
According to the discussion above, the right-hand side of Eq.~(\ref{tiltn}) is evaluated with $h_2$ depending on the crust thickness (Eq.~(\ref{knhnstatic})).
This effect accounts for the concavity of the tilt factor curves in Fig.~\ref{FigTiltThickness}.
Membrane estimates (solid curves) are in good agreement with SatStress predictions (dashed curves).
As noted for Fig.~\ref{FigTiltMembrane}, the tilt factor becomes smaller if the lower part of the crust becomes fluid-like.

\begin{figure}
   \centering
   \includegraphics[width=6cm]{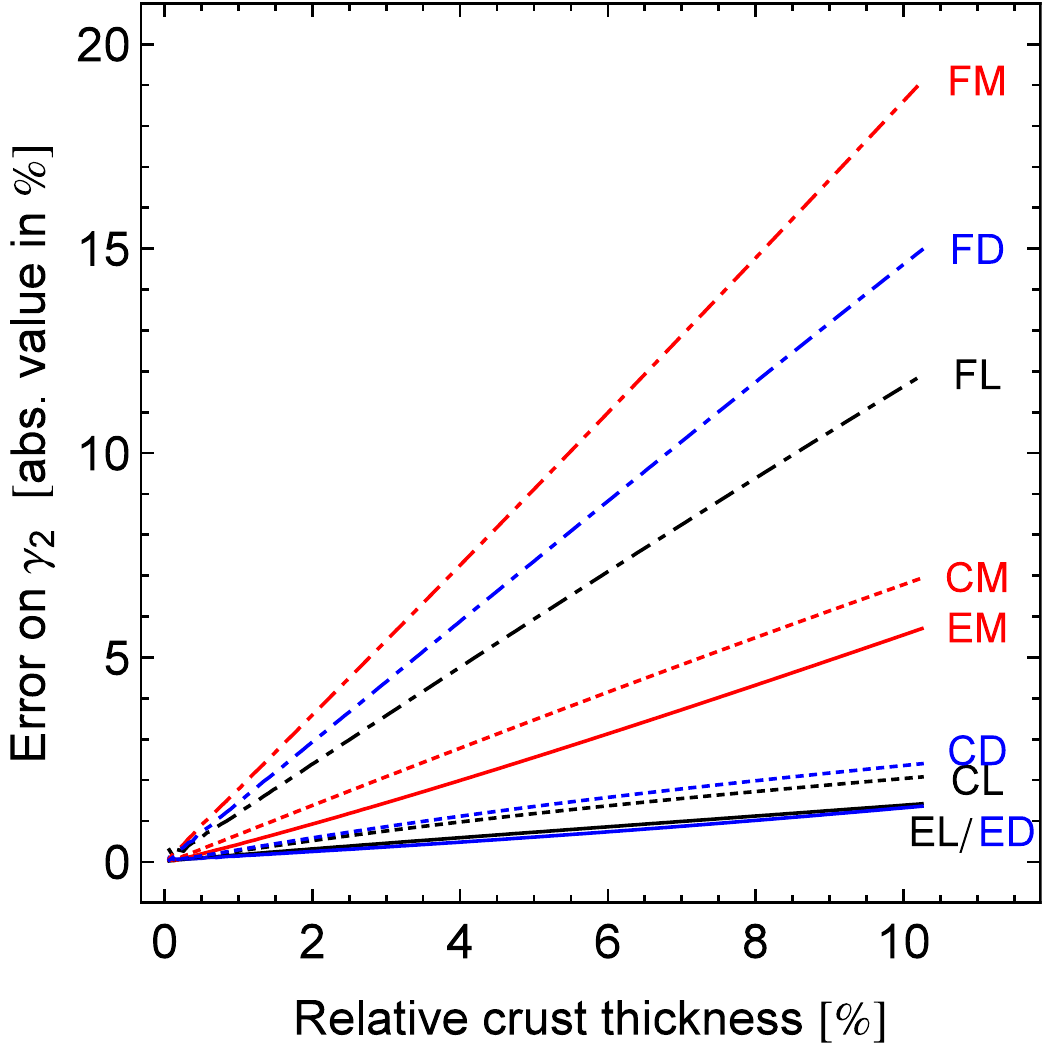}
   \caption[Relative error on Europa's degree-two tilt factor]
   {Relative error (absolute value) on Europa's degree-two tilt factor in the membrane approach.
   The error is defined as the relative difference between membrane and SatStress predictions.
   These errors are a bit larger than those deduced from Fig.~\ref{FigTiltThickness} because they include the imaginary part.
   Solid, dotted, and dash-dotted curves correspond to the elastic-like (E), critical (C), and fluid-like (F) cases, respectively.
    L/M/D identify the density models as in Fig.~\ref{FigTiltThickness}.
   The error curves EL and ED nearly coincide.}
   \label{FigTiltError}
\end{figure}

Fig.~\ref{FigTiltError} shows the error for the various density and rheology models.
The error is smallest (resp.\ largest) if the rheology is elastic-like (resp.\ fluid-like) and the crust-ocean density contrast is small (resp.\ large), because next-to-leading corrections are larger for density terms than for elastic terms.
The error on the tilt factor is one order of magnitude larger than the error on Love numbers because the tilt factor is a small quantity (of ${\cal O}(\varepsilon)$) resulting from the cancellation of the dominant terms (of ${\cal O}(1)$) in the Love numbers.

\section{Influence of deep interior}
\label{DeepInterior}

\subsection{Fluid-crust models of Europa and Titan}
\label{FluidCrustModelsEuropaTitan}

In the membrane approach, the deep interior affects the Love numbers through the fluid-crust Love numbers $h_n^\circ$ and $k_n^\circ$.
Fluid-crust models are thus sufficient when studying the influence of the deep interior.
It is easy to add the effect of the crust by using the membrane formulas (Eq.~(\ref{knhnstatic})).
In the fluid-crust limit, large icy satellites can be roughly described with four layers (not necessarily all present): high-density core, silicate mantle, layer of high-pressure ice, and surface ocean.
There are however only two observational constraints on the interior structure: the total mass and the moment of inertia (Table~\ref{TableBulkOrbital}).

For Europa, I consider three-layer models made of an iron core (solid or liquid, light or dense), a silicate mantle and a pure water ocean.
Similar models have been used by \citet{anderson1998}, \citet{moore2000}, \citet{tobie2005}, and \citet{schubert2009}.
The density of the three layers are fixed as in Table~\ref{TableDeepEuropa} and the radii of the interfaces are computed so that the constraints of total mass and moment of inertia are satisfied (this makes sense if the solid crust to be added to the fluid-crust model has the same density as the ocean).
When studying the influence of the liquid core, I will also consider a range of densities for the silicate mantle.

For Titan, I consider three-layer models made of a silicate core, a mantle of high-pressure ice, and an ocean.
As the total mass and moment of inertia are not very sensitive to the thickness of the high-pressure ice layer, I construct fluid-crust models on the basis of the interior models of \citet{mitri2014} (pure water ocean and high-density ocean, both with a 50~km-thick crust, see their Fig.~1).
I approximate the core, mantle and ocean as homogeneous layers and I replace the crust by an ocean layer (the total mass and moment of inertia thus slightly differ from the observed values).
The density structure of the two models is given in Table~\ref{TableDeepTitan}.

Regarding elastic parameters, all layers are incompressible so that only the shear modulus must be specified.
The shear modulus of Europa's iron core is zero if it is liquid, $75\rm\,GPa$ if it is made of solid FeS (low density), and $100\rm\,GPa$ if it is made of solid Fe, respectively \citep{sohl2002}.
The silicate mantle has a shear modulus between $40\rm\,GPa$ (hydrated silicates of low density) and $80\rm\,GPa$ (olivine).
High-pressure ice has a shear modulus between $6\rm\,GPa$ (Ice V) and $8\rm\,GPa$ (ice VI) \citep{sotin1998}.

Tidal deformations are computed with the propagator matrix method.
The resulting formulas for the three-layer models are complicated and will not be given here.
I will show that it is a good approximation to consider the core and mantle together as a layer of uniform density, so that one can resort to much simpler two-layer formulas (Eq.~(\ref{hn0visco})).
For this purpose, the mean density of the core-mantle system is given in Tables~\ref{TableDeepEuropa} and \ref{TableDeepTitan}.

The viscoelastic effect of the deep interior being rather small, it is interesting to define the change of the Love numbers with respect to the rigid mantle model, i.e. the quantities
\begin{equation}
\left( \epsilon_k \, , \, \epsilon_h \right) = \left(  \frac{k_2^\circ-k_{2r}^\circ}{k_{2r}^\circ} \, , \, \frac{h_2^\circ-h_{2r}^\circ}{h_{2r}^\circ} \right) \, ,
\label{epsh}
\end{equation}
where $(k_{2r}^\circ,h_{2r}^\circ)$ are given by Eq.~(\ref{LoveRigid}).
Since $h_{2}^\circ=k_{2}^\circ+1$ and $k_{2r}^\circ=(3\xi^\circ/5)h_{2r}^\circ$, $\epsilon_k$ and $\epsilon_h$ are related by
\begin{equation}
\epsilon_k = \frac{5}{3\xi^\circ} \, \epsilon_h \, .
\end{equation}
It is thus sufficient to compute either $\epsilon_k$ or $\epsilon_h$.
For Europa, $\epsilon_k\sim5\epsilon_h$ whereas $\epsilon_k\sim(5/2)\epsilon_h$ for Titan.

\begin{table}[ht]\centering
\ra{1.3}
\small
\caption[Fluid-crust models of Europa]{\small
Fluid-crust models of Europa: density structure.
Crust and ocean are assumed to have the same density so that the bulk density and the moment of inertia of the fluid-crust model are equal to the observed values.
}
\begin{tabular}{@{}lllll@{}}
\hline
& \multicolumn{2}{c}{Light core} & \multicolumn{2}{c}{Dense core}  \\
&  radius & density & radius & density \\
& (km) & ($\rm{}kg/m^3$) & (km) & ($\rm{}kg/m^3$) \\
\hline
Core                    & 727     & 5150 & 518 & 8000  \\
Mantle                 & 1445    & 3300  & 1449  & 3300  \\
Core-mantle system  & 1445   &  3535 &1449   & 3515 \\
Ocean                 & 1560.8  & 1000 & 1560.8 & 1000
\vspace{0.3mm} \\
\hline
\end{tabular}
\label{TableDeepEuropa}
\end{table}

\begin{table}[ht]\centering
\ra{1.3}
\small
\caption[Fluid-crust models of Titan]{\small
Fluid-crust models of Titan: density structure.
 The two models result from the homogeneous layer approximation of the end member models of \citet{mitri2014}.
Since the ocean layer that replaces the crust ($d=50\,$km, $\rho_c=935\rm\,kg/m^3$) does not have the same density as the crust, the bulk density and the moment of inertia (not given here) slightly differ from the observed values.
}
\begin{tabular}{@{}lllll@{}}
\hline
& \multicolumn{2}{c}{Light ocean} & \multicolumn{2}{c}{Dense ocean}  \\
&  radius & density & radius & density \\
& (km) & ($\rm{}kg/m^3$) & (km) & ($\rm{}kg/m^3$) \\
\hline
Core                    & 2094     & 2542  & 1968 & 2645  \\
Mantle                 & 2134    & 1367  & 2183  & 1373  \\
Core-mantle system   & 2134    &  2477 & 2183   &  2305  \\
Ocean                 & 2574.8  & 1118 & 2574.8 & 1270  \\               
Whole body  & 2574.8  & 1892 & 2574.8 & 1901
\vspace{0.3mm} \\
\hline
\end{tabular}
\label{TableDeepTitan}
\end{table}

\subsection{How good is the rigid mantle approximation?}
\label{RigidMantleApproximation}

At several occasions in this paper, I use the rigid mantle approximation, i.e.\ the limit of infinite mantle rigidity.
Though the mantle in icy satellites is not more rigid than Earth's mantle in an absolute sense, it is very rigid with respect to the ocean layer that surrounds it.
The largest part of the tidal response occurs in the ocean layer and in the crust.
I will quantify this statement in two ways, by computing the effect of an elastic mantle on the deformations of the mantle and crust.
As an illustration, I choose the solid light-core model of Europa (Table~\ref{TableDeepEuropa}) and the dense-ocean model of Titan (Table~\ref{TableDeepTitan}).
For Europa, the shear modulus is the same in the core and mantle.
For Titan, the shear modulus in the rocky core and in the high-pressure ice layer differ by a factor of ten.

\begin{figure}
   \hspace{-10mm}
   \includegraphics[width=16cm]{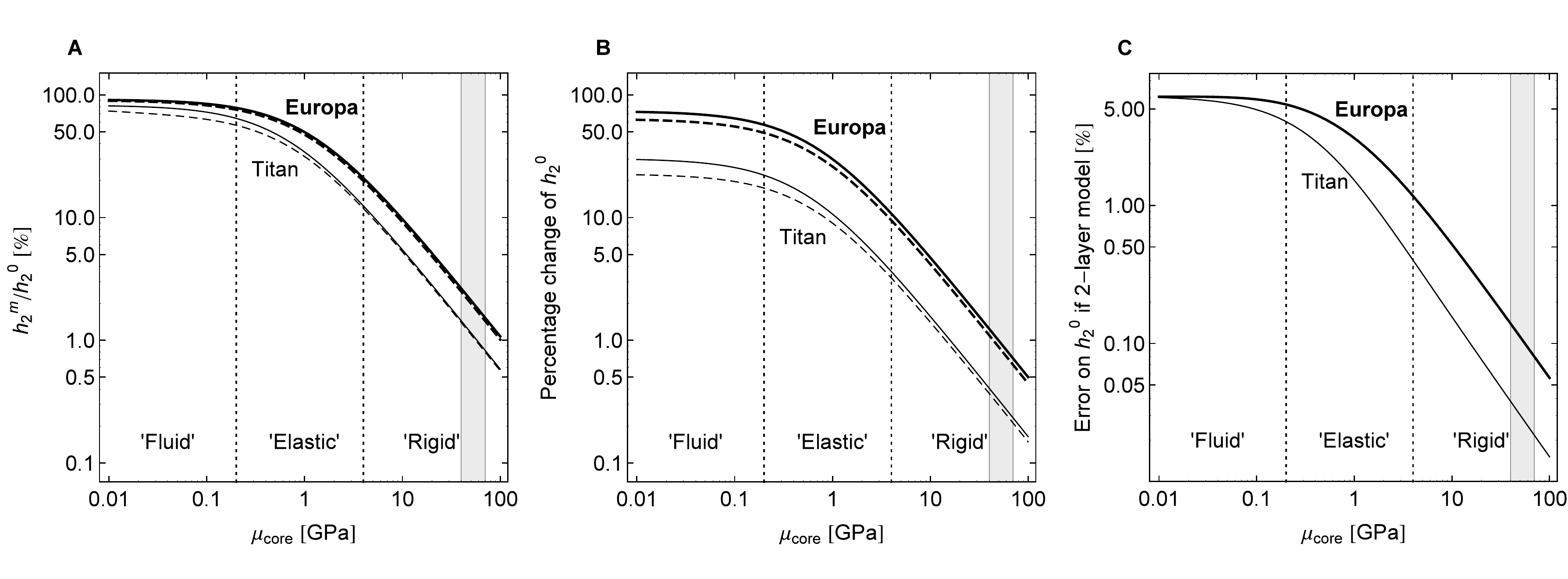}
   \caption[Influence of core-mantle elasticity on tidal deformations]
   {Influence of core-mantle elasticity on tidal deformations of Europa and Titan:
   (A) relative deformation of the top of the mantle with respect to the surface,
   (B) change of the surface deformation with respect to the rigid mantle model (Eq.~(\ref{epsh})), and
   (C) error on $h_2$ between two- and three-layer models.
   Thick curves refer to Europa (light-core model) while thin curves refer to Titan (dense-ocean model).
   Solid (resp.\ dashed) curves show results for the two-layer (resp.\ three-layer) model.
   All models have a fluid crust.
   Shaded bands indicate plausible ranges for the core rigidity.
   Dotted vertical lines separate fluid, elastic, and rigid regimes for the core.
   In panel~(C), the convergence of the curves in the fluid limit is a coincidence.
   See Section~\ref{RigidMantleApproximation} for details.}
   \label{FigRigidMantleApprox}
\end{figure}

First, I compute the relative deformation of the top of the mantle with respect to the surface, i.e.\ the ratio $h_2^m/h_2^0$ (given for the two-layer model by Eq.~(\ref{hnmvisco})).
Fig.~\ref{FigRigidMantleApprox}A shows how $h_2^m/h_2^0$ varies with the core rigidity.
Solid (resp.\ dashed) curves show results for the two-layer (resp.\ three-layer) model.
Plausible values for the shear modulus of the core range from 40 to $70\rm\,GPa$ (shaded band) so that $h_2^m/h_2^0$ is about 1\% for Titan and 2\% for Europa. 
The deformation of the mantle becomes large (say $h_2^m/h_2^0>20\%$) if the shear modulus of the core decreases below the value for ice~I (between 2 and $4\rm\,GPa$).
Below a shear modulus of about $0.2\rm\,GPa$, $h_2^m/h_2^0$ tends to a constant value corresponding to the deformation of a fluid body.

Second, I compute the change of the surface deformation with respect to the rigid mantle model, i.e.\ the quantity $\epsilon_h$ given by Eq.~(\ref{epsh}).
Fig.~\ref{FigRigidMantleApprox}B shows how this quantity varies with the shear modulus of the core.
Conventions are the same as in Fig.~\ref{FigRigidMantleApprox}A: solid (resp.\ dashed) curves show results for the two-layer (resp.\ three-layer) model.
For plausible values of the core rigidity (shaded band), the elasticity of the mantle increases the surface deformation by 0.2-0.4\% (for Titan) and by 0.7-1.2\% (for Europa).
Similarly to what was observed for mantle deformation, the elasticity of the mantle starts to have a large effect when the shear modulus of the core decreases below the value for ice~I.
Below a shear modulus of about $0.2\rm\,GPa$, $\epsilon_h$ tends to a constant because $h_2^\circ$ tends to its fluid Love number value.
Fig.~\ref{FigRigidMantleApprox}C shows the error made by using the two-layer model instead of the three-layer model.
For plausible values of the core rigidity (shaded band), the error made by treating the core and mantle as one homogeneous layer is one order of magnitude smaller than the correction due to the elasticity of the core and mantle.

It thus makes sense to consider the core of a large icy satellite as being infinitely rigid if its rigidity is larger than $\sim4\rm\,GPa$, as fluid if its rigidity is less than $\sim0.2\rm\,GPa$, and as elastic in-between.
For smaller icy satellites, the approximation of an infinitely rigid core is even better because Love numbers depend on the nondimensional shear modulus of the mantle which is inversely proportional to the product of the surface radius and surface gravity (Eqs.~(\ref{hn0visco})-(\ref{defyximu})).
The other lesson of Fig.~\ref{FigRigidMantleApprox} is that the two- and three-layer models give very similar results, especially for large (and thus plausible) values of the core rigidity.
The density contrast between core and mantle is thus a secondary factor, as is the presence within Titan of a high-pressure ice layer with a low shear modulus (the layer is too thin to affect much the response of the core).

\subsection{Influence of the liquid core}
\label{InfluenceLiquidCore}

If Europa's iron core is liquid, how much will it increase the surface deformation?
For the model with a solid core, I concluded that the density contrast between core and mantle is a secondary effect (compare solid and dashed curves in Fig.~\ref{FigRigidMantleApprox}).
This observation suggests the following approximation: neglect density effects by assuming that core and mantle have the same density $\rho_{cm}$.
In that case, the core-mantle boundary does not interact gravitationally with the ocean, so that one can treat the core and mantle as one layer (core-mantle system) with an \textit{equivalent shear modulus} denoted $\mu_{cm}$.

How is $\mu_{cm}$ related to $\mu_m$, the shear modulus of the mantle?
Let us first ignore the ocean.
The liquid core-solid mantle system is  a two-layer body of density $\rho_{cm}$, surface gravity $g_m$ and radius $R_m$.
The radial Love number of this model is given by Eq.~(\ref{LoveRigidHCH}):
\begin{equation}
h_2 = \frac{5}{2} \, \frac{1}{1+\frac{5}{2} \, X_a \, \hat\mu_m} \, ,
\label{h2LS}
\end{equation}
where $\hat\mu_m=\mu_m/(\rho_{cm}g_mR_m)$ and $X_a$ is a function of $x=R_c/R_m$ defined by Table~\ref{TableGeoFac}.

\begin{figure}[h]
   \includegraphics[width=14cm]{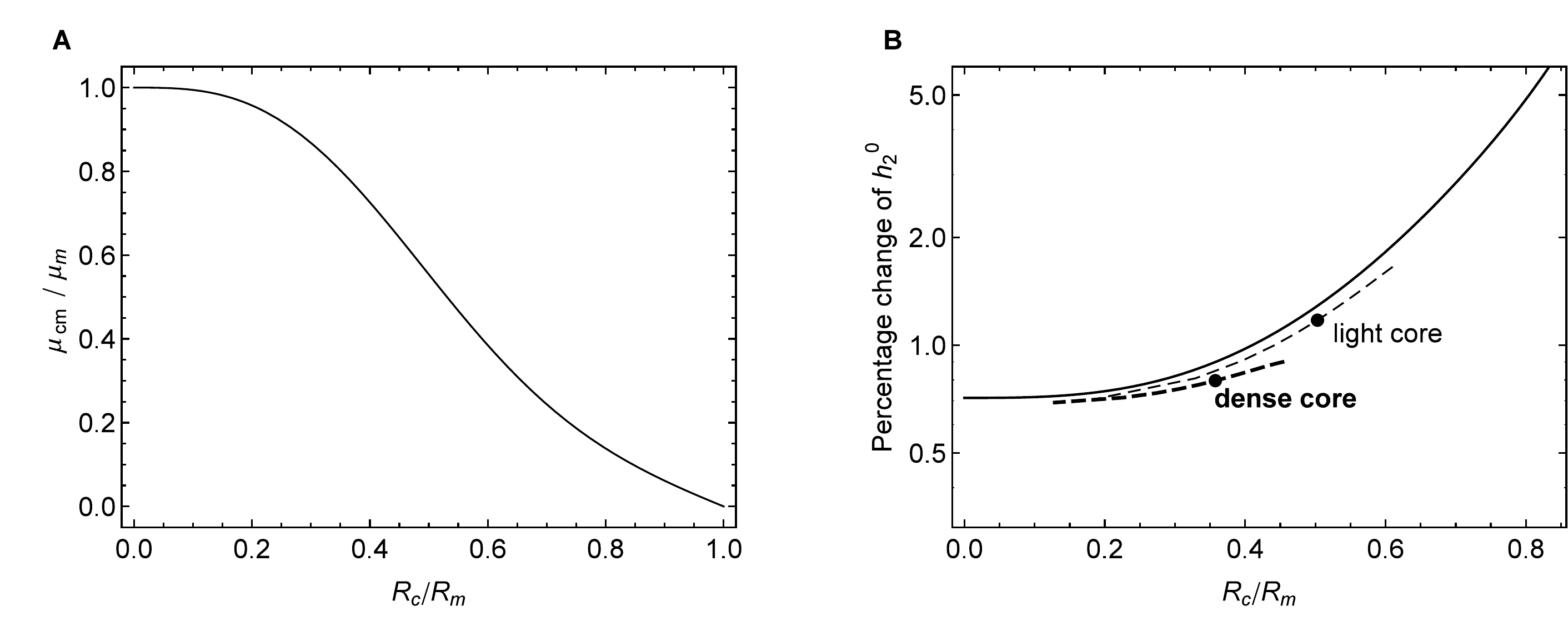}
   \caption[Influence of the liquid core on the surface deformation of Europa]
   {Influence of a liquid core on the surface deformation of Europa:
   (A) equivalent shear modulus of the core-mantle system, and
   (B) change in surface deformation.
   The $x$-axis variable is the relative size of the liquid core with respect to the mantle.
   In panel~(A), the $y$-axis variable is the equivalent shear modulus of the core-mantle system (Eq.~(\ref{equivmu})) divided by the mantle shear modulus.
   In panel~(B), the $y$-axis variable is the change in surface deformation with respect to the rigid mantle model (Eq.~(\ref{epsh})).
   The solid curve shows the two-layer model made of a viscoelastic mantle below an ocean (Eq.~(\ref{hn0visco})) in which the shear modulus of the mantle is replaced by the equivalent shear modulus of the core-mantle system.
   Big dots show the results for the three-layer models of Table~\ref{TableDeepEuropa}, while
   dashed curves show the results of three-layer models for a wide range of mantle densities.
   All models have a fluid crust.
   See Section~\ref{InfluenceLiquidCore} for details.}
   \label{FigLiquidCoreEffect}
\end{figure}

Equating Eq.~(\ref{h2LS}) with the Kelvin-Love formula, $h_2=(5/2)/(1+(19/2)\hat\mu_{cm})$ (see Eq.~(\ref{KelvinLove})), I define the equivalent shear modulus of the core-mantle system by
\begin{equation}
\mu_{cm} = \frac{5}{19} \, X_a \, \mu_m \, .
\label{equivmu}
\end{equation}
Fig.~\ref{FigLiquidCoreEffect}A shows the ratio $\mu_{cm}/\mu_m$ in terms of the relative size of the core with respect to the mantle ($R_c/R_m$).
It varies between one (no liquid core) and zero (no mantle).
If the core extends to about one-half the radius of the mantle, $\mu_{cm}\sim0.55\mu_m$ so that the shear modulus of the mantle ($70\rm\,GPa$) is reduced to an equivalent shear modulus of about $40\rm\,GPa$.

The concept of equivalent shear modulus remains valid if there is an ocean above the mantle, as long as the core and the mantle have the same density (this can be proven with the propagator matrix method).
Fig.~\ref{FigLiquidCoreEffect}B shows the change of the surface deformation with respect to the rigid mantle model (see Eq.~(\ref{epsh})) as a function of the relative size of the liquid core.
The solid curve corresponds to the results of the two-layer model made of a viscoelastic mantle below an ocean (Eq.~(\ref{hn0visco})) in which the shear modulus of the mantle ($70\rm\,GPa$) is replaced by the equivalent shear modulus of the core-mantle system.
Big dots show the results for the three-layer models of Table~\ref{TableDeepEuropa} with the shear modulus of the mantle equal to $70\rm\,GPa$.
Dashed curves show the results for three-layer models satisfying the constraints of total mass and moment of inertia, in which the mantle density varies between $2500\rm\,kg/m^3$ (large core) and $3800\rm\,kg/m^3$ (small core); the core and ocean densities are the same as in Table~\ref{TableDeepEuropa} (these models are inspired by the models of Table~3 in \citet{schubert2009}).
If there is no liquid core, $\epsilon_h\sim0.7\%$ (as in Fig.~\ref{FigRigidMantleApprox}B with $\mu_{core}=70\rm\,GPa$).
The presence of a liquid core increases the surface deformation up to $\epsilon_h\sim1.7\%$ if the core and the mantle are both light.
This figure also shows that the results of three-layer models do not differ much from those of the two-layer model using the equivalent shear modulus.
As in the solid core case, the density stratification of the core-mantle system is a secondary factor.

\subsection{Ocean stratification and screening effect}
\label{OceanStratification}

For a satellite with a subsurface ocean, the major parameter determining the magnitude of Love numbers is the ocean-to-bulk density ratio $\xi$ (or $\xi^\circ$ in the corresponding fluid-crust model).
This is easily seen with the two-layer body made of a rigid mantle and an incompressible homogeneous ocean, for which
Eq.~(\ref{LoveRigid}) gives
\begin{equation}
k_{2r}^\circ = \frac{3\xi^\circ}{5-3\xi^\circ} \, ,
\label{k20r}
\end{equation}
This formula yields values in the range [0.25, 0.35] for Europa and in the range [0.47, 0.71] for Titan (assuming an ocean density between 1000 and $1300\rm\,kg/m^3$).

What happens if the ocean is stratified in density?
For a hydrostatic body, increasing density concentration always decreases (fluid) Love numbers: this \textit{mass concentration effect} can be proven with the Radau relationship which relates the gravitational fluid Love number to the moment of inertia \citep[e.g.][]{schubert2004}.
By contrast, \citet{mitri2014} observe that density stratification (mainly due to pressure) within Titan's ocean leads to a 3 to 4\% increase in $k_2$.
Therefore density stratification has different effects in hydrostatic bodies and in bodies with elastic layers.
The difference can be attributed to a \textit{screening effect}, which I will first explain qualitatively before proceeding to numerical estimates.

\begin{figure}
\centering
   \includegraphics[width=12cm]{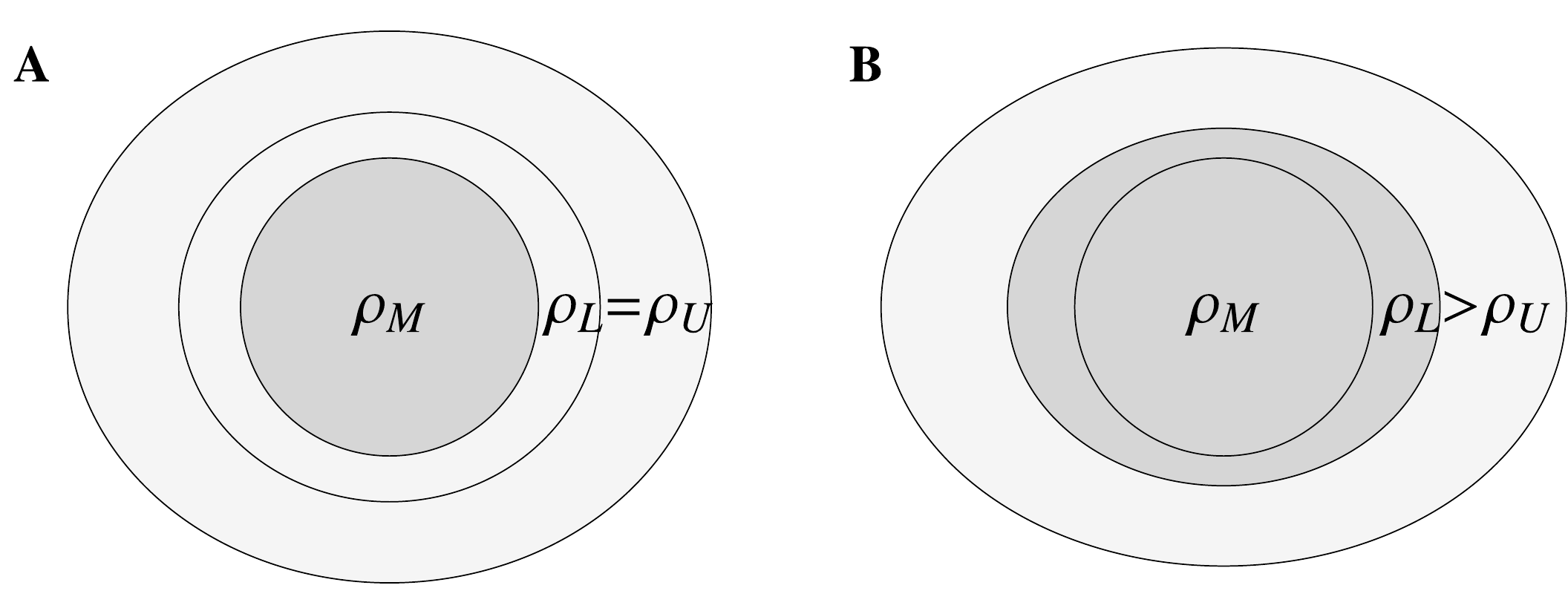}
   \caption[Screening effect due to ocean density stratification]
   {Screening effect due to ocean density stratification.
   The model is a three-layer body made of an infinitely rigid mantle \textit{M} below two ocean layers \textit{L} and \textit{U}.
   In panel~(A), the ocean layers have the same density and the rigid mantle acts as a gravitational brake on the surface deformation.
   In panel~(B), the lower ocean layer has the same density as the mantle and screens its gravitational braking effect.}
   \label{FigScreening}
\end{figure}

Consider the two-layer body made of a viscoelastic mantle and a surface ocean whose Love numbers are given in Appendix~\ref{FluidCrustModels}.
One expects that the surface deformation gets smaller if the mantle rigidity gets larger, and this can be indeed proven with Eq.~(\ref{hn0visco}).
The reason is that a more rigid mantle deforms less and thus contributes less to the induced geoid perturbation: it acts as a `gravitational brake' on the deformation of the ocean layer.
The effect vanishes if the mantle and the ocean have the same density, in which case the mantle has no effect whatsoever on the surface deformation.
This observation suggests a way to cancel the gravitational braking effect by redistributing mass from the top to the bottom of the ocean.
For example, insert at the mantle-ocean boundary a thin liquid layer having the same density as the mantle; decrease also slightly the ocean density so that the total mass remains constant.
The density distribution is nearly unchanged so that the mass concentration effect is not significant.
The deformation of the thin liquid layer completely \textit{screens} the gravitational braking of the mantle, effectively transforming it into a fluid layer (Fig.~\ref{FigScreening}).
As a result, the surface deformation increases and so do the Love numbers.
In more realistic models (Section~\ref{FluidCrustModelsEuropaTitan}), the core-mantle system is not of uniform density and the density at the bottom of the ocean cannot be larger than the mantle density.
Therefore the screening effect is partial: what matters is the density contrast between the mean density of the core-mantle system and the bottom layer of the ocean.

The screening effect can be quantified with the three-layer body made of an infinitely rigid layer $M$ (for mantle) and two ocean layers \textit{L}  (for lower) and \textit{U} (for upper).
Layer densities are denoted $(\rho_M, \rho_L, \rho_U)$ while layer radii are denoted $(R_M, R_L, R_U)$.
This model is equivalent to the three-layer fluid-crust model of Appendix~\ref{FluidCrustModels}, in which the upper ocean layer takes the place of the fluid crust (Eq.~(\ref{h20threelayers})).
If the densities of the two ocean layers are equal, the Love numbers reduce to those of a two-layer body with a rigid mantle (Eq.~(\ref{k20r})).
I parameterize the density stratification as follows:
\begin{eqnarray}
\rho_L &=& \rho + \delta\rho \, ,
\nonumber \\
\rho_U &=& \rho - \alpha \, \delta\rho \, ,
\label{rhoLU}
\end{eqnarray}
where $\delta\rho>0$ and $\alpha>0$ are free parameters.
Imposing that $\rho$ is the average density of the whole ocean, I obtain a constraint on the boundary between the upper and lower ocean layers:
\begin{equation}
x_L = \left( \frac{ x_M + \alpha }{1+\alpha} \right)^{1/3} \, ,
\label{xL}
\end{equation}
where $x_L=R_L/R_U$ and $x_M=R_M/R_U$.
If $\alpha\ll1$, the boundary is located close to the mantle while $\alpha=1$ corresponds to ocean layers of equal volume.

I consider two stratification models: the first one is an nearly linear increase in density with depth through the whole ocean, due to a pressure effect, as in \citet{mitri2014}.
The value $\alpha=1$ is a staircase approximation of this variation.
Another kind of ocean stratification arises from the presence of a thin layer of higher density at the bottom of the ocean.
This case is modeled with the value $\alpha=0.01$.

\begin{figure}
   \includegraphics[width=14cm]{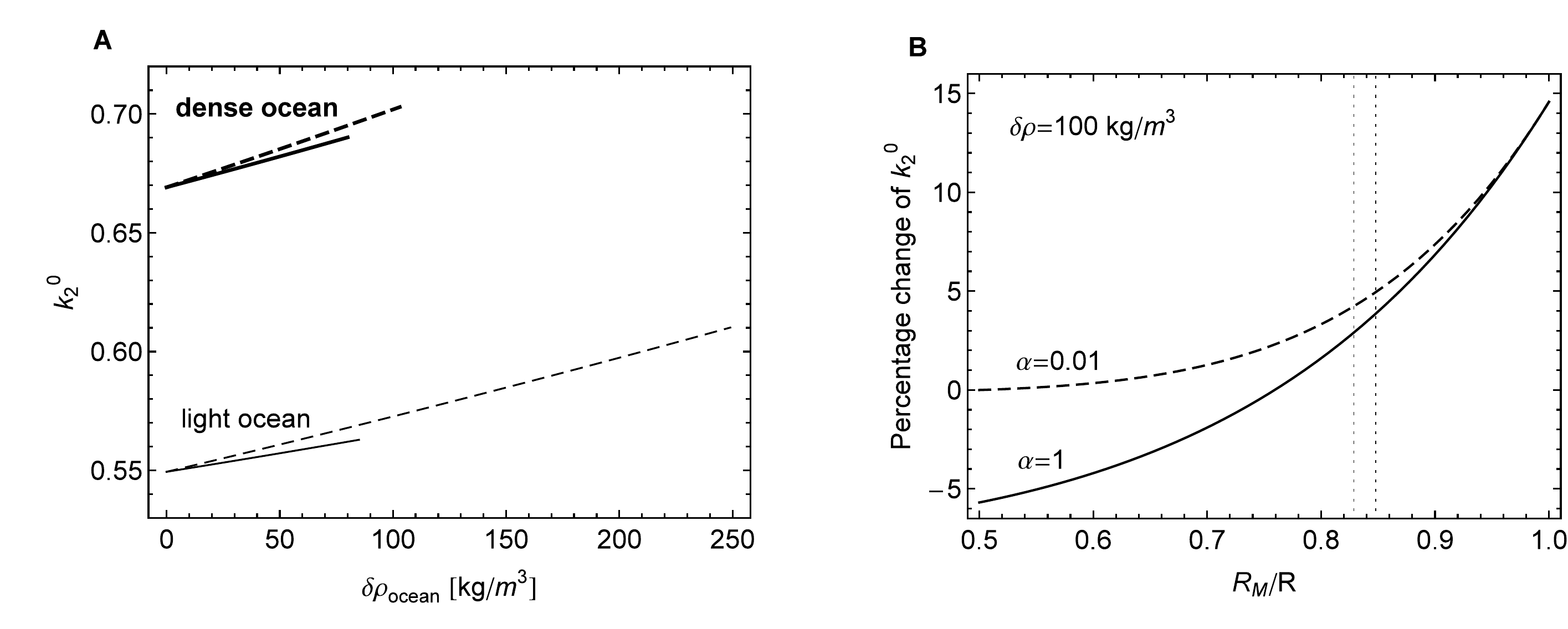}
   \caption[Influence of ocean density stratification on the gravitational Love number]
   {Influence of ocean density stratification on the gravitational Love number:
   (A) $k_2^\circ$ as a function of the density deviation of the lower ocean layer, and
    (B) change of $k_2^\circ$ as a function of the mantle radius if $\delta\rho=100\rm\,kg/m^3$.
   The Love number $k_2^\circ$ is computed with a three-layer model made of a rigid mantle and two ocean layers (Eq.~(\ref{h20threelayers})).
   The lower ocean layer is either thick ($\alpha=1$, solid curves) or thin ($\alpha=0.01$, dashed curves).
   Panel~(A) shows $k_2^\circ$ for the two models of Table~\ref{TableDeepTitan}.
   Panel~(B) shows the change in $k_2^\circ$ for a suite of models parameterized by their mantle radius (the mean ocean density is the same as in the light-ocean model);
   vertical dotted lines mark the position of the mantle radius for the light-ocean and dense-ocean models.
  See Section~\ref{OceanStratification} for details.}
   \label{FigStrat}
\end{figure}

The Love number $k_2^\circ$ can be computed with Eq.~(\ref{h20threelayers}) in which $(x,\xi^\circ, \bar\xi^\circ)$ are replaced by $(x_L,\rho_L/\rho_b, \rho_U/\rho_b)$ with $(x_L,\rho_L, \rho_U)$ given by Eqs.~(\ref{rhoLU})-(\ref{xL}).
Fig.~\ref{FigStrat}A shows the dependence of $k_2^\circ$ on the density deviation $\delta\rho$ for the two Titan models of Table~\ref{TableDeepTitan} (core and mantle are treated as one rigid layer).
Solid (resp. dashed) curves correspond to the case $\alpha=1$ (resp. $\alpha=0.01$).
The starting values of $k_2^\circ$ (at $\delta\rho=0$) differ greatly between the two models because the mean density of the ocean is different.
As the ocean becomes more stratified, $k_2^\circ$ increases with an nearly linear dependence on $\delta\rho$.
For these models, the slope of the curve does not depend much on the thickness of the lower layer (i.e.\ the parameter $\alpha$).
There is however an important difference between the cases $\alpha=1$ and $\alpha=0.01$.
In the former case, $\delta\rho$ is about $80-85\rm\,kg/m^3$ (see Fig.~1 of \citet{mitri2014}) and cannot be much larger otherwise either the density at the top of the ocean is lower than the crust density, or the density of the ocean bottom is higher than the mantle density.
This kind of density stratification thus increases $k_2^\circ$ by no more than 3\%.
In the latter case ($\alpha=0.01$, thin dense layer), $\delta\rho$ is constrained by the mantle density but not by the crust density.
In the light-ocean model, $\delta\rho$ can thus be as large as $250\rm\,kg/m^3$, which increases $k_2^\circ$ by 11\%.

The slope of the $k_2^\circ$ curve mainly depends on the relative radius of the mantle: it increases with $x_M$ so that it is slightly steeper for the dense-ocean model than for the light-ocean model.
Fig.~\ref{FigStrat}B shows the change of $k_2^\circ$ ($\epsilon_k$ given by Eq.~(\ref{epsh})) as a function of the relative mantle radius for $\delta\rho=100\rm\,kg/m^3$ (we known from Fig.~\ref{FigStrat}A that $k_2^\circ$ depends linearly on $\delta\rho$ so that it is easy to rescale the results for other $\delta\rho$ values).
Curves have been drawn for $\xi^\circ=0.59$ (light-ocean model) but are nearly the same for $\xi^\circ=0.67$ (dense-ocean model).
In the thin layer case ($\alpha=0.01$, dashed curve), the change of $k_2^\circ$ varies between 0\% and 15\%.
If $\alpha=1$, the change of $k_2^\circ$ can be positive or negative depending on the balance between the mass concentration and screening effects.
It is negative if $x_M\lesssim3/4$ because the mass concentration effect wins over the screening effect.

In conclusion, a thin and dense liquid layer at the bottom of a light ocean can increase the gravitational Love number by a significant amount (more than 10\% for Titan) depending on the radius of the mantle and on the density contrast with the core-mantle system.
This is good news for the light-ocean models of Titan which otherwise predict that $k_2$ should be much lower than what is measured with the Cassini spacecraft \citep{iess2012}.

\section{Dynamical Love numbers}
\label{LoveDynamic}

\subsection{Dynamical tides}

Tidal deformations of solid planetary bodies are most often computed in the static limit (equilibrium tides).
Tides are indeed slow in comparison with seismic velocities so that inertial terms are negligible in the viscoelastic-gravitational equations (Section~\ref{StaticDynamic}).
By contrast, the fluid nature of stars and giant gaseous planets requires the inclusion of various kinds of dynamical effects in tidal computations \citep[e.g.][]{ogilvie2014}.
The classical theory of ocean tides on Earth is also intrinsically dynamical \citep[e.g.][]{hendershott1981}.
Dynamical effects are thus potentially important in solid bodies with liquid layers, such as icy satellites with subsurface oceans.

Dynamical tides, however, is a vast subject extending much beyond the viscoelastic-gravitational equations considered in this paper.
First, rotation of the body leads to Coriolis and centrifugal effects which are of the same order of magnitude as inertial terms.
Second, viscoelastic-gravitational equations do not incorporate a correct description of fluid dynamics (Navier-Stokes equations).
These topics are clearly out of the scope of this paper.
The Coriolis effect, for example, breaks spherical symmetry so that tidal deformations cannot be completely described with three scalar Love numbers (see \citet{wahr1981} for the case of an oceanless Earth).
I thus restrict myself to computing dynamical effects on the Love numbers of a spherically symmetric and non-rotating body enclosing a global fluid layer of zero viscosity.
In this way, I hope to quantify the threshold beyond which dynamical effects become important for icy satellites.

Dynamical corrections to the $k_n-h_n$ formula are quantified by the parameter $\Lambda_\omega$ (Eq.~(\ref{LambdaOmega})) which depends on $y_3^F(R_\varepsilon)/y_1(R_\varepsilon)$, the ratio of lateral to radial displacement of the fluid at the crust-ocean boundary.
Computing $\Lambda_\omega$ requires solving the dynamical viscoelastic-gravitational problem within the ocean, but this cannot be done without solving the problem for the whole body at once.
In the dynamical compressible case, the solution is obtained by numerically integrating the viscoelastic-gravitational equations.
Numerical codes doing this job, however, usually diverge at low frequencies not far from the tidal frequencies of synchronously rotating satellites (note that the code included into SatStress is static).
I will solve here the dynamical viscoelastic-gravitational problem for the rigid mantle model of Section~\ref{RigidMantleLimit}.
As before, inertial terms are neglected within solid layers.

\subsection{Dynamical incompressible liquid layer}
\label{DynIncompLiqLayer}

\citet{denis1998} found a simple solution for dynamical tides in a homogeneous and incompressible fluid layer:
\begin{eqnarray}
y_1 &=& a \, r^{n-1} + b \, r^{-n-2} \, ,
\nonumber \\
y_3 &=&  a \, \frac{r^{n-1}}{n} - b \, \frac{r^{-n-2}}{n+1} \, ,
\nonumber \\
y_5 &=& \alpha \, r^n + \beta \, r^{-n-1} \, .
\label{y5dyn}
\end{eqnarray}
The constants $(a,b,\alpha,\beta)$ are fixed by the boundary conditions of the fluid layer.
As seen in Section~\ref{FluidLayer}, $y_4$ vanishes in the fluid layer, while $y_2$ is linearly related to the other variables by Eq.~(\ref{fluideqDyn}).
Finally, $y_6$ can be computed with Eq.~(\ref{EG5}).
The solution can be cast into the propagator matrix formalism:
\begin{equation}
\left( y_1 \, , \, y_2 \, , \, y_5 \, , \, y_6 \right)^T = \mathbf{Y}_{liq} \cdot \left( a \, , \, \alpha \, , \, b \, , \, \beta \right)^T \, ,
\label{liqdynpropagMain}
\end{equation}
where the propagator matrix $ \mathbf{Y}_{liq}$ is given by Eq.~(\ref{dill}) and its inverse by Eq.~(\ref{dillinv}).
Note that $y_3$, instead of $y_2$, is considered as a dependent variable because $y_3$ is discontinuous at the ocean boundaries.
Using Eq.~(\ref{liqdynpropagMain}) in the ocean and the static propagator matrix in the solid layers \citep[e.g.][]{sabadini2004}, one can solve for the tidal Love numbers of an incompressible body made of an arbitrary number of homogeneous spherical layers.
The ocean itself can be discretized in concentric layers of different densities.

\subsection{Rigid mantle model in its dynamical version}

The problem is further simplified by assuming an infinitely rigid mantle.
In that case, the radial displacement of the mantle-ocean boundary vanishes: $y_1(R_m)=0$, where $R_m$ is the mantle radius.
This constraint reduces the number of unknown constants in Eq.~(\ref{y5dyn}).
First, it yields a relation between $a$ and $b$:
\begin{equation}
b = - a \, R_m^{\,2n+1} \, .
\label{abrel}
\end{equation}
Combining Eqs.~(\ref{y5dyn}) and (\ref{abrel}), I compute the ratio of tangential to radial displacements within the ocean:
\begin{equation}
\frac{y_3}{y_1} = \frac{1}{n\left(n+1\right)} \, \frac{n+1+n \, (R_m/r)^{2n+1}}{1-(R_m/r)^{2n+1}} \, .
\label{y3y1ratio}
\end{equation}

The assumption of an infinitely rigid mantle also imposes that $\beta=0$ in Eq.~(\ref{y5dyn}).
This assertion can be proven by imposing the continuity of the variable $y_6$ (defined by Eq.~(\ref{EG5})) at the mantle-ocean boundary.
Intuitively, it can be understood as follows.
In Eq.~(\ref{y5dyn}), the term of $y_5$ in $r^n$ is an interior gravity solution caused by the deformation of the crust-ocean boundary and the layers above it.
By contrast, the term of $y_5$ in $r^{-n-1}$ is an exterior gravity solution caused by the deformation of the crust-mantle boundary and the layers beneath it.
Thus the term of $y_5$ in $r^{-n-1}$ vanishes if the mantle is infinitely rigid.
Therefore $y_5'$ can be expressed within the ocean in terms of $y_5$:
\begin{equation}
y_5' = \frac{n}{r} \, y_5 \, .
\label{y5primey5}
\end{equation}
This relation is at the basis of the $k_n-h_n$ proportionality derived in Appendix~\ref{knhnproportionality}.

\subsection{Tilt factor}
\label{TiltFactorDyn}

\begin{figure}
   \centering
      \includegraphics[width=7cm]{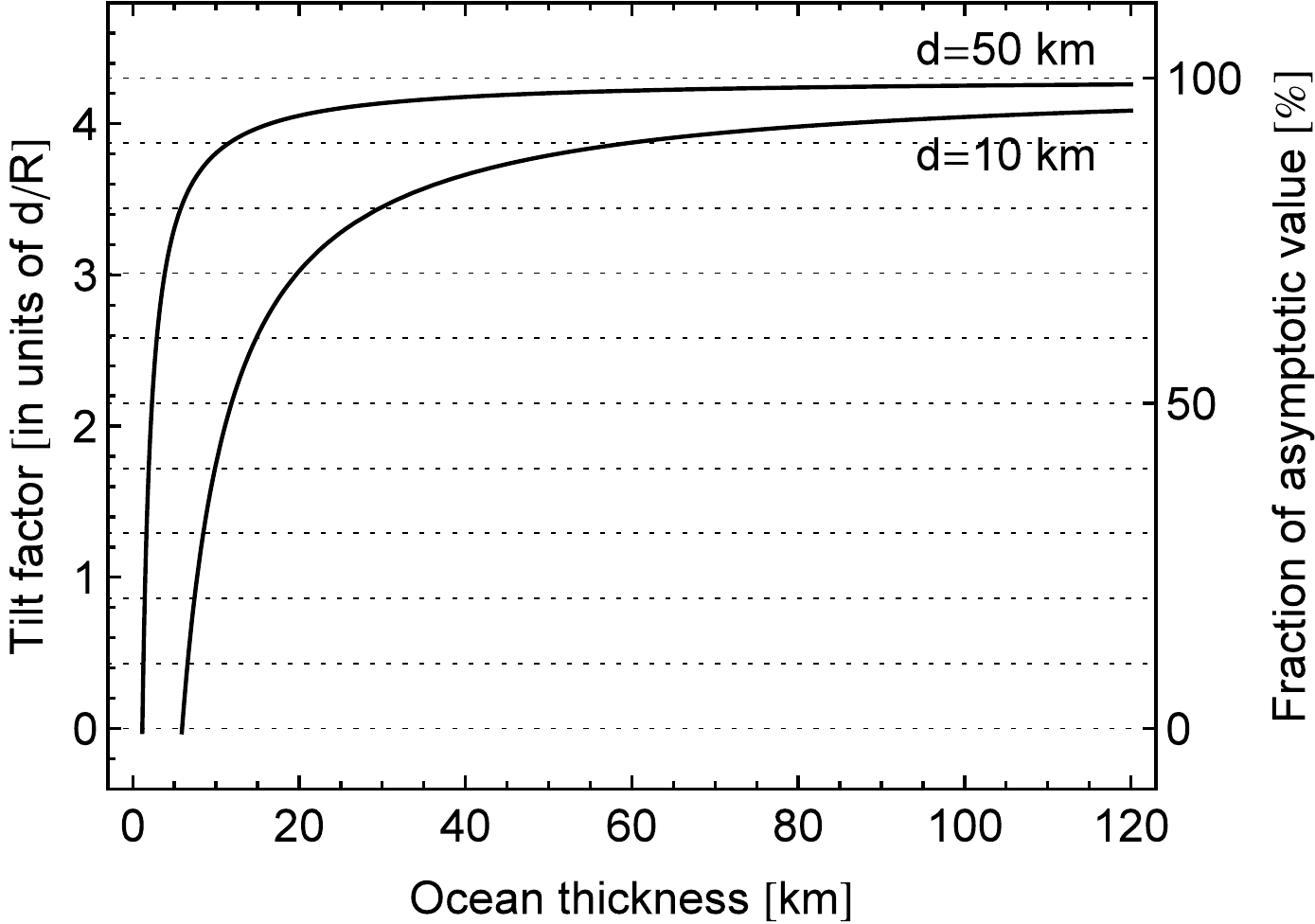}
   \caption[Dynamical corrections to the tilt factor of Europa]
   {Dynamical corrections to the tilt factor of Europa as a function of ocean thickness.
      The tilt factor is divided by the relative crust thickness $\varepsilon=d/R$.
   The asymptotic value is the tilt factor without dynamical corrections: $\gamma_2/\varepsilon\sim4.3$ (similar to the elastic-like case of Fig.~\ref{FigTiltMembrane} except that the imaginary part vanishes).
   See Section~\ref{TiltFactorDyn} for details.}
   \label{FigTiltDyn}
\end{figure}

The rigid mantle model, in its dynamical version, can be completely solved for the tilt factor and for the Love numbers.
Substituting Eq.~(\ref{y3y1ratio}) into Eq.~(\ref{LambdaOmega}) yields a simple formula for the dynamical correction to the tilt factor: 
\begin{equation}
\Lambda_\omega = - \frac{q_\omega}{n(n+1) } \, \frac{n+1+n z^{2n+1}}{1-z^{2n+1}} \, ,
\label{LambdaOmegaRigid}
\end{equation}
where $z=R_m/R_\varepsilon$.
The dynamical parameter $q_\omega$ is given in Table~\ref{TableBulkOrbital} for Europa and Titan.
If the ocean is shallow,
\begin{equation}
\Lambda_\omega \sim  - \frac{q_\omega}{n(n+1) } \,  \frac{R_\varepsilon}{D} \, ,
\label{LambdaOmegaRigidApprox}
\end{equation}
where $D=R_\varepsilon-R_m$ is the ocean thickness.
The main contributor to the tilt factor is typically the membrane spring constant $\Lambda_T\sim2\hat\mu\varepsilon$ (Fig.~\ref{FigTiltMembrane}).
If the ocean is shallow, the relative magnitude of degree-two dynamical corrections in the tilt factor (Eqs.~(\ref{tiltn})-(\ref{Lambdatot})) is
\begin{equation}
\frac{\Lambda_\omega}{\Lambda} \sim - \frac{1}{12} \, \frac{q_\omega}{\hat\mu} \, \frac{R^2}{Dd} \, .
\end{equation}
For Europa, $q_\omega\approx5\times10^{-4}$ and $\hat\mu=\bar\mu/(\rho{}gR)\sim1.71$ (if the crust is elastic, see Section~\ref{StaticDynamic}) so that $\Lambda_\omega/\Lambda\sim-2.4\times10^{-5}R^2/(Dd)$.
Fig.~\ref{FigTiltDyn} shows how the tilt factor decreases as the ocean becomes shallower.
The crust is elastic and either thick ($d=50\rm\,km$) or thin ($d=10\rm\,km$).
If the crust is viscoelastic, the crust thickness should be interpreted as the lithospheric thickness: $d\rightarrow{}d_{litho}=|\bar\mu/\mu_E|d$ (Eq.~(\ref{dlitho})).
The radius of the mantle varies with crust thickness and ocean depth.
Otherwise, Europa is modeled as in Table~\ref{TableCrustOceanDensity} (light density model) and Table~\ref{TableBodyRheology}.
If the ocean is deep (say $D=100\rm\,km$), the dynamical correction decreases the tilt factor by a few percents: 1\% if $d=50\rm\,km$ and 6\% if $d=10\rm\,km$.
The dynamical correction becomes important if the ocean is less than $20\rm\,km$ thick; neglecting it leads to underestimate the crust thickness.
The tilt factor can even become negative if the ocean is shallower than $1.2\rm\,km$ (if $d=50\rm\,km$) or $5.9\rm\,km$ (if $d=10\rm\,km$).
All in all, dynamical corrections are significant for Europa unless the ocean is very deep.
For Titan, dynamical corrections are much smaller because $q_\omega$ is smaller by one order of magnitude.

The rigid mantle model is based on rather restrictive assumptions. 
The hypothesis of an infinitely rigid mantle remains a good approximation for dynamical tides but can be dispensed with: the dynamical propagator matrix method is ideal for models with an arbitrary number of incompressible viscoelastic and liquid layers (Eq.~\ref{liqdynpropagMain}).
Ocean compressibility is expected to have a significant impact if the ocean is deep, but compressibility can be safely ignored in shallow oceans for which dynamical effects are maximum.

\subsection{Love numbers}
\label{LoveNumbersDyn}

Without the static assumption, I cannot use gravity scaling in order to obtain explicit formulas for Love numbers.
Instead, the $k_n-h_n$ proportionality derived in Appendix~\ref{knhnproportionality} provides a supplementary constraint.
Combining the $k_n-h_n$ relation (Eq.~(\ref{knhn})) with Eq.~(\ref{knhnRigid3}), I get
\begin{eqnarray}
k_{nr} +1 &=& \frac{ k_{nr}^\circ +1 }{ 1 + \frac{3\,\xi^\circ}{2n+1} \, \frac{1}{1+\Lambda+ \Lambda_\omega} \left( \left( k_{nr}^\circ +1 \right) \left( \Lambda  + \Lambda_\omega \right) + K_{\rho{r}}  \right) } \, ,
\nonumber \\
h_{nr} &=& \frac{ h_{nr}^\circ }{ 1 + \left( \Lambda + \Lambda_\omega \right) h_{nr}^\circ  + \Lambda_\chi + H_{\rho{r}} } \, ,
\label{knhnrigiddynamic}
\end{eqnarray}
where $(k_{nr}^\circ+1,h_{nr}^\circ)$ are the fluid-crust Love numbers if the fluid-crust density $\rho^\circ$ is equal to the ocean density $\rho$ (implying $\delta\rho^\circ=\delta\rho$).
Terms beyond ${\cal O}(\varepsilon)$ are neglected.
Density corrections are defined by
\begin{eqnarray}
 K_{\rho{r}} &=&
2 \, \Big( (n-1) \left( k_{nr}^\circ +1 \right) - (2 n + 1)\Big) \frac{\delta\rho}{\rho} \, \varepsilon \, ,
\nonumber \\
 H_{\rho{r}} &=& K_{\rho{r}} + \Lambda'_\rho \, ,
 \label{KHrhoRigidn}
\end{eqnarray}
where $\Lambda'_\rho$ is defined by Eq.~(\ref{defLambdarhoprime}).
Using the identity (\ref{RigidIdentity}), one can check that $(K_{\rho{r}},H_{\rho{r}})$ are the rigid mantle limits of the density corrections $(K_\rho,H_\rho)$ for the static case (Eq.~(\ref{KHrho})).

If $\Lambda_\omega=0$, the formulas for dynamical Love numbers (Eq.~(\ref{knhnrigiddynamic})) are equivalent to the formulas for static Love numbers in the rigid mantle limit (Eq.~(\ref{knhnstatic})), the only difference being of ${\cal O}(\varepsilon^2)$.
Conversely, static formulas (for the rigid mantle model) are transformed into dynamical formulas by substituting
\begin{equation}
\Lambda \rightarrow \Lambda + \Lambda_\omega \, .
\end{equation}

Being of different sign, the membrane spring constant $\Lambda$ and the dynamical correction $\Lambda_\omega$ have opposite effects (Fig.~\ref{FigTiltDyn}).
The dynamical correction partially or totally cancels the resistance of the crust to deformations, thus increasing Love numbers.
What happens if the ocean thickness decreases even further?
This question can be discussed with a simpler model in which the crust is incompressible and has the same density as the ocean.
With these assumptions, it is unnecessary to require that $\Lambda_\omega\sim{\cal O}(\varepsilon)$ as done in Section~\ref{TiltFactor}.
If $\Lambda_\chi=0$ and $\delta\rho=0$, the membrane formulas become
\begin{equation}
\left( k_{nr} , h_{nr} \right) = \frac{1}{1+\left( \Lambda+\Lambda_\omega \right) h_{nr}^\circ} \left( k_{nr}^\circ , h_{nr}^\circ \right) .
\label{khDyn}
\end{equation}
This formula coincides with the thin shell limit of the dynamical homogeneous crust model with rigid mantle (Eqs.~(\ref{zhlapprox})-(\ref{LoveRigidHCR})), providing another check on the dynamical membrane formula.
Eq.~(\ref{khDyn}) shows that Love numbers diverge if $\Lambda+\Lambda_\omega=-1/h_{nr}^\circ$: resonance occurs.
From Eq.~(\ref{LambdaOmegaRigid}), one sees that this cannot happen unless the ratio $z=R_m/R_\varepsilon$ is close to one, i.e.\ the ocean must be shallow.
The resonant ocean thickness is approximately equal to
\begin{equation}
D = \frac{q_\omega R}{n(n+1)} \, \left( 1 - \frac{3}{2n+1} \, \xi + Re(\Lambda) \right)^{-1} \, ,
\label{Dreson}
\end{equation}
which is obtained by substituting Eqs.~(\ref{LambdaOmegaRigidApprox}) and (\ref{LoveRigid}) into Eq.~(\ref{khDyn}).
I neglected terms of ${\cal O}(\varepsilon)$ except the membrane spring constant which can be of ${\cal O}(1)$ if the crust is hard (small or medium-sized satellite); other finite thickness corrections can be taken into account with the homogeneous crust model (Eq.~(\ref{LoveRigidHCR})) but do not play a significant role here.

In a large icy satellite, the crust is typically soft ($\Lambda\sim{\cal O}(\varepsilon)$, see Section~\ref{TiltFactor}) so that the resonant frequency  is nearly the same as the one obtained by neglecting the crust  ($\Lambda_\omega=-1/h_{nr}^\circ$):
\begin{equation}
\omega^2 =  n\left(n+1\right) \frac{1-z^{2n+1}}{n+1+n z^{2n+1}} \left( 1 - \frac{3}{2n+1} \, \xi \right) \frac{g}{R} \, ,
\label{EqLamb}
\end{equation}
where $z\cong{}R_m/R$.
This classical formula gives the frequency of the degree-$n$ free oscillation of an incompressible fluid enclosing a rigid core (or mantle) \citep[][p.~454]{lamb1932}.
In astronomy, this is called the \textit{f mode} (\textit{f} for fundamental) or the \textit{surface gravity mode} of oscillating stars and giant planets \citep{ogilvie2014}.
In seismology, one talks about the \textit{tsunami mode} for obvious reasons \citep{dahlen1999}.
In the model above (Eq.~(\ref{khDyn})), the resonance is damped by the viscoelasticity of the crust (proportional to \textit{Im}($\Lambda$)) but the damping is weak if the crust is soft.
In more realistic models, other sources of damping arise such as solid/liquid friction or the viscosity of the fluid itself.
\citet{ogilvie2014} illustrates the latter kind with the gravitational Love number of a homogeneous, incompressible, non-rotating, viscous fluid body: as in many classical oscillation problems, it amounts to the substitution $\omega^2\rightarrow\omega^2+i\alpha\omega$, where the coefficient $\alpha$ depends on the viscosity model (see his Eq.~(16)).

\begin{figure}
   \centering
      \includegraphics[width=7cm]{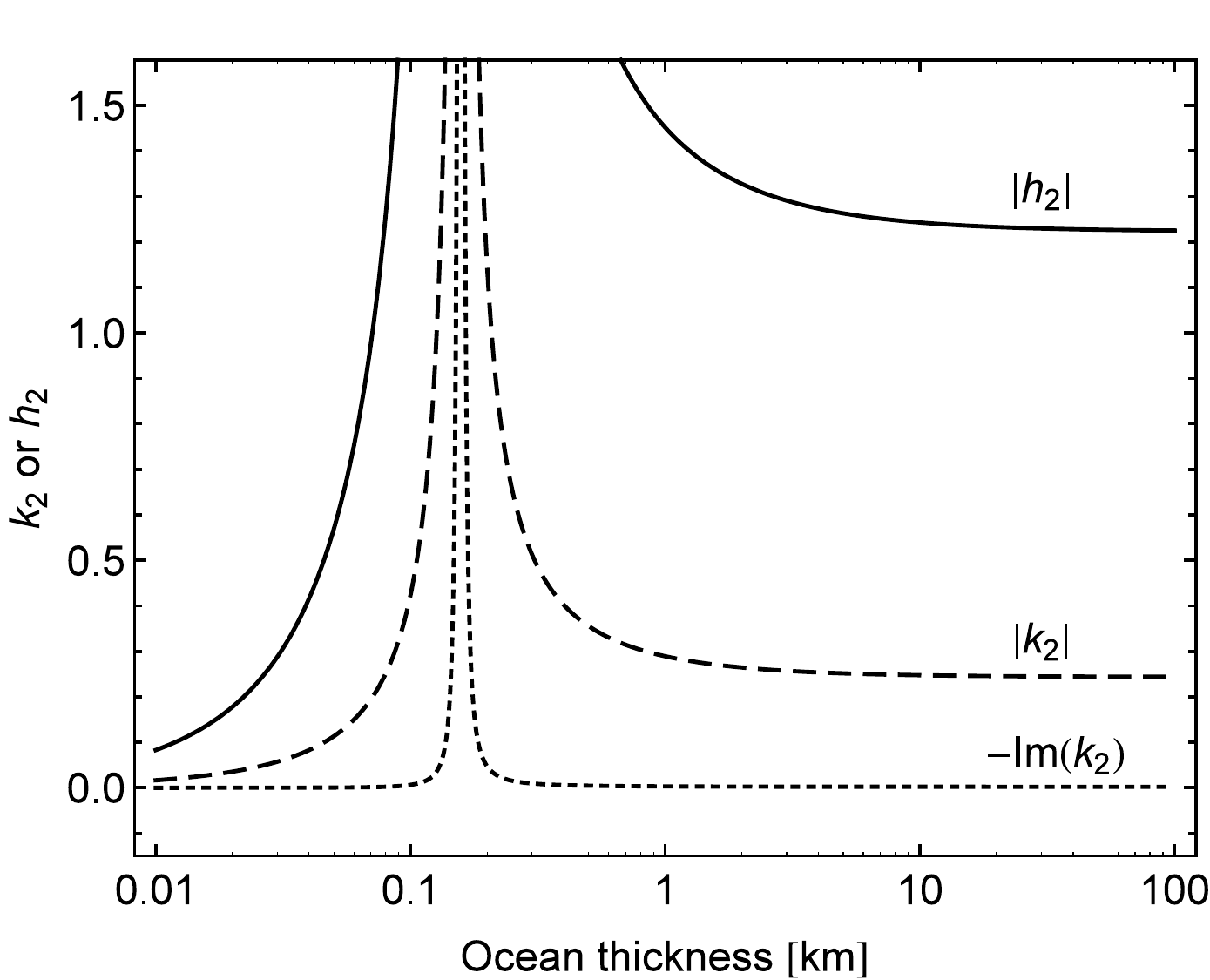}
   \caption[Dynamical resonance]
   {Dynamical resonance enhancing the Love numbers of Europa for a shallow ocean.
   The solid, dashed, and dotted curves show the absolute value of $h_2$, the absolute value of $k_2$, and the imaginary part of $k_2$, respectively.
   See Section~\ref{LoveNumbersDyn} for details.}
   \label{FigLoveDyn}
\end{figure}

Fig.~\ref{FigLoveDyn} shows the enhancement of Europa's Love numbers due to the dynamical resonance.
The crust is incompressible, is 10~km thick, and has the same density as the ocean ($1000\rm\,kg/m^3$).
Otherwise, Europa is modeled as in Tables~\ref{TableBodyRheology} and \ref{TableCrustRheology} (critical model).
Resonance occurs if the ocean is extremely shallow ($D\sim160\rm\,m$), which is an unlikely configuration except in a transiting state: the ocean is either just born or on the point of freezing.
At resonance, surface deformation (measured by $|h_2|$) and tidal dissipation within the crust (measured by \textit{Im}($k_2$)) both diverge.
If there is no dissipation in the core and mantle, tidal dissipation in the crust is related to surface deformation by Eq.~(\ref{Imk2}):
$\mbox{\textit{Im}}(k_2)=-(3/5)\xi|h_2|^2\mbox{\textit{Im}}(\Lambda)$.
Extremely large $k_2$ and $h_2$ values should not be taken at face value but are rather a sign that the theory breaks down.
In any case, viscoelastic-gravitational equations are only valid for small perturbations.
Furthermore, the present model does not take into account dissipation within the ocean and at its boundaries.

Contrary to other dynamical approaches, the model presented here includes a crust above the ocean.
However, it should not be considered as realistic for the analysis of dynamical resonances.
Besides the lack of dissipation within the ocean, its fundamental weakness is that other resonances appear once rotation is included, as is well-known from the analysis of Laplace tidal equations for a surface ocean \citep[e.g.][]{chen2014,matsuyama2014,tyler2014}.

\section{Summary}

In a previous paper, I developed the \textit{massless membrane approach} to tidal Love numbers of `membrane worlds', or icy satellites with subsurface oceans \citep{beuthe2014}.
This method, based on the classical equations of thin shell theory, leads to accurate formulas for Love numbers if the density contrast between the crust and the ocean is negligible.
The error due to this approximation is never large because the ice and ocean densities do not differ by more than thirty percents, whereas viscoelasticity easily reduces the rigidity of the whole crust by one order of magnitude.
The membrane approach has the advantage of easily taking into account crustal rheology through effective viscoelastic parameters.
Nevertheless, the neglect of the crust-ocean density contrast is an annoying feature of this approach, especially now that measurements of Titan's $k_2$ suggest that its ocean is very dense.
Besides, it has a significant effect on the tilt factor which is useful quantity for crust thickness estimates.
Massless membrane theory is also unsatisfactory regarding the inclusion of crust compressibility because of its assumption that the upper and lower shell surfaces deform in the same way (Appendix~\ref{MasslessMembrane}).
But let's not throw out the baby with the bath water:
massless membrane theory remains largely valid thanks to its flexible formulation in terms of Love numbers.
More precisely, the computation of Love numbers needs to be improved but the membrane formulas for tectonics and tidal dissipation are unchanged.

For these reasons, I have rebuilt from scratch a membrane theory based on the viscoel\-astic-gravitational theory used, among other things, for computing tidal Love numbers in thick shell theory.
The \textit{massive membrane approach} starts with a perturbative expansion of the viscoelastic-gravitational equations in the small parameter $\varepsilon=d/R$ ($d$ being the crust thickness): the variables of the problem are linearly propagated from the surface to the crust-ocean boundary.
After integration over crust thickness, the viscoelastic response of the crust depends on three effective viscoelastic parameters $\bar\mu$, $\bar\nu$, and $\bar\chi$ (see Table~\ref{TableNotation} for a list of parameters relevant to the membrane).
These three complex numbers are directly computable for any depth-dependent rheology, though I benchmarked the formulas for a two-layer crust with Maxwell rheology.
The compressibility factor $\bar\chi$ is new with respect to massless membrane theory and accounts for the compressibility problem mentioned above.

At the crust-ocean boundary, two constraints (free slip and fluid constraint) lead to two relations between tidal Love numbers, here given at degree two:
\begin{eqnarray}
l_2 &=& \frac{1+\bar\nu}{5+\bar\nu} \, h_2 \, ,
\nonumber \\
k_2 + 1 &=& \left( 1 + \Lambda_T \right) h_2 - 5 \, \frac{\delta\rho}{\rho} \, \varepsilon \, ,
\label{k2h2}
\end{eqnarray}
where $\delta\rho$ is the crust-ocean density contrast while $\Lambda_T$ is the sum of the membrane spring constant $\Lambda$, the compressibility correction $\Lambda_\chi$, the minor density correction $\Lambda_\rho$, and the dynamical correction $\Lambda_\omega$ (Eqs.~(\ref{Lambdatot})-(\ref{LambdaOmega})).
The $l_2-h_2$ and $k_2-h_2$ relations are of ${\cal O}(1)$ and ${\cal O}(\varepsilon)$, respectively, and could in principle be computed to the next order of perturbation.
These relations depend on the presence of a lithosphere (Eq.~(\ref{LithoCond2})).
There is thus no smooth transition, unless $\delta\rho=0$, between the $k_n-h_n$ relation and the hydrostatic relation $k_n^\circ+1=h_n^\circ$ (Eq.~(\ref{knhnhydrostat})).

In the static limit of equilibrium tides, one can write explicit formulas for Love numbers in terms of the \textit{fluid-crust Love numbers} $(k_2^\circ,h_2^\circ)$.
A fluid-crust model has the same internal structure as the physical model except that the crust is fluid and of arbitrary density $\rho^\circ$ (in practice, $\rho^\circ$ is either equal to the ocean density or to the original crust density, see Section~\ref{FluidCrustModel}).
The gravitational and radial Love numbers are given by Eq.~(\ref{knhnstatic}) or, at degree two, by
\begin{eqnarray}
k_2 +1 &=& \frac{ k_2^\circ + 1 }{1 + \frac{3}{5} \, \xi^\circ \left( \left( k_2^\circ + 1 \right) \frac{\Lambda}{1+\Lambda} + K_\rho \right) } \, ,
\nonumber \\
h_2  &=& \frac{ h_2^\circ }{1 + \left( 1 + \frac{3}{5} \, \xi^\circ \, h_2^\circ \right) \Lambda + \Lambda_\chi + H_\rho } \, ,
\label{k2h2static}
\end{eqnarray}
where $\xi^\circ=\rho/\rho_b^\circ$ is the ocean-to-bulk density ratio of the fluid-crust model (Eq.~(\ref{defbarxicirc})) and $(K_\rho,H_\rho)$ are density corrections proportional to $\delta\rho$ (Eq.~(\ref{KHrho})).
These formulas are valid at ${\cal O}(\varepsilon)$ if the crust is soft, as is generally the case for large icy satellites (Section~\ref{TiltFactor}).
If the crust is hard as could be the case for medium-sized and small icy satellites, $\Lambda$ is of ${\cal O}(1)$ instead of ${\cal O}(\varepsilon)$ and the formulas are only valid at ${\cal O}(1)$, meaning that density corrections should be neglected.

Fluid-crust Love numbers are most easily computed with the propagator matrix approach.
The important thing to keep in mind is that the fluid-crust model does not have in general the same bulk density as the physical model (Eq.~(\ref{rho0})).
A useful two-layer model consists of a viscoelastic core-mantle below a homogeneous ocean (Eq.~(\ref{hn0visco})).
Another simple model is the three-layer body with an infinitely rigid mantle below two homogeneous fluid layers (Eq.~(\ref{h20threelayers})) which is a good toy model for ocean stratification.
Nothing forbids, however, to compute fluid-crust Love numbers with more complicated models allowing for compressibility and continuous density variation.

Benchmarking the membrane formulas with a numerical code shows that the predictions of $k_2$, $h_2$ and $\gamma_2$ have an error less than 1\%, 0.5\%, and 10\%, respectively, assuming that the crust thickness is less than 5\% of the surface radius.
These error bounds correspond to worst-case scenarios in which the crust-ocean density contrast is maximum or the rheology of the crust is mostly fluid-like (Section~\ref{AccuracyLove}).

Besides their essential role in tidal deformations, tidal Love numbers serve to parameterize tidal heating.
The membrane formulas for tidal Love numbers satisfy the micro-macro equivalence: the global heat flow from the whole body (proportional to \textit{Im}($k_2$)) is the sum of the global heat flows  from the crust (proportional to $|h_2|^2\mbox{\textit{Im}}(\Lambda)$) and from the core-mantle (proportional to \textit{Im}($k_2^\circ$)) (Section~\ref{MicroMacro}).

In a sense, the membrane approach factorizes the contributions from the crust and deep interior.
The influence of the latter can thus be studied through fluid-crust Love numbers instead of physical Love numbers.
In this way, one sees that the elasticity of the mantle increases the surface deformation by less than one percent, while the presence of a liquid core in Europa has an effect of less than two percents.
It is thus a good approximation to consider the core-mantle system as being infinitely rigid (Sections~\ref{RigidMantleApproximation}).
This approximation is even better for smaller icy satellites because the nondimensional shear modulus of the mantle is inversely proportional to the surface radius and surface gravity (Eq.~(\ref{defyximu})).
The rigid mantle approximation, however, is not essential.
For example, the viscoelastic effect of a large liquid core can be taken into account in the two-layer fluid-crust model through the \textit{equivalent shear modulus} of the core-mantle system (Section~\ref{InfluenceLiquidCore}).
In this paper, I emphasized the rigid mantle approximation in order to show that tidal effects in membrane worlds mainly depend on the shallow interior (crust and ocean).
As long as crust thickness and rheology are not better known, it is besides the point to use sophisticated models of the deep interior to study tectonics and tidal dissipation.

As is well-known, the deep interior structure affects Love numbers chiefly through global density stratification.
For a hydrostatic body, Love numbers decrease if mass is more concentrated (as measured by the moment of inertia) but density stratification in an ocean sandwiched between elastic layers has a more complex effect.
In particular, ocean stratification increases Love numbers if the mantle radius is large, especially if the ocean density abruptly increases near the bottom by large amount.
This \textit{screening effect} occurs because the higher density layer at the bottom of the ocean screens the gravitational braking of the mantle.
For Titan, a thin and dense liquid layer at the bottom of a light ocean can increase $k_2$ by more than ten percents (Section~\ref{OceanStratification}). 

The membrane approach is not limited to equilibrium tides.
If the mantle is infinitely rigid and the ocean is homogeneous and incompressible, membrane formulas for dynamical Love numbers reveal the existence of a dynamical resonance for very shallow oceans (Section~\ref{LoveDynamic}).
Alternatively, dynamical Love numbers can be computed either with the dynamical homogeneous crust model (Appendix~\ref{HomogeneousCrustModel}), or more generally with the semi-dynamical propagator matrix method, where `semi' means that dynamical effects are taken into account in the ocean but not in the solid layers (Appendix~\ref{DynamicalPropagationMatrix}).
On Europa, the resonance significantly decreases the tilt factor which becomes negative if the ocean is less than a few km thick (the exact threshold depends on the crust thickness).
Dynamical effects should thus be taken into account when using the tilt factor for crust thickness estimates.
More speculatively, the dynamical resonance could strongly increase tidal deformations and tidal heating in the crust if the ocean thickness is of the order of a few hundred meters, though the effect also depends on the existence of other resonances appearing once the rotation of the satellite is taken into account.

Beyond tidal Love numbers, the membrane approach is well suited to the computation of load Love numbers which characterize the response of a spherically symmetric body to surface loading (Eq.~(\ref{hklLoad})).
Load Love numbers are related to tidal Love numbers by Eqs.~(\ref{SaitoMolo})-(\ref{hnLoad}).
If the crust is incompressible and has a negligible density contrast with the ocean, the static load Love numbers of degree two are given by
\begin{eqnarray}
k_2' &=& \Lambda \, h_2 -1 \, ,
\nonumber \\
h_2' &=& \frac{\Lambda \, h_2 - 5/(3\xi) }{1+\Lambda} \, .
\label{LoveLoad2}
\end{eqnarray}
If the crust is thin and soft, these load Love numbers are close to $k_2'\sim-1$ and $h_2'\sim-5/(3\xi)$.
If the crust is thick, these formulas can be transformed into thick shell formulas by the correspondence $\Lambda\leftrightarrow{}z_h\hat\mu$ (Appendix~\ref{HomogeneousCrustModel}).
This procedure bridges the gap between membrane formulas and the better-known Kelvin-Love formulas which are valid for a homogeneous body.
Load Love numbers of low degree are useful for planetary reorientation \citep{matsunimmo2014}, atmospheric loading \citep{tokano2011}, and ocean loading \citep{sohl1995,tokano2014}.

For an easy implementation of the membrane formulas, I provide upon request the \textit{Mathematica} notebook \textit{MembraneWorlds.nb} which illustrates the membrane formulas with examples taken from this paper.

\section*{Acknowledgments}
This work was financially supported by the Belgian PRODEX program managed by the European Space Agency in collaboration with the Belgian Federal Science Policy Office.
I thank Isamu Matsuyama for valuable comments on the manuscript.


\newpage
\begin{appendices}

\section{Maxwell rheology}
\label{MaxwellRheology}
\renewcommand{\theequation}{A.\arabic{equation}} 
\setcounter{equation}{0}  

The material in this section summarizes Appendix~C of \citet{beuthe2014}.
Under the assumption of zero bulk dissipation (\textit{Im}(\textit{K})), viscoelastic parameters can be related to elastic constants, viscosity $\eta$ and angular frequency $\omega$ for any linear rheology.
Once the dependence of the shear modulus $\mu$ on $(\eta,\omega)$ is known, the viscoelastic Poisson's ratio $\nu$ can be computed with
\begin{equation}
\nu = \frac{\mu_E \left(1 + \nu_E \right)-\mu \left(1-2\nu_E \right) }{2\mu_E \left( 1 + \nu_E \right) + \mu \left( 1 - 2\nu_E \right)} \, ,
\label{nuVisc1}
\end{equation}
 in which the subscript $E$ stands for `elastic'.
This relation is valid for any linear rheology with zero bulk dissipation.
Note that elastic incompressibility ($\nu_E=1/2$) implies that $\nu=1/2$.

For Maxwell rheology with no bulk dissipation, the viscoelastic shear modulus and Poisson's ratio read
\begin{eqnarray}
\mu &=&  \frac{\mu_E}{1-i\delta} \, ,
\nonumber \\
\nu &=& \frac{3\,\nu_E-i \left(1+\nu_E \right) \delta }{3-2 i \left(1+\nu_E \right) \delta} \, ,
\label{munuMaxwell}
\end{eqnarray}
where the dimensionless number $\delta$ is defined by
\begin{equation}
\delta = \frac{\mu_E}{\omega\eta} \, .
\label{defdelta}
\end{equation}
If the viscosity is high, $\delta\sim0$ so that $\mu\sim\mu_E$ and $\nu\sim\nu_E$: this is the \textit{elastic-like regime}.
If viscosity is low, $\delta$ becomes large so that $\mu\sim{i}\mu_E/\delta\sim {i\omega\eta}$ and $\nu\sim1/2$: this is the \textit{fluid-like regime} (Fig.~1 of \citet{beuthe2014}).

The transition between the elastic- and fluid-like regimes occurs at $\delta\sim1$, defining the \textit{critical state} for which \textit{Im}($\mu$) is maximum.
Since the dissipated power per unit volume is proportional to \textit{Im}($\mu$), dissipation is nearly maximum in the critical state (`nearly', because the dissipated power also depends on $h_2$, see Eq.~(\ref{Edot1})).
The \textit{critical viscosity} associated with the critical state is given by
\begin{equation}
\eta_{crit} = \frac{\mu_E}{\omega} \, .
\label{etacrit}
\end{equation}

\section{Massless membrane approach}
\label{MasslessMembrane}
\renewcommand{\theequation}{B.\arabic{equation}} 
\setcounter{equation}{0}  

The massless membrane approach of \citet{beuthe2014} is based on the membrane theory of thin elastic shells.
With respect to the massive membrane approach developed in the present paper, the massless membrane approach makes three supplementary approximations:
\begin{enumerate}
\item membrane of zero thickness,
\item no density contrast at the crust-ocean boundary,
\item static approximation in all layers.
\end{enumerate}
The zero thickness assumption imposes that all terms of ${\cal O}(\varepsilon)$ are negligible in Eqs.~(\ref{y1crust})-(\ref{y6crust}) except those associated with the intrinsic properties of the membrane, which are its rigidity $\Lambda$ and its surface mass density $\delta\rho\times{}d$ ($\bar\rho$ is thus replaced by $\delta\rho$).
In this limit, Eqs.~(\ref{y1crust})-(\ref{y7crust}) become (after substitution of Eqs.~(\ref{snhn})-(\ref{dnhn}))
\begin{eqnarray}
g y_1(R_\varepsilon) &=&  h_n \, ,
\label{COB1} \\
\frac{1}{\rho} \, y_2(R_\varepsilon) &=&
- \left( \Lambda + \Lambda_\rho \right) h_n + \varepsilon \left( 2n+1 \right) \, \frac{\delta\rho}{\rho} \, ,
\label{COB2} \\
g y_3(R_\varepsilon) &=& l_n \, ,
\label{COB3} \\
y_4(R_\varepsilon) &=& 0 \,  ,
\label{COB4} \\
y_5(R_\varepsilon) &=&
k_n+1 - 3 \, \varepsilon \, \frac{\delta\rho}{\rho_b} \, h_n \, ,
\label{COB5} \\
R y_6(R_\varepsilon) &=&
2n+1-  3 \varepsilon \, \frac{\delta\rho}{\rho_b} \left( n^2-1 \right) \frac{n+1-\bar\nu}{x_n+1+\bar\nu} \, h_n \, ,
\label{COB6} \\
R y_7(R_\varepsilon) &=& 2 n+1
- 3  \varepsilon \, \frac{\rho}{\rho_b} \, \Lambda \, h_n
- 3  \varepsilon \, \frac{\delta\rho}{\rho_b}  \left( 2\left( n-1 \right) h_n - (2n+1) \right) .
\label{COB7}
\end{eqnarray}
One can show that these simpler membrane equations lead to the same formulas for Love numbers as the method assuming a massive membrane of finite thickness, except that the compressibility factor $\bar\chi$ is absent ($\Lambda_\chi=0$).
For example, the $k_n-h_n$ relation obtained with these equations is $k_n+1=(1+\Lambda+\Lambda_\rho+\Lambda_\omega)h_n-(2n+1)(\delta\rho/\rho)\varepsilon$.
Comparing Eq.~(\ref{COB1}) to Eq.~(\ref{y1crust}), one sees that the absence of the compressibility term in thin shell theory results from the assumption that the upper and lower shell surfaces deform in the same way.

If there is no density contrast at the crust-ocean boundary, the membrane must be massless ($\delta\rho=0$), otherwise the membrane would carry a surface mass density in the limit of zero thickness.
In this limit, Eqs.~(\ref{COB1})-(\ref{COB6}) tend to the \textit{membrane boundary conditions} of the massless membrane approach:
\begin{itemize}
\item Displacements are constant through the crust: $y_1(R_\varepsilon)=y_1(R)$ and $y_3(R_\varepsilon)=y_3(R)$.
\item Gravity perturbations are constant through the crust: $y_5(R_\varepsilon)=y_5(R)$ and $y_6(R_\varepsilon)=y_6(R)$.
\item Stresses at the crust-ocean boundary are characterized by $y_4(R_\varepsilon)=0$ and $y_2(R_\varepsilon)=-\rho\Lambda{}gy_1(R)$.
\end{itemize}
To get a physical picture of the last equation, I rewrite it as
\begin{equation}
q =\left( \rho  g  \Lambda \right) w  \, ,
\label{COB2lim}
\end{equation}
where $U$ is the tidal potential, $w=y_1(R_\varepsilon) U$ is the membrane deflection and $q=-y_2(R_\varepsilon)U$ is the load acting on the bottom of the crust (\citet{beuthe2014}, Section~5.1).
Eq.~(\ref{COB2lim}) has the form a Hooke's law, justifying the name of \textit{membrane spring constant} for $\Lambda$ (\citet{beuthe2014}, Section~4.3).
Eq.~(\ref{COB2}) shows how the Hooke's law (Eq.~(\ref{COB2lim})) of the massless membrane approach is modified by density corrections at ${\cal O}(\varepsilon)$.

Finally the static approximation means that $\omega=0$ and $\Lambda_\omega=0$.
The three supplementary assumptions of the massless membrane approach thus lead to $\Lambda_\chi=\Lambda_\rho=\Lambda_\omega=\delta\rho=0$ so that the $k_n-h_n$ relation (Eq.~(\ref{knhn})) reduces to $k_n+1=(1+\Lambda)h_n$, which is the $k_n-h_n$ relation of the massless membrane approach \citep[][Eq.~(34)]{beuthe2014}.

\section{Love numbers of fluid-crust models}
\label{FluidCrustModels}
\renewcommand{\theequation}{C.\arabic{equation}} 
\setcounter{equation}{0}  

In Section~\ref{FluidCrustModel}, the fluid-crust model is defined as having the same internal structure as the body with a viscoelastic crust (physical model), except that the crust is fluid and of density $\rho^\circ$.
The fluid-crust density $\rho^\circ$ is equal either to the original crust density ($\rho^\circ=\bar\rho$), or to the density of the top layer of the ocean ($\rho^\circ=\rho$).
The latter choice makes the fluid-crust model simpler, but it changes the bulk density of the body from $\rho_b$ to $\rho^\circ_b$.
I compute the fluid-crust Love numbers with the propagator matrix method \citep[e.g.][]{sabadini2004} in the Fourier domain.
This method is applicable if tides are static and the body is made of incompressible homogeneous layers.
In the fluid layer, I use the variables $(y_5,y_7)$ as defined by \citet{saito1974}.

\subsection*{C.1. Rigid mantle and surface ocean}

If $\rho^\circ=\rho$, the simplest fluid-crust model (next to the fluid body) is the two-layer incompressible body (bulk density $\rho^\circ_b$) with an infinitely rigid mantle and a surface ocean (density $\rho$):
\begin{equation}
h_{nr}^\circ = k_{nr}^\circ + 1= \left( 1-\frac{3}{2n+1} \, \xi^\circ \right)^{-1} \, ,
\label{LoveRigid}
\end{equation}
where $\xi^\circ$ is the ocean-to-bulk density ratio for the fluid-crust model (Eq.~(\ref{defbarxicirc})).
This well-known formula is easily derived by combining the hydrostatic $k_n^\circ-h_n^\circ$ relation (Eq.~(\ref{knhnhydrostat})) with the proportionality relation (Eq.~(\ref{knhnRigid0}) in which $\xi\rightarrow\xi^\circ$) as done for example in Section~4.5 of \citep{beuthe2014}.
Note that Eq.~(\ref{LoveRigid}) does not depend explicitly on mantle parameters (radius and density): the mantle can be either large and light or small and dense.
A useful identity is
\begin{equation}
h_{nr}^\circ = 1 + \frac{3}{2n+1} \, \xi^\circ \, h_{nr}^\circ \, .
\label{RigidIdentity}
\end{equation}

\subsection*{C.2. Viscoelastic mantle and surface ocean}

If $\rho^\circ=\rho$, a more general fluid-crust model is the two-layer incompressible body (bulk density $\rho^\circ_b$, radius $R$, surface gravity $g$) with viscoelastic core (radius $R_m$, shear modulus $\mu_m$) and surface ocean.
The Love numbers of this model are given by
\begin{equation}
h_n^\circ = k_n^\circ + 1 = 
\frac{ A_n(y,\xi^\circ) + \left(2n+1\right) y^4 \, \hat\mu_m }{ B_n(y,\xi^\circ)   + \left(2n+1-3\,\xi^\circ \right) y^4 \, \hat\mu_m } \, ,
\label{hn0visco}
\end{equation}
where
\begin{equation}
\left( y, \xi^\circ, \hat\mu_m \right) = \left( \frac{R_m}{R} , \frac{\rho}{\rho^\circ_b} , \frac{\mu_m}{\rho^\circ_b{g^\circ}R} \right) ,
\label{defyximu}
\end{equation}
where $g^\circ$ is the surface gravity of the fluid-crust model.
If the mantle is infinitely rigid, Eq.~(\ref{hn0visco}) reduces to Eq.~(\ref{LoveRigid}) as expected.
The functions $A_n$ and $B_n$ are defined by
\begin{eqnarray}
A_n(y,\xi) &=& f_n \left(2n+1\right) \left(1-\xi \right) p_{A}(y,\xi,n) \, ,
\nonumber \\
B_n(y,\xi) &=& f_n \left(1-\xi \right) p_{B}(y,\xi,n) \, ,
\label{constAB}
\end{eqnarray}
in which the superscript ${}^\circ$ is omitted so as to simplify the notation.
The factor $f_n$ is given by
\begin{equation}
f_n = \frac{n}{2 \left(n-1\right) \left(3+4n+2n^2\right)} \, ,
\end{equation}
while the functions $p_A$ and $p_B$ are defined by
\begin{eqnarray}
\hspace{-2cm} && p_{A}(y,\xi,n) =
\left( 2 \left( n-1 \right) + 3 y^{2n+1} \right) \left(1-\xi\right) + \left(2n+1\right) y^3 \, \xi \, ,
\nonumber \\
\hspace{-2cm} && p_{B}(y,\xi,n) =
\left( 2n+1-3\xi \right) \Big( 2\left(n-1\right)\left(1-\xi\right)+\left(2n+1\right) y^3\,\xi \Big)
- \,\, 9 \left(1-\xi\right) y^{2n+1} \xi \, .
\label{pAB}
\end{eqnarray}

The displacement of the mantle-ocean boundary is given by
\begin{equation}
h_n^m=f_n \, (2n+1)^2 \, \frac{y^{n+2} \left(1-\xi^\circ \right)}{ B_n(y,\xi^\circ )  +  \left( 2n+1-3\xi^\circ  \right) y^4 \, \hat\mu_m } \, ,
\label{hnmvisco}
\end{equation}
where $h_n^m=g{}y_1(R_m)$.
For tides of degree two, these formulas are identical to Eqs.~(118)-(120) of \citet{beuthe2014}.
The various limits of this model (fluid or rigid core, uniform density, shallow ocean) are discussed in Appendix~B of \citet{beuthe2014}.
For example, $h_2^\circ=5/2$ in the uniform density limit, in which case the mantle does not influence the surface deformation.

\subsection*{C.3. Rigid mantle, ocean and fluid crust}

If $\rho^\circ=\bar\rho$,
a simple fluid-crust model is the three-layer incompressible body (bulk density $\rho^\circ_b$) made of an infinitely rigid mantle, a homogeneous ocean (radius $R-d$, density $\rho$), and a homogeneous fluid crust (radius $R$, density $\bar\rho$).
The Love numbers of this model are
\begin{equation}
h_n^\circ = k_n^\circ + 1
= \left(2n+1\right) \frac{p_C(x,\xi^\circ,\bar\xi^\circ,n) - 3 \left( \bar\xi^\circ - \xi^\circ  \right) x^{2n+4}}{\left( 2n+1- 3 \bar\xi^\circ \right) p_C(x,\xi^\circ,\bar\xi^\circ,n) + 9 \, \bar\xi^\circ \left( \bar\xi^\circ - \xi^\circ  \right) x^{2n+4}} \, ,
\label{h20threelayers}
\end{equation}
where
\begin{equation}
\left( x, \xi^\circ, \bar\xi^\circ \right) = \left( \frac{R-d}{R} , \frac{\rho}{\rho^\circ_b} ,  \frac{\bar\rho}{\rho^\circ_b} \right) ,
\end{equation}
and
\begin{equation}
p_C(x, \xi^\circ, \bar\xi^\circ,n) =  \left(2n+1\right) \left( 1 - \left(1-x^3\right) \bar\xi^\circ \,  \right) + 3 \left( \bar\xi^\circ - \xi^\circ \right) x^3 \, .
\end{equation}
If the densities of the fluid crust and ocean are equal ($\bar\xi^\circ=\xi^\circ$) or if there is no fluid crust ($x=1$), Eq.~(\ref{h20threelayers}) reduces to Eq.~(\ref{LoveRigid}) as expected.
Similarly to Eq.~(\ref{LoveRigid}),  Eq.~(\ref{h20threelayers}) does not depend explicitly on mantle parameters (radius and density).

If the densities of the mantle and ocean are equal, the total mass equation yields $\xi^\circ{x^3}+\bar\xi^\circ(1-x^3)=1$.
Using this constraint to eliminate $\xi^\circ$ in favour of $\bar\xi^\circ$, one can show that Eq.~(\ref{h20threelayers}) reduces to $A_n(x,\bar\xi^\circ)/B_n(x,\bar\xi^\circ)$, i.e.\ the formula for a two-layer fluid body (Eq.~(\ref{hn0visco}) with $\hat\mu_m=0$).
This result is expected because Love numbers do not depend on the properties of the mantle if there is no density contrast at the mantle-ocean boundary: this is the screening effect discussed in Section~\ref{OceanStratification}.

\section{Comparison with \citet{wahr2006}}
\label{LoveFormulasRigid}
\renewcommand{\theequation}{D.\arabic{equation}} 
\setcounter{equation}{0}  

In Section~\ref{RigidMantleLimit}, I showed that the degree-two membrane formula for the tilt factor agrees with the analytical model of \citet{wahr2006} in the rigid mantle limit.
I will now do a similar comparison for the Love numbers themselves.
Recall that the membrane Love numbers are expressed in terms of the fluid-crust Love numbers.
If the fluid-crust density is equal to the ocean density, the fluid-crust model is an incompressible body of bulk density $\rho^\circ_b\neq\rho_b$ made of an infinitely rigid mantle below a homogeneous ocean of density $\rho$ reaching the surface (Eq.~(\ref{LoveRigid})).
The fluid-crust Love numbers represent the first term, of ${\cal O}(\varepsilon^0)$, in a perturbative expansion in the relative crust thickness $\varepsilon$.

\citet{wahr2006} also expand their solution about a two-layer model with infinitely rigid mantle and surface ocean.
However, they choose a slightly different configuration for the density distribution at ${\cal O}(\varepsilon^0)$.
More precisely, the bulk density of their zeroth-order configuration is equal to the bulk density of the physical model (i.e.\ $\rho_b$) instead of $\rho^\circ_b$.
This choice means the mantle density is implicitly adjusted in their zeroth-order configuration so as to compensate the density change due to the replacement of the crust by an ocean layer.
The Love numbers of their zeroth-order configuration are thus given by Eq.~(\ref{LoveRigid}) in which $\xi^\circ=\rho^\circ/\rho_b$ is replaced by $\xi=\rho/\rho_b$:
\begin{equation}
k_0 + 1 = h_0 = \left( 1 - \frac{3 \, \xi}{2n+1} \right)^{-1} .
\end{equation}
Using Eq.~(\ref{rho0}), I can express at ${\cal O}(\varepsilon)$ the Love numbers of the fluid-crust model in terms of $h_0$:
\begin{equation}
k_{nr}^\circ + 1 = h_{nr}^\circ = h_0 \left( 1 - \frac{3\, \xi}{2n+1} \left(3 \, \xi \, h_0 \right) \frac{\delta \rho}{\rho} \, \varepsilon \right)^{-1}  .
\label{knr0}
\end{equation}
Next,  I substitute Eq.~(\ref{knr0}) into Eq.~(\ref{knhnstatic}) and expand to ${\cal O}(\varepsilon)$ in the denominator.
Reexpressing the density terms  with the help of the identity (\ref{RigidIdentity}), I can write the Love numbers as 
\begin{eqnarray}
k_{nr} + 1 &=&  \frac{k_0 + 1}{ 1 + \frac{3\,\xi}{2n+1} \left( \frac{\Lambda}{1+\Lambda} \left( k_0 +1 \right) + K_0 \right) } ,
\nonumber \\
h_{nr} &=&  \frac{h_0 }{ 1 +  \Lambda \left( 1 + \frac{3\,\xi}{2n+1} \, h_0 \right) + \Lambda_\chi + H_0 } ,
\label{knhnrigidstatic}
\end{eqnarray}
where $K_0$ and $H_0$ are given by
\begin{eqnarray}
K_0 &=& - \left( 3 h_0 + 2n+1 \right) \frac{\delta\rho}{\rho} \, \varepsilon \, ,
\nonumber \\
H_0  &=& \Lambda_\rho - \left( 2n-2 + 3 h_0 \right) \frac{\delta\rho}{\rho} \, \varepsilon \, .
\label{K0H0}
\end{eqnarray}
For tides of degree two,
\begin{eqnarray}
K_0 &=& -  5 \, \frac{8 - 3\,\xi }{5-3\,\xi} \, \frac{\delta \rho}{\rho} \, \varepsilon \, ,
\nonumber \\
H_0 &=& 
- 3 \, \frac{ 40 - 3 \left( 3 - \bar\nu \right) \xi}{\left(5+\bar\nu\right) \left(5-3\,\xi \right)} \, \frac{\delta \rho}{\rho} \, \varepsilon \, .
 \label{K0H0deg2}
\end{eqnarray}
These equations generalize the results of \citet{wahr2006} to a compressible crust with depth-dependent rheology.
If the crust is incompressible ($\bar\nu=1/2$) and homogeneous ($\bar\mu=\mu$), the expansions to ${\cal O}(\varepsilon)$ of Eq.~(\ref{knhnrigidstatic}) with $(K_0,H_0)$ given by Eq.~(\ref{K0H0deg2}) agree with Eqs.~(3)-(6) of \citet{wahr2006}.

\section{Dynamical homogeneous crust model}
\label{HomogeneousCrustModel}
\renewcommand{\theequation}{E.\arabic{equation}} 
\setcounter{equation}{0}  

The homogeneous crust model describes a body in which (1) the crust is homogeneous and incompressible, and (2) there is a subsurface ocean, the top layer of which has the same density as the crust.
Otherwise, the structure below the crust can be freely chosen.
The static version of the model (for tides of degree two) is discussed in detail in Appendix~A of \citet{beuthe2014}.
I extend it here to dynamical tides.
In comparison with static tides, there is much less freedom in choosing the deep interior structure because one cannot express the Love numbers in terms of arbitrary fluid-crust models.
More complex models can be built with the dynamical propagator matrix method (Appendix~F).

As elsewhere, $(\rho,\rho_b,g,R,d,\mu)$ denote the ocean (and crust) density, bulk density, surface gravity, surface radius, crust thickness and shear modulus of the crust, respectively.
The Love numbers are expressed in terms of dimensionless ratios, three of which are
\begin{equation}
\left( x, \xi, \hat\mu \right) = \left( \frac{R-d}{R} , \frac{\rho}{\rho_b} , \frac{\mu}{\rho{g}R} \right) ,
\label{ratios}
\end{equation}
the fourth one being the dynamical correction $\Lambda_\omega$ (Eq.~(\ref{LambdaOmega})).

I obtain relations between Love numbers with the propagator matrix method as explained in Appendix~A of \citet{beuthe2014}: I propagate the $y_i$ from the surface down to the crust-ocean boundary, where I apply the free-slip condition (Eq.~(\ref{freeslip})) and the dynamical fluid constraint (Eq.~(\ref{fluideqDyn})).
The $k_2-h_2$ and $l_2-h_2$ relations read
\begin{eqnarray}
k_2 +1 &=& \left( 1 + z_h \, \hat \mu \right) h_2 \, ,
\nonumber \\
l_2 &=& z_l \, h_2 \, ,
\label{relationsHC}
\end{eqnarray}
where
\begin{eqnarray}
z_h  &=& \frac{X_a \, \hat\mu + X_b \, \Lambda_\omega}{\hat\mu + X_e \, \Lambda_\omega} \, ,
\nonumber \\
z_l &=&  \frac{X_c \, \hat\mu + X_d \, \Lambda_\omega}{\hat\mu + X_e \, \Lambda_\omega} \, ,
\label{zhl}
\end{eqnarray}
in which $X_j$ are geometrical factors depending only on $x$ (Table~\ref{TableGeoFac}).

Load Love numbers satisfy similar relations:
\begin{eqnarray}
k_2' + 1 &=& \left( 1 + z_h \, \hat \mu \right) h_2' + \frac{5}{3\xi} \, ,
\nonumber \\
l_2' &=& z_l \, h_2' \, ,
\label{relationsHCLoad}
\end{eqnarray}
The gravitational Load number is related to the tidal Love numbers by the Saito-Molodensky relation (Eq.~(\ref{SaitoMolo})) which can be combined with the $k_2-h_2$ and $k_2'-h_2'$ relations in order to express $(k_2',h_2')$ in terms of $h_2$.
Load and tidal Love numbers are thus related by
\begin{eqnarray}
k_2' &=& z_h \, \hat \mu \, h_2 -1 \, ,
\nonumber \\
h_2' &=& \frac{z_h \, \hat \mu \, h_2 - 5/(3\xi) }{1+z_h \, \hat \mu} \, .
\label{LoveLoad2Thick}
\end{eqnarray}
Alternatively, these equations can be obtained from the membrane formulas (Eq.~(\ref{LoveLoad2})) by the substitution $\Lambda\rightarrow{}z_h\hat\mu$, as already noted in \citet{beuthe2014} (see his Table~8 and Appendix~A).

In the thin shell limit, the factors appearing in the Love number relations behave to leading order as
\begin{eqnarray}
z_h \hat\mu &\sim&  \Lambda + \Lambda_\omega \, ,
\nonumber \\
z_l &\sim& \frac{3}{11} \left( 1 + \frac{\Lambda_\omega}{6\hat\mu} \right) \, ,
\label{zhlapprox}
\end{eqnarray}
where $\Lambda=(24/11)\hat\mu\varepsilon$ is the incompressible membrane spring constant (Eq.~(\ref{springconst})).

\begin{table}[ht]\centering
\ra{1.3}
\small
\caption[Geometrical factors in the homogeneous crust model]{\small
Geometrical factors $X_j$ appearing in the Love number relations (Eq.~(\ref{zhl})).
In each case, the polynomial in the denominator is $P=24 + \, 40 \, x^3 - 45 \, x^7 - 19 \, x^{10}$.
The polynomials appearing below have a simple or double root at $x=1$, but factorizing them yields longer expressions.
`Thin shell' denotes the thin shell limit ($\varepsilon=1-x=d/R$); the coefficients of $\hat\mu$ and $\Lambda_\omega$ are expanded to next-to-leading and leading orders in $\varepsilon$, respectively.
`Homog.' denotes the homogeneous limit in which the shell fills the whole body, i.e.\ $x=0$.
The factors $X_a$ and $X_c$ are shown as functions of $x$ in Fig.~13 of \citet{beuthe2014}.
}
\begin{tabular}{@{}llll@{}}
\hline
 & Thick shell &  Thin shell & Homog.
 \\
\hline
$X_a$ & $\frac{24}{5} \left(19 - 75 \, x^3 + 112 \, x^5 - 75 \, x^7 + 19 \, x^{10}\right)/P$ & $\frac{24}{11} \, \varepsilon \left( 1+ \frac{4}{11} \, \varepsilon \right)$ & $\frac{19}{5}$
\vspace{1mm}  \\
$X_b$ & $x^2 \left(19+45x^3-40x^7-24x^{10}\right)/P$ & $1$ & $0$
\vspace{1mm}  \\
$X_c$ & $\frac{1}{5} \left( 36 - 100 \, x^3 + 308 \, x^5 - 225 \, x^7 - 19 \, x^{10}\right)/P$ & $\frac{3}{11}\left(1-\frac{32}{33} \, \varepsilon \right)$ & $\frac{3}{10}$
\vspace{1mm}  \\
$X_d$ & $\frac{1}{2} \, x^2\left(3+5x^3-10x^7+2x^{10}\right)/P$ & $\frac{1}{22}$ & $0$
\vspace{1mm}  \\
$X_e$ & $5 \, x^2\left(1-x^3-x^7+x^{10}\right)/P$ & $\frac{3}{11} \, \varepsilon$ & $0$
\vspace{1mm}  \\
\hline
\end{tabular}
\label{TableGeoFac}
\end{table}%

Some assumptions about the deep interior are required in order to determine $\Lambda_\omega$.
In the rigid mantle model, $\Lambda_\omega$ is given by Eq.~(\ref{LambdaOmegaRigid})
and $k_2$ is proportional to $h_2$ (Eq.~(\ref{knhnRigid0})).
Combining the latter constraint with the $k_2-h_2$ relation, I get explicit formulas for dynamical Love numbers:
\begin{equation}
\left( k_2 \, , h_2 \, , k_2' \, , h_2' \right) = \frac{1}{ 1 + h_{2r}^\circ \, z_h \, \hat \mu } \left( k_{2r}^\circ \, , h_{2r}^\circ \, , -1 \, , -\frac{5}{3\xi} \right) ,
\label{LoveRigidHCR}
\end{equation}
where $k_{2r}^\circ+1=h_{2r}^\circ=5/(5-3\xi)$ (Eq.~(\ref{LoveRigid})).
Alternatively, one can derive this formula with the dynamical propagator matrix method (Appendix~F).

Additionally, if tides are static and the body is of uniform density, $z_h=X_a$ and $h_2^\circ=k_2^\circ+1=5/2$.
Such a model describes a two-layer body made of a liquid core surrounded by a solid layer:
\begin{equation}
\left( k_2 \, , h_2 \, , k_2' \, , h_2' \right) = \frac{1}{ 1 + \frac{5}{2} \, X_a \, \hat \mu } \left( \frac{3}{2} \, , \frac{5}{2} \, -1 \, , -\frac{5}{3}\right) .
\label{LoveRigidHCH}
\end{equation}
This equation reduces to the well-known Kelvin-Love formula if the solid layer extends to the center of the body ($X_a=19/5$ if $x=0$):
\begin{equation}
\left( k_2 \, , h_2 \, , k_2' \, , h_2' \right) = \frac{1}{ 1 + \frac{19}{2} \, \hat \mu }\left( \frac{3}{2} \, , \frac{5}{2} \, -1 \, , -\frac{5}{3}\right)  ,
\label{KelvinLove}
\end{equation}
as quoted by \citet{lambeck1980} (his Eq.~(2.1.9)).

\section{Dynamical propagator matrix}
\label{DynamicalPropagationMatrix}
\renewcommand{\theequation}{F.\arabic{equation}} 
\setcounter{equation}{0}  

In the propagator matrix method, one usually assumes that tides are static and the body is incompressible.
If either of these assumptions does not hold, the elastic-gravitational problem can be solved in terms of spherical Bessel functions if $g_r/r$ is constant within the body, $g_r$ being the gravitational acceleration at radius $r$.
\citet{love1911} and \citet{pekeris1958} computed in this way the deformations of a compressible body of uniform density.
Assuming that $g_r/r$ is approximately constant inside each layer, \citet{gilbert1968} proposed a propagator matrix method, in the Fourier domain, for the oscillations and dynamical tides of an elastic stratified compressible body.
\citet{vermeersen1996} later considered the viscoelastic problem in the Laplace domain in order to model deformations at the geological time scale (see also Appendix~A of \citet{sabadini2004}).
This method has not become popular for several reasons.
First, the approximation of constant $g_r/r$ in each layer introduces errors that possibly offset any gain due to the inclusion of compressibility, at least in models with few layers (benchmarking has yet to be done, as observed by \citet{vermeersen1996}).
Second, spherical Bessel functions are a bit tricky to use, the more so in the viscoelastic generalization of the problem.
Third, Bessel functions obscure the dependence of Love numbers on physical parameters, so that the analytical formulas cannot be interpreted without being first numerically evaluated.
Fourth, the static and incompressible limits are problematic so that comparisons with simpler cases are difficult.
In conclusion, the benefits brought by this rather complicated analytical method are not obvious: you can as well resort to numerical integration.

By contrast, the dynamical solution within a homogeneous incompressible liquid layer takes a very simple power form (Section~\ref{DynIncompLiqLayer}).
Since dynamical effects are much smaller in solid than in liquid layers, I propose to use the static and dynamical solutions within the solid and liquid layers, respectively.
The static propagator matrix appears in the literature (e.g.\ \citet{sabadini2004} with other $y_i$ conventions) but the dynamical propagator matrix for an incompressible liquid layer does not seem to have been written down before.
This propagator matrix does not involve the variables $y_3$ and $y_4$: the former is discontinuous at the fluid-solid interfaces and is considered as a dependent variable (through Eq.~(\ref{fluideqDyn})) whereas the latter vanishes everywhere within the fluid. 
Combining Eq.~(\ref{y5dyn}) with Eqs.~(\ref{EG6}) and (\ref{fluideqDyn}) gives the following propagation formula: 
\begin{equation}
\left( y_1 \, , \, y_2 \, , \, y_5 \, , \, y_6 \right)^T = \mathbf{Y}_{liq} \cdot \left( a \, , \, \alpha \, , \, b \, , \, \beta \right)^T \, ,
\label{liqdynpropagApp}
\end{equation}
where $(a,\alpha,b,\beta)$ are free constants.
In the conventions of \citet{sabadini2004}, the left-hand side is replaced by $(y_1,y_3,-y_5, -y_6)^T$.
$\mathbf{Y}_{liq}$ is the $4\times4$ propagator matrix defined by
\begin{equation}
\mathbf{Y}_{liq} =
\mathbf{\widehat  Y}_{liq}
\cdot
\mathbf{C}_{liq}
\, ,
\label{dill}
\end{equation}
where $\mathbf{C}_{liq}$ is a diagonal matrix given by
\begin{equation}
\mathbf{C}_{liq} = {\rm diag} \left( r^{n-1} \, , \, r^n \, , \, r^{-n-2} \, , \, r^{-n-1} \right) ,
\end{equation}
while
$\mathbf{\widehat Y}_{liq}$ is given by
\begin{equation}
\mathbf{\widehat Y}_{liq} =
\left(
\begin{array}{cccc}
1 & 0 & 1 & 0 \\
\rho\left(g_r-\frac{\omega^2r}{n}\right) & -\rho & \rho\left(g_r+\frac{\omega^2r}{n+1}\right) & -\rho \\
 0 & 1 & 0 & 1 \\
-4\pi{}G\rho & \frac{2n+1}{r} & -4\pi{}G\rho & 0
\end{array}
\right) ,
\label{dillbar}
\end{equation}
where $\omega$ is the tidal angular frequency, $\rho$ is the uniform density of the layer, and $g_r$ is the unperturbed gravitational acceleration at radius $r$.

The inverse propagator matrix reads
 \begin{equation}
\mathbf{Y}_{liq}^{-1} =
\mathbf{D}_{liq}
\cdot
\mathbf{\overline Y}_{liq}
\, ,
\label{dillinv}
\end{equation}
where $\mathbf{D}_{liq}$ is a diagonal matrix given by
\begin{equation}
\mathbf{D}_{liq} =  \frac{1}{2n+1} \, {\rm diag} \left( \frac{n(n+1)}{\rho \, \omega^2} \, r^{-n} \, , \, r^{-n+1} \, , \, \frac{n(n+1)}{\rho \, \omega^2} \, r^{n+1} \, , \, r^{n+2} \right) ,
\end{equation}
while
$\mathbf{\overline Y}_{liq}$ is given by
\begin{equation}
\mathbf{\overline Y}_{liq} =
\left(
\begin{array}{cccc}
\rho\left(g_r+\frac{\omega^2r}{n+1}\right) & -1 & -\rho & 0 \\
 4\pi{}G\rho & 0 & 0 & 1 \\
-\rho\left(g_r-\frac{\omega^2r}{n}\right) & 1 & \rho & 0 \\
-4\pi{}G\rho & 0 & \frac{2n+1}{r} & -1
\end{array}
\right) .
\label{dillinvbar}
\end{equation}
As an illustration, consider a three-layer body with viscoelastic core, ocean and crust.
The viscoelastic-gravitational problems depends on 13 free constants (3 in the core, 4 in the ocean and 6 in the crust).
Besides the 3 surface boundary conditions (Eq.~(\ref{boundcond})), there are 8 constraints due to the continuity of the variables $(y_1,y_2,y_5,y_6)$ at the core-ocean and crust-ocean boundaries and 2 constraints due to $y_4=0$ at the same boundaries.
The problem can thus be solved by propagating the general solution from the center to the surface and applying the various boundary conditions \citep[e.g.][]{roberts2008}.
If the ocean density increases with depth, one can discretize the ocean in layers of uniform density.

\section{$k_n-h_n$ proportionality}
\label{knhnproportionality}
\renewcommand{\theequation}{G.\arabic{equation}} 
\setcounter{equation}{0}  

The rigid mantle model is a three-layer body made of an infinitely rigid mantle, a homogeneous and incompressible ocean, and a crust.
In this model, the deep interior does not contribute to the gravity perturbation: $k_n$ is only due to the crust deformation and to the ocean bulge just below the crust.
If the crust is incompressible and has the same density $\rho$ as the ocean, the induced geoid perturbation is proportional to the surface deformation \citep{beuthe2014}:
\begin{equation}
k_{nr} = \frac{3\,\xi}{2n+1} \, h_{nr}  \, ,
\label{knhnRigid0}
\end{equation}
where $\xi=\rho/\rho_b$ as before and the subscript $r$ denotes the rigid mantle model.

What is the analogue of Eq.~(\ref{knhnRigid0}) if the crust is compressible and does not have the same density as the ocean?
Although there is no similar relation for a thick crust, I will show how to extend Eq.~(\ref{knhnRigid0}) if the crust is thin.
For this purpose, I evaluate $y_6$ in terms of $(y_1,y_5)$ on the ocean side of the crust-ocean boundary with Eq.~(\ref{EG5}) in which I substitute Eq.~(\ref{y5primey5}):
\begin{equation}
y_6(R_\varepsilon) = \frac{2n+1}{R(1-\varepsilon)} \, y_5(R_\varepsilon) - \frac{3\,\xi}{R} \, gy_1(R_\varepsilon) \, .
\end{equation}
Since $(y_1,y_5,y_6)$ are continuous across the crust-ocean boundary, I can replace these variables by their values on the crust side of the boundary (Eqs.~(\ref{y1crust})-(\ref{y6crust})), before substituting Eqs.~(\ref{snhn})-(\ref{dnhn}) for $s_n$ and $\delta_n$.
Expanding to ${\cal O}(\varepsilon)$, I obtain the analog of Eq.~(\ref{knhnRigid0}) for a massive membrane of finite thickness:
\begin{equation}
k_{nr} = \frac{3\,\xi}{2n+1} \left( 1 + \Lambda_\chi + \Lambda_\rho + 3 \, \frac{\delta\rho}{\rho} \, \varepsilon \right) h_{nr} \,  ,
\label{knhnRigid1}
\end{equation}
where $\Lambda_\chi$ and $\Lambda_\rho$ are defined by Eqs.~(\ref{LambdaChi})-(\ref{LambdaRho}).

As a last step, I express the $k_n-h_n$ proportionality in terms of the Love numbers of the fluid-crust model, with $\rho^\circ=\rho$ (implying $\delta\rho^\circ=\delta\rho$).
Using Eq.~(\ref{rho0}), I rewrite Eq.~(\ref{knhnRigid1}) so that it depends on $\xi^\circ=\rho/\rho^\circ_b$ instead of $\xi=\rho/\rho_b$.
This substitution is only necessary in the term of ${\cal O}(1)$ because $\xi$ and $\xi^\circ$ are interchangeable in terms of ${\cal O}(\varepsilon)$.
The $k_n-h_n$ proportionality becomes
\begin{equation}
k_{nr} = \frac{3\, \xi^\circ}{2n+1} \left( 1 + \Lambda_\chi + \Lambda'_\rho \right) h_{nr} \,  ,
\label{knhnRigid2}
\end{equation}
where
\begin{equation}
\Lambda'_\rho = \Lambda_\rho + 3 \left( 1- \xi \right) \frac{\delta\rho}{\rho} \, \varepsilon \, .
\label{defLambdarhoprime}
\end{equation}
Finally I substitute Eq.~(\ref{LoveRigid}) into Eq.~(\ref{knhnRigid2}) so that $h_{nr}^\circ$ appears explicitly.
In its final form, the $k_n-h_n$ proportionality reads
\begin{equation}
k_{nr} = \left( 1 - (h_{nr}^\circ)^{-1}\right) \left( 1 + \Lambda_\chi + \Lambda'_\rho \right) h_{nr} \,  .
\label{knhnRigid3}
\end{equation}
Though the induced geoid perturbation is still proportional to the radial displacement, Eqs.~(\ref{knhnRigid0}) and (\ref{knhnRigid3}) differ by two types of terms:
(1) $\Lambda_\chi$ is the compressibility correction appearing when the membrane is of finite thickness, and (2) $\Lambda'_\rho$ is the correction due to the density contrast at the crust-ocean boundary.

\end{appendices}


\renewcommand{\baselinestretch}{0.5}

\end{document}